\documentclass{llncs} 
\usepackage{llncsdoc} 
\usepackage{times}
\usepackage{amssymb}
\usepackage{amsmath}
\usepackage{epsfig}
\usepackage{algorithm}
\usepackage{algorithmic}
\usepackage{multirow}
\usepackage{threeparttable}
\usepackage{boxedminipage}
\usepackage{slashbox}
\usepackage{subfigure}

\newcommand{\topktuple}[2]{top_{#1, #2}}
\newcommand{\ptopkset}[2]{Ans_{#1, #2}}
\newcommand{\ptopktuple}[2]{all_{#1, #2}}

\newcommand{\rollback}{\emph{Rollback}}
\newcommand{\rollbacksort}{\emph{RollbackSort}}
\newcommand{\simplebasic}{\emph{Basic}}
\newcommand{\simpleta}{\emph{TA}}
\newcommand{\generalbasic}{\emph{Reduction}}

\newlength{\figwidth}
\begin{document}

\title{Semantics and Evaluation of Top-k Queries in Probabilistic Databases\thanks{Research partially supported by NSF grant IIS-0307434. An earlier version of some of the results in this paper was presented in \cite{DBLP:conf/icde/ZhangC08}.}}

\author{Xi Zhang \and Jan Chomicki}

\institute{Department of Computer Science and Engineering \\
University at Buffalo, SUNY, U.S.A. \\
\email{\{xizhang,chomicki\}@cse.buffalo.edu}}

\maketitle

\begin{abstract}
We study here  fundamental issues involved in top-$k$ query evaluation in probabilistic databases. We consider {\em simple} probabilistic databases in which probabilities are associated with individual tuples, and {\em general} probabilistic databases in which, additionally, exclusivity
relationships between tuples can be represented. In contrast to other recent research in this area, we do not limit ourselves to injective scoring functions.
We formulate three intuitive postulates for the semantics of top-$k$ queries in probabilistic databases, and introduce a new semantics, Global-Top$k$, that satisfies those postulates to a large degree. We also show how to evaluate queries under the Global-Top$k$ semantics. For simple databases we design dynamic-programming based algorithms. For general databases we show polynomial-time reductions to the simple cases, and provide effective heuristics to speed up the computation in practice. For example, we demonstrate that for a fixed $k$ the time complexity of top-$k$ query evaluation is as low as linear, under the assumption that probabilistic databases are simple and scoring functions are injective.
\end{abstract}

\section{Introduction}
The study of incompleteness and uncertainty in databases has long been an interest of the database community \cite{DBLP:journals/jacm/ImielinskiL84,DBLP:conf/vldb/CavalloP87,DBLP:journals/ai/Halpern90,abiteboul94databases,DBLP:journals/tois/FuhrR97,DBLP:journals/tcs/Zimanyi97,DBLP:journals/tods/LakshmananLRS97}.
Recently, this interest has been rekindled by an increasing demand for managing rich data, often incomplete and uncertain, emerging from scientific data management, sensor data management, data cleaning, information extraction etc.  \cite{DBLP:journals/vldb/DalviS07} focuses on query evaluation in traditional probabilistic databases; ULDB \cite{DBLP:conf/vldb/BenjellounSHW06} supports uncertain data and data lineage in Trio \cite{DBLP:conf/cidr/Widom05};  MayBMS \cite{MayBMS} uses the vertical World-Set representation of uncertain data \cite{DBLP:journals/tcs/OlteanuKA08}. The standard semantics adopted in most works is the \emph{possible worlds} semantics \cite{DBLP:journals/jacm/ImielinskiL84,DBLP:journals/tois/FuhrR97,DBLP:journals/tcs/Zimanyi97,DBLP:conf/vldb/BenjellounSHW06,DBLP:journals/vldb/DalviS07,DBLP:journals/tcs/OlteanuKA08}.

On the other hand, since the seminal papers of Fagin \cite{DBLP:journals/jcss/Fagin99,DBLP:conf/pods/FaginLN01}, the top-$k$ problem has been extensively studied in multimedia databases \cite{DBLP:conf/vldb/NatsevCSLV01}, middleware systems \cite{DBLP:journals/tods/MarianBG04}, data cleaning \cite{DBLP:conf/vldb/GuhaKMS04}, core technology in relational databases \cite{DBLP:conf/vldb/IlyasAE02,DBLP:conf/vldb/IlyasAE03} etc.
In the top-$k$ problem, each tuple is given a \emph{score}, and users are interested in $k$ tuples with the highest scores.

More recently, the top-$k$ problem has been studied in probabilistic databases \cite{DBLP:conf/icde/SolimanIC07,DBLP:journals/tods/SolimanIC08,DBLP:conf/icde/ReDS07}.
Those papers, however, are solving two essentially different top-$k$ problems. Soliman et al. \cite{DBLP:conf/icde/SolimanIC07,DBLP:journals/tods/SolimanIC08} assumes the existence of a scoring function to rank tuples. 
Probabilities provide information on how likely tuples will appear in the database.  In contrast, in  \cite{DBLP:conf/icde/ReDS07}, the ranking criterion for top-$k$ is the probability associated with each query answer. 
In many applications, it is necessary to deal with tuple probabilities and scores at the same time. Thus, in this paper, we use the model of  \cite{DBLP:conf/icde/SolimanIC07,DBLP:journals/tods/SolimanIC08}. Even in this model, different semantics for top-$k$ queries are possible, so a part of the challenge is to categorize different semantics.

As a motivating example, let us consider the following graduate admission example.

\begin{example}\label{expl_hiring}
A graduate admission committee needs to select two winners of a fellowship. They narrow the candidates down to the following short list:

\smallskip
\begin{center}
\begin{tabular}{cc}
\begin{tabular}{|c|c|}
\hline
\small{Name}
&\small{Overall Score}\\
\hline
\small{Aidan} & $0.65$ \\
\hline
\small{Bob} & $0.55$ \\
\hline
\small{Chris} & $0.45$ \\
\hline
\end{tabular}
\begin{tabular}{c}
\small{Prob. of Coming} \\
\hline
$0.3$\\
$0.9$\\
$0.4$
\end{tabular}
\end{tabular}
\end{center}
\smallskip
where the {\em overall score} is the normalized score of each candidate based on their qualifications, and the {\em probability of acceptance} is derived from historical statistics on candidates with similar qualifications and background.

The committee want to make offers to the best two candidates who will take the offer. This decision problem can be formulated as a top-$k$ query over the above probabilistic relation, where $k=2$.
\end{example}

In Example \ref{expl_hiring}, each tuple is associated with an \emph{event}, which is that the candidate will accept the offer. The probability of the event is shown next to each tuple. In this example, all the events of tuples are independent, and tuples are therefore said to be \emph{independent}.
Such a relation is said to be \emph{simple}. In contrast, Example \ref{expl_sensor} illustrates a more general case.

\begin{example}\label{expl_sensor}In a sensor network deployed in a habitat, each sensor reading comes with a confidence value \emph{Prob}, which is the probability that the reading is valid. The following table shows the temperature sensor readings at a given sampling time. These data are from two sensors, Sensor 1 and Sensor 2, which correspond to two \emph{parts} of the relation, marked $C_1$ and $C_2$ respectively. Each sensor has only one \emph{true} reading at a given time, therefore tuples from the same part of the relation correspond to exclusive events.
\smallskip
\begin{center}
\begin{tabular}{ccc}
 & \multirow{5}{*}{
\begin{tabular}{|c|}
\hline
Temp.$^{\circ}$F (Score)\\
\hline
$22$\\
$10$\\
\hline
$25$\\
$15$\\
\hline
\end{tabular}
}& \multirow{5}{*}{
\begin{tabular}{c}
Prob\\
\hline
$0.6$\\
$0.4$\\
$0.1$\\
$0.6$
\end{tabular}
}\\
\multirow{2}{*}{$C_1$} & & \\
& &\\
\multirow{2}{*}{$C_2$} & & \\
& & 
\end{tabular}
\end{center}
\smallskip

Our question is: ``What is the temperature of the warmest spot?''

The question can be formulated as a top-$k$ query, where $k=1$, over a probabilistic relation containing the above data. The scoring function is the temperature. However, we must take into consideration that the tuples in each part $C_i, i=1,2$, are exclusive.
\end{example}

Our contributions in this paper are the following:
\begin{itemize}
\item[$\bullet$] We formulate three intuitive semantic postulates and use them to analyze and categorize different top-$k$ semantics in probabilistic databases (Section \ref{sec_postulates});
\item[$\bullet$] We propose a new semantics for top-$k$ queries in probabilistic databases, called Global-Top$k$, which satisfies the above postulates to a large degree (Section \ref{sec_globaltopksem});
\item[$\bullet$] We exhibit polynomial algorithms for evaluating top-$k$ queries under the Global-Top$k$ semantics in \emph{simple} probabilistic databases (Section \ref{topk_TO_Ind}) and general probabilistic databases, under injective scoring functions (Section \ref{topk_TO_IndEx}).
\item[$\bullet$] We generalize Global-Top$k$ semantics to general scoring functions, where ties are allowed, by introducing the notion of {\em allocation policy}. We propose dynamic programming based algorithms for query evaluation under the {\em Equal} allocation policy (Section \ref{sec_WO}). 
\item[$\bullet$] We provide theoretical time/space analysis for the algorithms proposed. In some cases, we design efficient heuristics to improve the performance of the basic algorithms (Section \ref{sec_TA}, Section \ref{sec_r_rs}). Experiments are carried out to demonstrate the efficacy of those optimizations (Section \ref{sec_exp}).
\end{itemize}

\section{Background}\label{sec_background}

\subsection{Probabilistic Relations}
To simplify the discussion in this paper, we assume that a probabilistic database contains a single \emph{probabilistic relation}. We refer to a traditional database relation as a \emph{deterministic relation}. A deterministic relation $R$ is a set of tuples. A \emph{partition} $\mathcal{C}$ of $R$ is a collection of non-empty subsets of $R$ such that every tuple belongs to one and only one of the subsets. That is, $\mathcal{C}=\{C_1, C_2,\ldots, C_m\}$ such that $C_1\cup C_2\cup\ldots\cup C_m=R$ and $C_i\cap C_j=\emptyset, 1\leq i\neq j\leq m$. Each subset $C_i, i=1,2,\ldots, m$ is a \emph{part} of the partition $\mathcal{C}$. A \emph{probabilistic relation} $R^p$ has three components, a \emph{support (deterministic) relation} $R$, a probability function $p$ and a partition $\mathcal{C}$ of the support relation $R$. The probability function $p$ maps every tuple in $R$ to a probability value in $(0,1]$. The partition $\mathcal{C}$ divides $R$ into subsets such that the tuples within each subset are exclusive and therefore their probabilities sum up to at most $1$. In the graphical presentation of $R$, we use horizontal lines to separate tuples from different parts.

\begin{definition}[Probabilistic Relation]\label{def_probdb}
A probabilistic relation $R^p$ is a triplet $\langle R, p, \mathcal{C}\rangle$, where $R$ is a support deterministic relation, $p$ is a probability function $p:R\mapsto (0,1]$ and $\mathcal{C}$ is a partition of $R$ such that
$\forall C_i\in \mathcal{C}, \sum_{t\in C_i}{p(t)}\leq 1$.
\end{definition}

In addition, we make the assumption that tuples from different parts of of $\mathcal{C}$ are independent, and tuples within the same part are exclusive. Definition \ref{def_probdb} is equivalent to the model used in Soliman et al. \cite{DBLP:conf/icde/SolimanIC07,DBLP:journals/tods/SolimanIC08} with exclusive tuple generation rules.  R\'e et al. \cite{DBLP:conf/icde/ReDS07} proposes a more general model, 
however only a restricted model with a fixed scoring function is used in top-$k$ query evaluation.

Example \ref{expl_sensor} shows an example of a probabilistic relation whose partition has two parts. 
Generally, each part corresponds to a real world entity, in this case, a sensor. 
Since there is only one true state of an entity, tuples from the same part are exclusive. 
Moreover, the probabilities of all possible states of an entity sum up to at most $1$. 
In Example \ref{expl_sensor}, the sum of the probabilities of tuples from Sensor $1$ is $1$, while that from Sensor $2$ is $0.7$. 
This can happen for various reasons. In the above example, we might encounter a  physical difficulty in collecting the sensor data, and end up with partial data. 

\begin{definition}[Simple Probabilistic Relation]
A probabilistic relation $R^p=\langle R, p, \mathcal{C}\rangle$ is simple iff the partition $\mathcal{C}$ contains only singleton sets.
\end{definition}

The probabilistic relation in Example \ref{expl_hiring} is {\em simple} (individual parts not illustrated). Note that in this case, $|R|=|\mathcal{C}|$.

We adopt the well-known \emph{possible worlds} semantics for probabilistic relations \cite{DBLP:journals/jacm/ImielinskiL84,DBLP:journals/tois/FuhrR97,DBLP:journals/tcs/Zimanyi97,DBLP:conf/vldb/BenjellounSHW06,DBLP:journals/vldb/DalviS07,DBLP:journals/tcs/OlteanuKA08}.

\begin{definition}[Possible World] Given a probabilistic relation $R^p=\langle R, p, \mathcal{C}\rangle$,
a deterministic relation $W$ is a \emph{possible world} of $R^p$ iff
\begin{enumerate}
\item $W$ is a subset of the support relation, i.e., $W\subseteq R$;

\item For every part $C_i$ in the partition $\mathcal{C}$, at most one tuple from $C_i$ is in $W$, i.e., $\forall C_i\in \mathcal{C}, |C_i\cap W|\leq 1$;

\item The probability of $W$ (defined by Equation (\ref{eqn_pwdprob})) is positive, i.e., $Pr(W)>0$.

\begin{equation}\label{eqn_pwdprob}
Pr(W)=\prod_{t\in W}p(t)
\prod_{C_i\in \mathcal{C'}}(1-\sum_{t\in C_i}p(t))
\end{equation}
\noindent where $\mathcal{C'}=\{C_i\in \mathcal{C} | W \cap C_i=\emptyset\}$.

\end{enumerate}
\end{definition}

Denote by $pwd(R^p)$ \emph{the set of all possible worlds} of $R^p$. 



\subsection{Total order v.s. Weak order}

A binary relation $\succ$ is 
\begin{itemize}
\item irreflexive: $\forall x.~x\not\succ x,$
\item asymmetric:$\forall x,y.~x\succ y \Rightarrow y\not\succ x,$
\item transitive: $\forall x,y,z.~ (x\succ y\wedge y\succ z) \Rightarrow x\succ z,$
\item negatively transitive: $\forall x,y,z.~(x\not\succ y \wedge y\not\succ z)\Rightarrow x\not\succ z,$
\item connected: $\forall x,y.~x\succ y\vee y\succ x \vee x=y.$
\end{itemize}

A {\em strict partial order} is an irreflexive, transitive (and thus asymmetric) binary relation.
A {\em weak order} is a negatively transitive strict partial order.
A {\em total order} is a connected strict partial order.

\subsection{Scoring function} 
A \emph{scoring function over a deterministic relation} $R$ is a function from $R$ to real numbers, i.e., $s:R\mapsto \mathbb{R}$. The function $s$ induces a \emph{preference relation} $\succ_s$ and an \emph{indifference relation} $\sim_s$ on $R$. For any two distinct tuples  $t_i$ and $t_j$ from $R$,

\[
\begin{array}{l}t_i\succ_s t_j \text{ iff } s(t_i)>s(t_j);\\
t_i\sim_s t_j \text{ iff } s(t_i)=s(t_j).
\end{array}\]

A \emph{scoring function over a probabilistic relation} $R^p=\langle R, p, \mathcal{C}\rangle$ is a scoring function $s$ over its support relation $R$.
In general, a scoring function establishes a {\em weak order} over $R$, where tuples from $R$ can tie in score. However, when the scoring function $s$ is \emph{injective}, $\succ_s$ is a \emph{total order}. In such a case, no two tuples tie in score.

\subsection{Top-k Queries}
\begin{definition}[Top-$k$ Answer Set over a Deterministic Relation]\label{def_topkdete}
Given a deterministic relation $R$, a non-negative integer $k$ and a scoring function $s$ over $R$, a top-$k$ answer set in $R$ under $s$ is a set $T$ of tuples such that

\begin{tabular}{l}
1. $T\subseteq R$;\\
2. If $|R|<k$, $T=R$, otherwise $|T|=k$;\\
3. $\forall t\in T~\forall t'\in R-T.~t\succ_s t'$ or $t\sim_s t'$.
\end{tabular}
\end{definition}

According to Definition \ref{def_topkdete}, given $k$ and $s$, there can be more than one top-$k$ answer set in a deterministic relation $R$. The evaluation of a top-$k$ query over $R$ returns one of them nondeterministically, say $S$. However, if the scoring function $s$ is injective, $S$ is unique, denoted by $\topktuple{k}{s}(R)$.

\section{Semantics of Top-k Queries}\label{topk_TO_sem}
In the following two sections, we restrict our discussion to {\em injective} scoring functions. We will discuss the generalization to general scoring functions in Section \ref{sec_WO}.

\subsection{Semantic Postulates for Top-$k$ Answers}\label{sec_postulates}
Probability opens the gate for various possible semantics for top-$k$ queries. As the semantics of a probabilistic relation involves a set of worlds, it is to be expected that there may be more than one top-$k$ answer set, even under an injective scoring function.
The answer to a top-$k$ query over a probabilistic relation $R^p=\langle R, p, \mathcal{C}\rangle$ should clearly be a set of tuples from its support relation $R$. 
We formulate below three desirable \emph{postulates}, which serve as a benchmark to categorize different semantics. 

In the following discussion, 
denote by $\ptopkset{k}{s}(R^p)$ the collection of all top-$k$ answer sets of $R^p$ under the function $s$.


\smallskip

\noindent \emph{\textbf{Postulates}}

\begin{itemize}
\item \textbf{Static Postulates}
\begin{enumerate}
\item \emph{Exact $k$}: When $R^p$ is sufficiently large ($|\mathcal{C}|\geq k$), the cardinality of every top-$k$ answer set $S$ is exactly $k$;
\[|\mathcal{C}|\geq k \Rightarrow [\forall S\in \ptopkset{k}{s}(R^p).~|S|=k].\]

\item \emph{Faithfulness}: For every top-$k$ answer set $S$ and any two tuples $t_1,t_2\in R$, if both the score and  the probability of $t_1$ are higher than those of $t_2$ and $t_2\in S$, then $t_1\in S$;
\[\forall S\in \ptopkset{k}{s}(R^p)~\forall t_1, t_2\in R.~s(t_1)>s(t_2)\wedge p(t_1)>p(t_2)\wedge t_2\in S\Rightarrow t_1\in S.\]




\end{enumerate}
\item \textbf{Dynamic Postulate}

\begin{enumerate}
\item[]
$\cup~\ptopkset{k}{s}(R^p)$ denotes the union of all top-$k$ answer sets of $R^p=\langle R, p, \mathcal{C}\rangle$ under the function $s$. For any $t\in R$,
\[\begin{array}{l}
t\textrm{ is a {\em winner} iff } t\in \cup~\ptopkset{k}{s}(R^p)\\
t\textrm{ is a {\em loser} iff } t\in R-\cup~\ptopkset{k}{s}(R^p)\\
\end{array}\]

\item[3.] \emph{Stability}:

\begin{itemize}
\item Raising the score/probability of a winner will not turn it into a loser;

\begin{enumerate}
\item If a scoring function $s'$ is such that $s'(t)>s(t)$ and for every $t'\in R-\{t\}$, $s'(t')=s(t')$, then
\[t\in \cup~\ptopkset{k}{s}(R^p)\Rightarrow t\in \cup~\ptopkset{k}{s'}(R^p).\]

\item If a probability function $p'$ is such that $p'(t)>p(t)$ and for every $t'\in R-\{t\}$, $p'(t')=p(t')$, then
\[t\in \cup~\ptopkset{k}{s}(R^p)\Rightarrow t\in \cup~\ptopkset{k}{s}((R^p)'),\]
where $(R^p)'=\langle R, p', \mathcal{C}\rangle$. 
\end{enumerate}

\item Lowering the score/probability of a loser will not turn it into a winner.

\begin{enumerate}
\item If a scoring function $s'$ is such that $s'(t)<s(t)$ and for every $t'\in R-\{t\}$, $s'(t')=s(t')$, then
\[t\in R-\cup~\ptopkset{k}{s}(R^p)\Rightarrow t\in R-\cup~\ptopkset{k}{s'}(R^p).\]
\item If a probability function $p'$ is such that $p'(t)<p(t)$ and for every $t'\in R-\{t\}$, $p'(t')=p(t')$, then
\[t\in R-\cup~\ptopkset{k}{s}(R^p)\Rightarrow t\in R-\cup~\ptopkset{k}{s}((R^p)'),\]
where $(R^p)'=\langle R, p', \mathcal{C}\rangle$. 
\end{enumerate}

\end{itemize}

\end{enumerate}
\end{itemize}

All of those postulates reflect certain requirements of top-$k$ answers.

\emph{Exact $k$} expresses user expectations about the size of the result. Typically, a user issues a top-$k$ query in order to restrict the size of the result and get a subset of cardinality $k$ (cf.~Example \ref{expl_hiring}). Therefore, $k$ can be a crucial parameter specified by the user that should be complied with.

\emph{Faithfulness} reflects the significance of score and probability in a static environment. It plays an important role in designing efficient query evalution algorithms. The satisfaction of {\em Faithfulness} admits a set of pruning techniques based on {\em monotonicity}.

\emph{Stability} reflects the significance of score and probability in a dynamic environment. In a dynamic world, it is common that user might update score/probability on-the-fly. {\em Stability} requires that the consequences of such changes should not be counterintuitive.

 


\subsection{Global-Top$k$ Semantics}\label{sec_globaltopksem}

We propose here a new top-$k$ answer semantics in probabilistic relations,  namely \textbf{Global-Top$k$}, which satisfies the postulates formulated in Section \ref{sec_postulates} to a large degree:

\begin{enumerate}
\item[$\bullet$]\textbf{Global-Top$k$}: return $k$ highest-ranked tuples according to their probability of being in the top-$k$ answers in possible worlds.

\end{enumerate}

Considering a probabilistic relation $R^p=\langle R, p, \mathcal{C}\rangle$ under an injective scoring function $s$, any $W\in pwd(R^p)$ has a unique top-$k$ answer set $\topktuple{k}{s}(W)$. 
Each tuple from the support relation $R$ can be in the top-$k$ answer set (in the sense of Definition \ref{def_topkdete}) in zero, one or more possible worlds of $R^p$.
Therefore, the sum of the probabilities of those possible worlds provides a global ranking criterion.

\begin{definition}[Global-Top$k$ Probability]\label{def_topkprob_TO} Assume a probabilistic relation $R^p=\langle R, p, \mathcal{C}\rangle$, a non-negative integer $k$ and an injective scoring function $s$ over $R^p$. 
For any tuple $t$ in $R$, the Global-Top$k$ probability of $t$, denoted by $P^{R^p}_{k,s}(t)$, is the sum of the probabilities of all possible worlds of $R^p$ whose top-$k$ answer set contains $t$.

\begin{equation}\label{eqn_topkprob_TO}
P^{R^p}_{k,s}(t)=\sum_{
\begin{subarray}{l}
W\in pwd(R^p)\\
t\in \topktuple{k}{s}(W)
\end{subarray}}Pr(W).
\end{equation}
\end{definition}

For simplicity, we skip the superscript in $P^{R^p}_{k,s}(t)$, i.e., $P_{k,s}(t)$, when the context is unambiguous.

\begin{definition}[Global-Top$k$ Answer Set over a Probabilistic Relation]\label{def_topkprob} Given a probabilistic relation $R^p=\langle R, p, \mathcal{C}\rangle$, a non-negative integer $k$ and an injective scoring function $s$ over $R^p$, a Global-Top$k$ answer set in $R^p$ under $s$ is a set $T$ of tuples such that

\begin{tabular}{l}
1. $T\subseteq R$;\\
2. If $|R|<k$, $T=R$, otherwise $|T|=k$;\\
3. $\forall t\in T, \forall t'\in R-T, P_{k,s}(t)\geq P_{k,s}(t')$.
\end{tabular}
\end{definition}

Notice the similarity between Definition \ref{def_topkprob} and Definition \ref{def_topkdete}. In fact, the probabilistic version only changes the last condition, which restates the preferred relationship between two tuples by taking probability into account. This semantics preserves the nondeterministic nature of Definition \ref{def_topkdete}. For example, if two tuples are of the same Global-Top$k$ probability, and there are $k-1$ tuples with a higher Global-Top$k$ probability, Definition \ref{def_topkprob} allows one of the two tuples to be added to the top-$k$ answer set nondeterministically. Example \ref{expl_globaltopk} gives an example of the Global-Top$k$ semantics.

\begin{example}\label{expl_globaltopk}
Consider the top-$2$ query in Example \ref{expl_hiring}. 
Clearly, the scoring function here is the {\em Overall  Score} function. The following table shows all the possible worlds and their probabilities. For each world, the names of the people in the top-$2$ answer set of that world are underlined.
\begin{center}
\begin{tabular}{l c}
  Possible World &Prob\\
  \hline
  $W_1=\emptyset$ & $0.042$\\
  $W_2=\{\underline{Aidan}\}$ & $0.018$\\
  $W_3=\{\underline{Bob}\}$ & $0.378$\\
  $W_4=\{\underline{Chris}\}$ & $0.028$\\
  $W_5=\{\underline{Aidan}, \underline{Bob}\}$ & $0.162$\\
  $W_6=\{\underline{Aidan}, \underline{Chris}\}$ & $0.012$\\
  $W_7=\{\underline{Bob}, \underline{Chris}\}$ & $0.252$\\
  $W_8=\{\underline{Aidan}, \underline{Bob}, Chris\}$ & $0.108$\\
  \hline
\end{tabular}
\end{center}
\smallskip

Chris is in the top-$2$ answer of $W_4, W_6, W_7$, so the top-$2$ probability of Chris is $0.028+0.012+0.252=0.292$. Similarly, the top-$2$ probability of Aidan and Bob are $0.9$ and $0.3$ respectively. $0.9>0.3>0.292$, therefore Global-Top$k$ will return $\{Aidan, Bob\}$.

\end{example}

Note that top-$k$ answer sets may be of cardinality less than $k$ for some possible worlds. We refer to such possible worlds as \emph{small} worlds. In Example \ref{expl_globaltopk}, $W_{1\ldots 4}$ are all small worlds.

\subsection{Other Semantics}

We present here the most well-established top-$k$ semantics in the literature before 2008 (inclusive).

Soliman et al.~\cite{DBLP:conf/icde/SolimanIC07} proposes two semantics for top-$k$ queries in probabilistic relations.
\begin{enumerate}
\item[$\bullet$]\textit{U-Top$k$}: return the most probable top-$k$ answer set that belongs to possible world(s);

\item[$\bullet$]\textit{U-$k$Ranks}: for $i=1,2,\ldots,k$, return the most probable $i^{th}$-ranked tuples across all possible worlds.

\end{enumerate}

Hua et al.~\cite{DBLP:conf/sigmod/HuaPZL08} independently proposes PT-$k$, a semantics based on Global-Top$k$ probability as well. PT-$k$ takes an additional parameter: probability threshold $p_{\tau}\in (0,1]$.

\begin{enumerate}
\item[$\bullet$]\textit{PT-$k$}: return every tuple whose probability of being in the top-$k$ answers in possible worlds is at least $p_{\tau}$.
\end{enumerate}

\begin{example}\label{expl_twosemantics}
Continuing Example \ref{expl_globaltopk}, under U-Top$k$ semantics, the probability of top-$2$ answer set $\{Bob\}$ is $0.378$, and that of $\{Aidan, Bob\}$ is $0.162+0.108=0.27$. Therefore, $\{Bob\}$ is more probable than $\{Aidan, Bob\}$ under U-Top$k$. In fact, $\{Bob\}$ is the most probable top-$2$ answer set in this case, and will be returned by U-Top$k$.

Under U-$k$Ranks semantics, Aidan is in $1^{st}$ place in the top-$2$ answer of $W_2$, $W_5$, $W_6$, $W_8$, therefore the probability of Aidan being in $1^{st}$ place in the top-$2$ answers in possible worlds is $0.018+0.162+0.012+0.108=0.3$. However, Aidan is not in $2^{nd}$ place in the top-$2$ answer of any possible world, therefore the probability of Aidan being in $2^{nd}$ place is $0$. In fact, we can construct the following table.
\begin{center}
\begin{tabular}{l c c c}
& Aidan & Bob & Chris \\
\hline
Rank 1 & $0.3$ & \underline{$0.63$} & $0.028$ \\
Rank 2 & 0 & \underline{$0.27$} & $0.264$ \\
\hline
\end{tabular}
\end{center}
\smallskip

U-$k$Ranks selects the tuple with the highest probability at each rank (underlined) and takes the union of them. In this example, Bob wins at both Rank 1 and Rank 2. Thus, the top-$2$ answer returned by U-$k$Ranks is $\{Bob\}$.

PT-$k$ returns every tuple with its Global-Top$k$ probability above the user specified threshold $p_{\tau}$, therefore the answer depends on $p_{\tau}$. Say $p_{\tau}=0.6$, then PT-$k$ return $\{Aidan\}$, as it is the only tuple with a Global-Top$k$ probability at least $0.6$.
\end{example}

The postulates introduced in Section \ref{sec_postulates} lay the ground for analyzing different semantics. 
In Table \ref{tab_postulates}, a single ``$\checkmark$'' (resp.~``$\times$'') indicates that postulate is (resp.~is not) satisfied under that semantics. ``$\checkmark/\times$'' indicates that, the postulate is satisfied by that semantics in \emph{simple} probabilistic relations, but not in the general case.


\begin{center}
\begin{threeparttable}[b]
\begin{tabular}{|l|c|c|c|c|}
  \hline
  Semantics & Exact $k$ & Faithfulness & Stability \\
  \hline
  Global-Top$k$ & \checkmark & \checkmark/$\times$  & \checkmark \\
  PT-$k$ & $\times$ & \checkmark/$\times$ & \checkmark  \\
  U-Top$k$ & $\times$ & \checkmark/$\times$  & \checkmark \\
  U-$k$Ranks & $\times$ & $\times$  & $\times$ \\
  \hline
\end{tabular}
\caption{Postulate Satisfaction for Different Semantics}
\label{tab_postulates}
\end{threeparttable}
\end{center}


For \emph{Exact k}, Global-Top$k$ is the only semantics that satisfies this postulate. Example \ref{expl_twosemantics} illustrates the case where U-Top$k$, U-$k$Ranks and PT-$k$ violate this postulate. It is not satisfied by U-Top$k$ because a \emph{small} possible world with a high probability could dominate other worlds. In this case, the dominating possible world might not have enough tuples. It is also violated by U-$k$Ranks because a single tuple can win at multiple ranks in U-$k$Ranks. In PT-$k$, if the threshold parameter $p_{\tau}$ is set too high, then less than $k$ tuples will be returned (as in Example \ref{expl_twosemantics}). As $p_{\tau}$ decreases, PT-$k$ return more tuples. In the extreme case when $p_{\tau}$ approaches $0$, any tuple with a positive Global-Top$k$ probability will be returned.

For \emph{Faithfulness}, Global-Top$k$ violates it when exclusion rules lead to a highly restricted distribution of possible worlds, and are combined with an unfavorable scoring function (\emph{see} Appendix A (5)). PT-$k$ violates \emph{Faithfulness} for the same reason (\emph{see} Appendix A (6)).
U-Top$k$ violates \emph{Faithfulness} since it requires all tuples in a top-$k$ answer set to be compatible. This postulate can be violated when a high-score/probability tuple could be dragged down arbitrarily by its compatible tuples which are not very likely to appear (\emph{see} Appendix A (7)).
U-$k$Ranks violates both \emph{Faithfulness} and \emph{Stability}. 
Under U-$k$Ranks, instead of a set, a top-$k$ answer is an ordered vector, where ranks are significant. A change in a tuple's probability/score might have unpredictable consequence on ranks, therefore those two postulates are not guaranteed to hold (\emph{see} Appendix A (8)(12)).

\emph{Faithfulness} is a postulate which can lead to significant pruning in practice. Even though it is not fully satisfied by any of the four semantics, some degree of satisfaction can still be beneficial, as it will help us find pruning rules. For example, our optimization in Section \ref{sec_TA} explores the \emph{Faithfulness} of Global-Top$k$ in simple probabilistic databases. Another example: one of the pruning techniques in \cite{DBLP:conf/sigmod/HuaPZL08} explores the \emph{Faithfulness} of exclusive tuples in general probabilistic databases as well.

See Appendix A for the proofs of the results in Table \ref{tab_postulates}.

It worths mentioning here that the intention of Table \ref{tab_postulates} is to provide a list of semantic postulates, so that users would be able to choose the appropriate postulates for an application. 
For example, in a government contract bidding, only $k$ companies from the first round will advance to the second round. The score is inverse to the price offered by a company, and the probability is the probability that company will complete the task on time. The constraint of $k$ is hard, and thus \emph{Exact k} is a must for the top-$k$ semantics chosen. 
In contrast, during college admission, where the score reflects the qualification of an applicant and the probability is the probability of offer acceptance, while we intend to have a class of $k$ students, there is usually room for fluctuation. In this case, \emph{Exact k} is not a must. 
It is the same story with \emph{Faithfulness} and \emph{Stability}: \emph{Faithfulness} is required in applications such as auctions, where the score is the value of an item and the probability is the availability of the item. In this case, it is a natural to aim at the ``best deals'', i.e., items with high value and high availability. \emph{Stability} is a common postulate required by many dynamic applications. For example, we want to maintain a best $k$ seller list, where the score is inverse to the price of an item and the probability is its availability. It is to be expected that a discounted price and improved availability of an item should not have an adverse influence on the item's stand on the best $k$ seller list\footnote{In real life, we sometimes observe cases when \emph{stability} does not hold: a cheaper Wii console with improved availability does not make it more popular than it was. The reason could be psychological.}.

In short, we are not advertising that a specific semantics is superior/inferior to any other semantics using Table \ref{tab_postulates}. Rather, with the help of Table \ref{tab_postulates}, users will be able to search for the most appropriate semantics based on the right combination of postulates for their applications.

\section{Query Evaluation under Global-Top$k$}

\subsection{Simple Probabilistic Relations}\label{topk_TO_Ind}

We first consider a \emph{simple} probabilistic relation $R^p=\langle R, p, \mathcal{C}\rangle$ under an injective scoring function $s$. 

\begin{proposition}\label{prop_recursion}
Given a simple probabilistic relation $R^p=\langle R, p, \mathcal{C}\rangle$ and an injective scoring function $s$ over $R^p$, if $R=\{t_1$, $t_2$, $\ldots$, $t_n\}$ and $t_1\succ_s t_2\succ_s \ldots \succ_s t_n$, the following recursion on Global-Top$k$ queries holds:

\begin{equation}\label{eqn_recursion}
q(k,i) = \left\{ \begin{array}{lr}
0  & k=0 \\
p(t_i) & 1\leq i\leq k \\
(q(k,i-1)\dfrac{\bar{p}(t_{i-1})}{p(t_{i-1})}+q(k-1,i-1))p(t_i)&\textrm{ otherwise}
\end{array} \right.
\end{equation}
where $q(k,i)=P_{k,s}(t_i)$ and $\bar{p}(t_{i-1})=1-p(t_{i-1})$.
\end{proposition}

\noindent \emph{Proof}.~~~\emph{See} Appendix B.

Notice that Equation (\ref{eqn_recursion}) involves probabilities only, while the scores are used to determine the order of computation.
\begin{example}\label{expl_dp}Consider a simple probabilistic relation $R^p=\langle R, p, \mathcal{C}\rangle$, where
$R=\{t_1$, $t_2$, $t_3$, $t_4\}$, $p(t_i)=p_i, 1\leq i \leq 4$, $\mathcal{C}=\{\{t_1\}, \{t_2\}, \{t_3\}, \{t_4\}\}$, and an injective scoring function $s$ such that $t_1 \succ_s t_2 \succ_s t_3 \succ_s t_4$.
The following table shows the Global-Top$k$ probability of $t_i$, where $0\leq k\leq 2$.
\begin{center}
\begin{tabular}{|c|l l l l|}
  \hline
  $k$ & $t_1$ & $t_2$ & $t_3$ & $t_4$ \\
  \hline
  0 & $0$ & $0$ & $0$ & $0$ \\
  1 & $p_1$ & $\bar{p}_1 p_2$ & $\bar{p}_1\bar{p}_2 p_3$ &  $\bar{p}_1\bar{p}_2\bar{p}_3p_4$\\
  2 & $\mathbf{p_1}$ & $\mathbf{p_2}$ & $\mathbf{(\bar{p}_2+\bar{p}_1 p_2)p_3}$ & $\mathbf{((\bar{p}_2+\bar{p}_1 p_2)\bar{p}_3}$\\
&&&&$\mathbf{+\bar{p}_1\bar{p}_2 p_3)p_4}$\\
  \hline
\end{tabular}
\end{center}

Row 2 (bold) is each $t_i$'s Global-Top$2$ probability. Now, if we are interested in a top-$2$ answer in $R^p$, we only need to pick the two tuples with the highest value in Row 2.
\end{example}

\begin{theorem}[Correctness of Algorithm \ref{alg_ind}]\label{thm_ind}
Given a simple probabilistic relation $R^p=\langle R, p$, $\mathcal{C}\rangle$, a non-negative integer $k$ and an injective scoring function $s$, Algorithm \ref{alg_ind} correctly computes a Global-Top$k$ answer set of $R^p$ under the scoring function $s$. 
\end{theorem}

\begin{proof}
Algorithm \ref{alg_ind} maintains a priority queue to select the $k$ tuples with the highest Global-Top$k$ value. Notice that the nondeterminism is reflected in Line 6 in the algorithm for maintaining the priority queue in the presence of tying elements. As long as Line 2 in Algorithm \ref{alg_ind} correctly computes the Global-Top$k$ probability of each tuple in $R$, Algorithm \ref{alg_ind} returns a valid Global-Top$k$ answer set.
By Proposition \ref{prop_recursion}, Algorithm \ref{alg_indsub} correctly computes the Global-Top$k$ probability of tuples in $R$.
\end{proof}

Algorithm \ref{alg_ind} is a one-pass computation over the probabilistic relation, which can be easily implemented even if secondary storage is used. The overhead is the initial sorting cost (not shown in Algorithm \ref{alg_ind}), which would be amortized by the workload of consecutive top-$k$ queries.

\begin{algorithm}
\caption{\textbf{(Ind\_Topk)} Evaluate Global-Top$k$ Queries in a Simple Probabilistic Relation under an Injective Scoring Function}
\label{alg_ind}
\begin{algorithmic}[1]
\REQUIRE $R^p=\langle R, p, \mathcal{C}\rangle, k$
\ENSURE tuples in $R$ are sorted in the decreasing order based on the scoring function $s$
\STATE Initialize a fixed cardinality $(k+1)$ priority queue $Ans$ of $\langle t, prob\rangle$ pairs, which compares pairs on $prob$, i.e., the Global-Top$k$ probability of $t$;
\STATE Calculate Global-Top$k$ probabilities using Algorithm \ref{alg_indsub}, i.e.,
\[q(0\ldots k, 1\ldots |R|)=\textrm{Ind\_Topk\_Sub}(R^p, k);\]\label{ln_ind_dptable}
\FOR{$i=1$ to $|R|$}
\STATE Add $\langle t_i, q(k, i)\rangle$ to $Ans$;\label{ln_ind_update}
\IF{$|Ans|>k$}
\STATE remove the pair with the smallest $prob$ value from $Ans$;
\ENDIF
\ENDFOR
\RETURN $\{t_i|\langle t_i,q(k,i)\rangle\in Ans\}$;
\end{algorithmic}
\end{algorithm}

\begin{algorithm}
\caption{\textbf{(Ind\_Topk\_Sub)} Compute Global-Top$k$ Probabilities in a Simple Probabilistic Relation under an Injective Scoring Function}
\label{alg_indsub}
\begin{algorithmic}[1]
\REQUIRE $R^p=\langle R, p, \mathcal{C}\rangle, k$
\ENSURE tuples in $R$ are sorted in the decreasing order based on the scoring function $s$
\STATE $q(0,1)=0$;
\FOR{$k'=1$ to $k$}
\STATE $q(k',1)=p(t_1)$;
\ENDFOR
\FOR{$i=2$ to $|R|$}
\FOR{$k'=0$ to $k$}
\IF{$k'=0$} 
\STATE $q(k',i)=0;$
\ELSE
\STATE $q(k',i)=p(t_i)(q(k',i-1)\dfrac{\bar{p}(t_{i-1})}{p(t_{i-1})}+q(k'-1,i-1))$;
\ENDIF
\ENDFOR
\ENDFOR
\RETURN $q(0\ldots k, 1\ldots |R|)$;
\end{algorithmic}
\end{algorithm}

Algorithm \ref{alg_indsub} takes $O(kn)$ to compute the dynamic programming (DP) table. In addition, Algorithm \ref{alg_ind} uses a priority queue to maintain the $k$ highest values, which takes $O(n\log k)$. Altogether, Algorithm \ref{alg_ind} takes $O(kn)$. 

The major space use in Algorithm \ref{alg_ind} is the bookkeeping of the DP table in Line \ref{ln_ind_dptable} (Algorithm \ref{alg_indsub}). A straightword implementation of Algorithm \ref{alg_ind} and Algorithm \ref{alg_indsub} takes $O(kn)$ space. However, notice that in Algorithm \ref{alg_indsub}, the column $q(0\ldots k, i)$ depends on the column $q(0\ldots k, i-1)$ only, and for the column $q(0\ldots k, i-1)$, only the $k$th value $q(k,i-1)$ will be used in updating the priority queue in Line \ref{ln_ind_update} of Algorithm \ref{alg_ind} later.
Therefore, in practice, we can reduce the space complexity to $O(k)$ by moving the update of the priority queue in Algorithm \ref{alg_ind} to Algorithm \ref{alg_indsub}, and using a vector of size $k+1$ to keep track of the previous column in the DP table. To be more specific, in Algorithm \ref{alg_indsub}, each time we finish computing the current column based on the previous column in the DP table, we add the $k$th value in the current column to the priority queue and update the previous column with the current column. For readability, we present here the original algorithms without this optimization for space. 

\subsection{Threshold Algorithm Optimization}\label{sec_TA}

Fagin \cite{DBLP:conf/pods/FaginLN01} proposes \emph{Threshold Algorithm (TA)} for processing top-$k$ queries in a middleware scenario. In a middleware system, an \emph{object} has $m$ attributes. For each attribute, there is a sorted list ranking objects in the decreasing order of its score on that attribute. An \emph{aggregation function} $f$ combines the individual attribute scores $x_i$, $i$$=$$1,2,\ldots,m$ to obtain the overall object score $f(x_1,x_2,\ldots,x_m)$. An aggregation function is \emph{monotonic} iff $f(x_1,x_2,\ldots,x_m)\leq f(x'_1,x'_2,\ldots,x'_m)$ whenever $x_i\leq x'_i$ for every $i$. Fagin \cite{DBLP:conf/pods/FaginLN01} shows that TA is cost-optimal in finding the top-$k$ objects in such a system.

Denote $T$ and $P$ for the list of tuples in the decreasing order of score and probability respectively. Following the convention in  \cite{DBLP:conf/pods/FaginLN01}, $\underline{t}$ and $\underline{p}$ are the last value seen in $T$ and $P$ respectively.

\begin{boxedminipage}[b]{0.9\textwidth}
\textbf{Algorithm $1^{\textit{TA}}$ (TA\_Ind\_Topk)}
\begin{enumerate}
\item[(1)] Go down $T$ list, and fill in entries in the DP table. Specifically, for $\underline{t}=t_j$, compute the entries in the $j^{th}$ column up to the $k^{th}$ row.
Add $t_j$ to the top-$k$ answer set $Ans$, if any of the following conditions holds:

\begin{itemize}
\item[(a)]$Ans$ has less than $k$ tuples, i.e., $|Ans|<k$;

\item[(b)]The Global-Top$k$ probability of $t_j$, i.e., $q(k,j)$, is greater than the lower bound of $Ans$, i.e., $LB_{Ans}$, where $LB_{Ans}=\min_{t_i\in Ans}q(k,i)$.

\end{itemize}
In the second case, we also need to drop the tuple with the lowest Global-Top$k$ probability in order to preserve the cardinality of $Ans$.

\item[(2)] After we have seen at least $k$ tuples in $T$, we go down $P$ list to find the first
$p$ whose tuple $t$ has not been seen. Let $\underline{p}=p$, and we can use $\underline{p}$ to estimate the \emph{threshold}, i.e., upper bound ($UP$) of the Global-Top$k$ probability of any unseen tuple. Assume $\underline{t}=t_i$,

\[UP=(q(k,i)\frac{\bar{p}(t_i)}{p(t_i)}+q(k-1,i)) \underline{p}.\]

\item[(3)] If $UP>LB_{Ans}$, $Ans$ might be updated in the future, so
go back to (1). Otherwise, we can safely stop and report $Ans$.
\end{enumerate}
\end{boxedminipage}

\begin{theorem}[Correctness of Algorithm $1^{\textit{TA}}$]\label{thm_ta}
Given a simple probabilistic relation $R^p=\langle R,p,\mathcal{C}\rangle$, a non-negative integer $k$ and an injective scoring function $s$ over $R^p$, the above TA-based algorithm correctly finds a Global-Top$k$ answer set.
\end{theorem}

\noindent \emph{Proof}.~~~\emph{See} Appendix B.

The optimization above aims at an early stop. Bruno et al. \cite{DBLP:journals/tkde/BrunoW07} carries out an extensive experimental study on the effectiveness of applying TA in RDMBS. They consider various aspects of query processing. One of their conclusions is that if at least one of the indices available for the attributes\footnote{Probability is typically supported as a special attribute in DBMS.} is a \emph{covering index}, that is, it is defined over all other attributes and we can get the values of all other attributes directly without performing a primary index lookup, then the improvement by TA can be up to two orders of magnitude. The cost of building a useful set of indices once would be amortized by a large number of top-$k$ queries that subsequently benefit form such indices. Even in the lack of covering indices, if the data is highly correlated, in our case, that means high-score tuples having high probabilities, TA would still be effective.
 
TA is guaranteed to work as long as the aggregation function is monotonic. For a simple probabilistic relation, if we regard \emph{score} and \emph{probability} as two special attributes, Global-Top$k$ probability $P_{k,s}$ is an aggregation function of \emph{score} and \emph{probability}. The \emph{Faithfulness} postulate in Section \ref{sec_postulates} implies the monotonicity of Global-Top$k$ probability in simple probabilistic relations. 
Consequently, assuming that we have an index on probability as well, we can guide the dynamic programming (DP) in Algorithm \ref{alg_indsub} by TA. Now, instead of computing all $kn$ entries for DP, where $n=|R|$, the algorithm can be stopped as early as possible.
A subtlety is that Global-Top$k$ probability $P_{k,s}$ is \emph{only} well-defined for $t\in R$, unlike in \cite{DBLP:conf/pods/FaginLN01}, where an aggregation function is well-defined over the domain of all possible attribute values. 
Therefore, compared to the original TA, we need to achieve the same behavior without referring to virtual tuples which are not in $R$.

U-Top$k$ satisfies \emph{Faithfulness} in simple probabilistic relations. An adaptation of the TA algorithm in this case is available in \cite{DBLP:journals/tods/SolimanIC08}.
TA is not applicable to U-$k$Ranks. Even though we can define an aggregation function per $rank$, $rank=1,2,\ldots,k$, for tuples under U-$k$Ranks, the violation of \emph{Faithfulness} in Table \ref{tab_postulates} suggests a violation of monotonicity of those $k$ aggregation functions.
PT-$k$ computes Global-Top$k$ probabilities as well, and is therefore a natural candidate for TA in simple probabilistic relations.

\subsection{Arbitrary Probabilistic Relations}\label{topk_TO_IndEx}

\subsubsection{Induced Event Relation}
In the general case of probabilistic relations (Definition \ref{def_probdb}), each part of the partition $\mathcal{C}$ can contain more than one tuple. 
The crucial \emph{independence} assumption in Algorithm \ref{alg_ind} no longer holds.
However, even though tuples from one part of the partition $\mathcal{C}$ are not independent, tuples from different parts are. In the following definition, we assume an identifier function $id$. For any tuple $t$, $id(t)$ identifies the part where $t$ belongs.

\begin{definition}[Induced Event Relation]\label{def_ied}
Given a probabilistic relation $R^p=\langle R, p, \mathcal{C}\rangle$, an injective scoring function $s$ over $R^p$ and a tuple $t\in C_{id(t)}\in \mathcal{C}$, the event relation induced by $t$, denoted by $E^p=\langle E, p^E, \mathcal{C}^E\rangle$, is a probabilistic relation whose support relation $E$ has only one attribute, $Event$. The relation $E$ and the probability function $p^E$ are defined by the following two generation rules: 

\begin{itemize}
\item Rule 1:\hspace{0.2in} $t_{e_t}\in E \textrm{~and~} p^E(t_{e_t})=p(t)$;

\item Rule 2:\hspace{0.2in} $\forall C_i\in \mathcal{C}\wedge C_i\neq C_{id(t)}$. 
\[(\exists t'\in C_i \wedge t'\succ_s t)\Rightarrow
(t_{e_{C_i}}\in E) \textrm{ and } p^E(t_{e_{C_i}})=
\sum_{\begin{subarray}{l}
t'\in C_i\\
t'\succ_s t
\end{subarray}}p(t').\]
\end{itemize}
No other tuples belong to $E$. The partition $\mathcal{C}^E$ is defined as the collection of singleton subsets of $E$.
\end{definition}

Except for one special tuple generated by \emph{Rule 1}, each tuple in the induced event relation (generated by \emph{Rule 2}) represents an event $e_{C_i}$ associated with a part $C_i \in \mathcal{C}$. 
Given the tuple $t$, the \emph{event} $e_{C_i}$ is defined as ``there is a tuple from the part $C_i$ with a score higher than that of $t$''.
The probability of this event, denoted by $p(t_{e_{C_i}})$, is the probability that $e_{C_i}$ occurs. 

The role of the special tuple $t_{e_t}$ and its probability $p(t)$ will become clear in Proposition \ref{prop_iedscore}. Let us first look at an example of an induced event relation.

\begin{example}\label{expl_ied}
Given $R^p$ as in Example \ref{expl_sensor}, we would like to construct the induced event relation $E^p=\langle E, p^E, \mathcal{C}^E\rangle$ for tuple $t$=(Temp: $15$) from $C_2$.
By Rule 1, we have $t_{e_t}\in E$, $p^E(t_{e_t})=0.6$. By Rule 2, since $t\in C_2$, we have $t_{e_{C_1}}\in E$ and $p^E(t_{e_{C_1}})=\sum_{\begin{subarray}{l}
t'\in C_1\\
t'\succ_s t
\end{subarray}}p(t')=p((\textrm{Temp: }22))=0.6$. Therefore,
\begin{center}
\begin{tabular}{ll}
$E$: & $p^E$:\\
\begin{tabular}{|c|}
  \hline
  Event\\
  \hline
  $t_{e_t}$ \\
  \hline
  $t_{e_{C_1}}$\\
  \hline
\end{tabular}
&
\begin{tabular}{c}
Prob\\
\hline
$0.6$\\
$0.6$
\end{tabular}
\end{tabular}
\end{center}

\end{example}

\begin{proposition}\label{ppt_iedind}
An induced event relation in Definition \ref{def_ied} is a simple probabilistic relation.
\end{proposition}

\subsubsection{Evaluating Global-Top$k$ Queries}
With the help of \emph{induced event relations}, we can reduce Global-Top$k$ in the general case to Global-Top$k$ in simple probabilistic relations. 

\begin{lemma}\label{thm_ied}
Let $R^p=\langle R, p, \mathcal{C}\rangle$ be a probabilistic relation, $s$ an injective scoring function, $t\in R$, and $E^p=\langle E,p^E,\mathcal{C}^E\rangle$ the event relation induced by $t$.
Define $Q^p=\langle E-\{t_{e_t}\},p^E,\mathcal{C}^E-\{\{t_{e_t}\}\}\rangle$.
Then, the Global-Top$k$ probability of $t$ satisfies the following:
\[
P^{R^p}_{k,s}(t)=p(t)\sum_{\begin{subarray}{l}
W_e\in pwd(Q^p)\\
|W_e|< k
\end{subarray}}Pr(W_e).\]
\end{lemma}

\noindent \emph{Proof}.~~~\emph{See} Appendix B.

\begin{proposition}\label{prop_iedscore}
Given a probabilistic relation $R^p=\langle R, p, \mathcal{C}\rangle$ and an injective scoring function $s$, for any $t\in R^p$, the Global-Top$k$ probability of $t$ equals the Global-Top$k$ probability of $t_{e_t}$ when evaluating top-$k$ in the induced event relation $E^p=\langle E, p^E, \mathcal{C}^E\rangle$ under the injective scoring function $s^E:E\rightarrow \mathbb{R}, s^E(t_{e_t})=\frac{1}{2}$ and $s^E(t_{e_{C_i}})=i$:
\[P^{R^p}_{k,s}(t)=P^{E^p}_{k,s^E}(t_{e_t}).\]
\end{proposition}

\noindent \emph{Proof}.~~~\emph{See} Appendix B.

In Proposition \ref{prop_iedscore}, the choice of the function $s^{E}$ is rather arbitrary. In fact, any injective function giving $t_{e_t}$ the lowest score will do. 
Every tuple other than $t_{e_t}$ in the induced event relation corresponds to an event that a tuple with a score higher than that of $t$ occurs.
We want to track the case that at most $k-1$ such events happen.
Since any induced event relation is simple (Proposition \ref{ppt_iedind}), Proposition \ref{prop_iedscore} illustrates how we can reduce the computation of $P^{R^p}_{k,s}(t)$ in the original probabilistic relation to a top-$k$ computation in a simple probabilistic relation, where we can apply the DP technique described in Section \ref{topk_TO_Ind}.
The complete algorithms are shown as Algorithm \ref{alg_indEx} and Algorithm \ref{alg_indExsub}.

\begin{algorithm}
\caption{\textbf{(IndEx\_Topk)} Evaluate Global-Top$k$ Queries in a General Probabilistic Relation under an Injective Scoring Function}
\label{alg_indEx}
\begin{algorithmic}[1]
\REQUIRE $R^p=\langle R, p, \mathcal{C}\rangle, k, s$
\STATE Initialize a fixed cardinality $k+1$ priority queue $Ans$ of $\langle t, prob\rangle$ pairs, which compares pairs on $prob$, i.e., the Global-Top$k$ probability of $t$;
\FOR{$t\in R$}
\STATE Calculate $P^{R^p}_{k,s}(t)$ using Algorithm \ref{alg_indExsub}, i.e.,
\[P^{R^p}_{k,s}(t)=\textrm{IndEx\_Topk\_Sub}(R^p, k, s, t);\]
\STATE Add $\langle t,P^{R^p}_{k,s}(t)\rangle$ to $Ans$;
\IF{$|Ans|>k$}
\STATE remove the pair with the smallest $prob$ value from $Ans$;
\ENDIF
\ENDFOR
\RETURN $\{t|\langle t,P^{R^p}_{k,s}(t)\rangle\in Ans\}$;
\end{algorithmic}
\end{algorithm}

\begin{algorithm}
\caption{\textbf{(IndEx\_Topk\_Sub)} Calculate $P^{R^p}_{k,s}(t)$ using an induced event relation}
\label{alg_indExsub}
\begin{algorithmic}[1]
\REQUIRE $R^p=\langle R, p, \mathcal{C}\rangle, k, s, t\in R$
\STATE Find the part $C_{id(t)}\in \mathcal{C}$ such that $t\in C_{id(t)}$;
\STATE $E=\{t_{e_t}\}$, where $p^E(t_{e_t})=p(t)$;\label{ln_init_ep}
\FOR{$C_i\in \mathcal{C}$ and $C_i\neq C_{id(t)}$}\label{ln_reduction_start}
\STATE \vspace{0.05in}\hspace{0.9in} $p(e_{C_i})=\sum_{\begin{subarray}{l}
t'\in C_i\\
t' \succ_s t
\end{subarray}}p(t')$;\vspace{0.05in} 
\IF{$p(e_{C_i})>0$}
\STATE $E=E\cup{\{t_{e_{C_i}}\}}$, where $p^E(t_{e_{C_i}})=p(e_{C_i})$;
\ENDIF
\ENDFOR\label{ln_reduction_end}
\STATE Use Algorithm \ref{alg_indsub} to compute Global-Top$k$ probabilities in $E^p=\langle E, p^{E}, \mathcal{C}^{E}\rangle$, i.e., 
\[q(0\ldots k, 1\ldots |E|)=\textrm{Ind\_Topk\_Sub}(E^p, k)\]\label{ln_call_simple}
\STATE $P^{R^p}_{k,s}(t)=P^{E^p}_{k,s^E}(t_{e_t})=q(k,|E|)$;
\RETURN  $P^{R^p}_{k,s}(t)$;
\end{algorithmic}
\end{algorithm}

In Algorithm \ref{alg_indExsub}, we first find the part $C_{id(t)}$ where $t$ belongs. In Line \ref{ln_init_ep}, we initialize the support relation $E$ of the induced event relation with the tuple generated by Rule 1 in Definition \ref{def_ied}. 
For any part $C_i$ other than $C_{id(t)}$, we compute the probability of the event $e_{C_i}$ according to Definition \ref{def_ied} (Line 4), and add it to $E$ if its probability is non-zero (Lines 5-7). Since all tuples from the same part are exclusive, this probability is the sum of the probabilities of all qualifying tuples in that part. If no tuple from $C_i$ qualifies, this probability is zero. In this case, we do not care whether any tuple from $C_i$ will be in the possible world or not, 
since it does not have any influence on whether $t$ will be in top-$k$ or not. The corresponding event tuple is therefore excluded from $E$. Note that, by default, any probabilistic database assumes that any tuple not in the support relation is with probability zero. Line \ref{ln_call_simple} uses Algorithm \ref{alg_indsub} to compute $P^{E^p}_{k,s}(t_{e_t})$. Note that Algorithm \ref{alg_indsub} requires all tuples be sorted on score. Since we already know the scoring function $s^E$, we simply need to organize tuples based on $s^E$ when generating $E$. No extra sorting is necessary.

\begin{theorem}[Correctness of Algorithm \ref{alg_indEx}]\label{thm_indEx}
Given a probabilistic relation $R^p=\langle R, p$, $\mathcal{C}\rangle$, a non-negative integer $k$ and an injective scoring function $s$, Algorithm \ref{alg_indEx} correctly computes a Global-Top$k$ answer set of $R^p$ under the scoring function $s$. 
\end{theorem}

\begin{proof}
The top-level structure of Algorithm \ref{alg_indEx} resembles that of Algorithm \ref{alg_ind}. Therefore, as long as Line 3 in Algorithm \ref{alg_indEx} correctly computes the Global-Top$k$ probability of each tuple in $R$, Algorithm \ref{alg_indEx} returns a valid Global-Top$k$ answer set.
Lines 1-8 in Algorithm \ref{alg_indExsub} compute the event relation induced by the tuple $t$. By Proposition \ref{prop_iedscore}, Lines 9-10 in Algorithm \ref{alg_indExsub} correctly compute the Global-Top$k$ probability of $t$.
\end{proof}

In Algorithm \ref{alg_indExsub}, Lines \ref{ln_reduction_start}-\ref{ln_reduction_end} take $O(n)$ time to build $E$ (we need to scan all tuples within each part). The call to Algorithm \ref{alg_indsub} in Line \ref{ln_call_simple} takes $O(k|E|)$, where $|E|$ is no more than the number of parts in partition $\mathcal{C}$, which is in turn no more than $n$. So Algorithm \ref{alg_indExsub} takes $O(kn)$.
Algorithm \ref{alg_indEx} make $n$ calls to Algorithm \ref{alg_indExsub} to compute $P^{R^p}_{k,s}(t)$ for every tuple $t\in R$. Again, Algorithm \ref{alg_indEx} uses a priority queue to select the final answer set, which takes $O(n\log k)$. The entire algorithm takes $O(kn^2+n\log k)=O(kn^2)$.

A straightforward implementation of Algorithm \ref{alg_indEx} and Algorithm \ref{alg_indExsub} take $O(kn)$ space, as the call to Algorithm \ref{alg_indsub} in Algorithm \ref{alg_indExsub} could take up to $O(k|E|)$ space. However, by using a spatially optimized version of Algorithm \ref{alg_indsub} mentioned in Section \ref{topk_TO_Ind}, this DP table computation in Algorithm \ref{alg_indExsub} can be completed in $O(k)$ space. Algorithm \ref{alg_indExsub} still needs $O(|E|)$ space to store the induced event relation computed between Lines \ref{ln_reduction_start}-\ref{ln_reduction_end}. As $|E|$ has an upper bound $n$, the total space is therefore $O(k+n)$. 

\subsection{Optimizations for Arbitrary Probabilistic Relations}\label{sec_r_rs}
In the previous section, we presented the basic algorithms to compute Global-Top$k$ probabilities in general probabilistic relations. In this section, we provide two heuristics, \rollback\ and \rollbacksort, to speed up this computation.
Our optimizations are similar to {\em prefix sharing} optimizations in \cite{DBLP:conf/sigmod/HuaPZL08}, although the assumptions and technical details are different. In our terminology, the {\em aggressive} and {\em lazy} prefix sharing in \cite{DBLP:conf/sigmod/HuaPZL08} assume the ability to ``look ahead'' in the input tuple stream to locate the next tuple belonging to every part. In contrast, \rollback\ assumes no extra information, and \rollbacksort\ assumes the availability of aggregate statistics on tuples.

\rollback\ and \rollbacksort\ take advantage of the following two facts in the basic algorithms:
\begin{enumerate}
\item[\emph{Fact 1:}] The overlap of the event relations induced by consecutive tuples;

\rollback\ and \rollbacksort\ are based on the following ``incremental'' computation of induced event relations for tuples in $R$. By Definition \ref{def_ied}, for any tuple $t\in R$, only tuples with a higher score will have an influence on $t$'s induced event relation.
Given a scoring function $s$, consider two adjacent tuples $t_i$, $t_{i+1}$ in the decreasing order of scores. Denote by $E_i$ and $E_{i+1}$ their induced event relations under the function $s$ respectively. 

\begin{description}
\item[Case 1:] $t_i$ and $t_{i+1}$ are exclusive. 

Then $t_i$ and $t_{i+1}$ have the same induced event relation except for the one tuple generated by Rule 1 in each induced event relation. 
\begin{equation}\label{eqn_rbcase1}
E_i-\{t_{e_{t_i}}\}=E_{i+1}-\{t_{e_{t_{i+1}}}\}.
\end{equation}

\item[Case 2:] $t_i$ and $t_{i+1}$ are independent, and $t_{i+1}$ is independent of $t_1,\ldots, t_{i-1}$ as well.

Recall that any tuple $t_j  \in C_{id(t_i)}, 1\leq j \leq i-1$, where $C_{id(t_i)}$ is the part containing $t_i$, does not contribute to $E_i$ due to the existence of $t_{e_{t_{i}}}$ in $E_i$. 
Tuple $t_{i+1}$ is independent of such tuple $t_j$. In $E_{i+1}$, instead of $t_{e_{t_i}}$, there is an event tuple $t_{e_{C_{id(t_i)}}}$, which corresponds to the event that one tuple from $C_{id(t_i)}$ appears.
The second condition guarantees that there is no tuple in $E_i-\{t_{e_{t_i}}\}$ which is incompatible with the event tuple $t_{e_{t_{i+1}}}$ generated by Rule 1 in $E_{i+1}$. Therefore, all event tuples in $E_i-\{t_{e_{t_i}}\}$ should be retained in $E_{i+1}$. Consequently,

\begin{equation}\label{eqn_rbcase2}
E_i-\{t_{e_{t_i}}\}=E_{i+1}-\{t_{e_{C_{id(t_i)}}}, t_{e_{t_{i+1}}}\}.
\end{equation}

\item[Case 3:] $t_i$ and $t_{i+1}$ are independent, and $t_{i+1}$ is incompatible with at least one tuple from $t_1,\ldots, t_{i-1}$.

In this case, like in Case 2, the first condition guarantees the existence of $t_{C_{id(t_i)}}$ in $E_{i+1}$.
However, the second condition essentially states that some tuple from $C_{id(t_{i+1})}$ has a score higher than that of $t_i$.  Thus, there is an event tuple $t_{e_{C_{id(t_{i+1})}}}$ in $E_{i}$, which is incompatible with $t_{e_{t_{i+1}}}$ generated by Rule 1 in $E_{i+1}$. As a result, besides the one tuple generated by Rule 1 in each induced event relation, $E_{i+1}$ and $E_i$ also differ in the event tuple $t_{e_{C_{id(t_i)}}}$ and $t_{e_{C_{id(t_{i+1})}}}$.
\begin{equation}\label{eqn_rbcase3}
E_i-\{t_{e_{C_{id(t_{i+1})}}},t_{e_{t_i}}\}=E_{i+1}-\{t_{e_{C_{id(t_i)}}}, t_{e_{t_{i+1}}}\}.
\end{equation}

\end{description}
\item[\emph{Fact 2:}] The arbitrary choice of the scoring function $s^{E}$ in Proposition \ref{prop_iedscore}. 

As we can see from Proposition \ref{prop_iedscore}, the event tuple $t_{e_t}$ has the same Global-Top$k$ probability in the induced event relation under two distinctive scoring functions as long as they both give $t_{e_t}$ the lowest score.

\end{enumerate}
\subsubsection{Rollback}

In \rollback, 
we use an annotated $(k+1)\times n$ table $T^{a}$ to support two major operations for each induced event relation: (1) the creation of the induced event relation, and (2) the computation in the dynamic programming (DP) table to calculate the Global-Top$k$ probability of the tuple inducing it. Each column in $T^a$ is annotated with $(part\_id, prob)$ of an event tuple in the current induced event relation. Each entry (row) in the column corresponds to an entry in the DP table when calculating the Global-Top$k$ probabilities.

By \emph{Fact 1}, it is clear that the creation of induced event relations is incremental if we do it for tuples in the decreasing order of scores. Fortunately, the decreasing order of scores is also used in computing the Global-Top$k$ probability in each induced event relation. \rollback\ exploits this alignment in order and piggybacks the creation of the induced event relation to the computation in the DP table.

By \emph{Fact 2}, we can reuse the scoring function to the greatest extent between two consecutive induced event relations, and therefore avoid the recomputation of a part of the DP table.

Without loss of generality, assume $t_1\succ t_2\succ \ldots\succ t_n$, and the tuple just \emph{processed} is $t_i$, $1\leq i \leq n$. By ``processed'', we mean that there is a DP table for computing the Global-Top$k$ probability, denoted by $DP_i$, where each column is associated with an event tuple in $t_i$'s induced event relation $E_i$. Assume $|E_i|=l_i$, then there are $l_i$ columns in $DP_i$. $l_i\leq i$, since only $t_1, t_2, \ldots, t_i$ can contribute to $E_i$.
In fact, $l_i=i$ when all $i$ tuples are independent. In this case, each tuple corresponds to a distinct event tuple in $E_i$. When there are exclusive tuples, $l_i<i$. Because in this case, if a tuple from $t_1, t_2, \ldots, t_{i-1}$ is incompatible with $t_i$, it is ignored due to the existence of $t_{e_{t_i}}$ in $E_i$. For other exclusive tuples, the tuples from the same part collapse into a single event tuple in $E_i$. Moreover, the probability of such event tuple is the sum of the probabilities of all exclusive tuples contributing to it. 

Now, consider the next tuple to be processed, $t_{i+1}$, its induced event relation $E_{i+1}$, and the DP table $DP_{i+1}$ to compute the Global-Top$k$ probability in $E_{i+1}$. If the current situation is of Case 1, then $E_{i}$ and $E_{i+1}$ only differ in the event tuple generated by Rule 1. Recall that the only requirement on the scoring function used in an induced event relation is to assign the lowest score to the event tuple generated by Rule 1. This requirement is translated into the computation in the DP table as associating the tuple generated by Rule 1 with the last column. Therefore, we can take the first $l_i-1$ columns from $DP_i$ and reuse them in $DP_{i+1}$. In other words, by reusing the scoring function in $DP_i$ as much as possible based on {\em Fact 2}, the resulting $DP_{i+1}$ table differs from $DP_i$ only in the last column. In practice, $DP_{i+1}$ is computed incrementally by modifying the last column of $DP_i$ in place. Denoted by $col_{cur}$ the current last column in $DP_i$. In $DP_{i+1}$, $col_{cur}$ should be reassociated with the event tuple $t_{e_{t_{i+1}}}$, i.e.,
\begin{eqnarray*}
&&col_{cur}.part\_id = id(t_{i+1}),\\
&& col_{cur}.prob = p(t_{i+1}). 
\end{eqnarray*}
It is easy to see that the incremental computation cost is the cost of computing $k+1$ entries in $col_{cur}$. 

Similarly for Case 2, the first $l_i-1$ columns in $DP_i$ can be reused. The two new event tuples in $DP_{i+1}$ are $t_{e_{C_{id(t_i)}}}$ and $t_{e_{t_{i+1}}}$. To compute $DP_{i+1}$, we need to change the association of two columns, $col_{cur}$ and $col_{cur+1}$.
The last column in $DP_i$ ($col_{cur}$) is reassociated with $t_{e_{C_{id(t_i)}}}$: \begin{eqnarray*}
&&col_{cur}.part\_id = id(t_i),\\
&&col_{cur}.prob = \sum_{\begin{subarray}{l}t_{j''}\in C_{id(t_i)}\\ 1\leq j'' \leq i\end{subarray}} p(t_{j''}). 
\end{eqnarray*}
The last column in $DP_{i+1}$ ($col_{cur+1}$) is associated with $t_{e_{t_{i+1}}}$:
\begin{eqnarray*}
&&col_{cur+1}.part\_id = id(t_{i+1}),\\
&& col_{cur+1}.prob=p(t_{i+1}).
\end{eqnarray*}

\begin{example}\label{expl_rollback}
Consider the following data\footnote{We explicitly include partition information into the representation, and thus the horizontal lines do not represent partition here.}, and a top-$2$ query. 
\smallskip
\begin{center}
\begin{tabular}{ccc}
\begin{tabular}{c}
 \\
$t_1$\\
$t_2$\\
$t_3$\\
$t_4$\\
$t_5$
\end{tabular}
\begin{tabular}{|c|c|}
\hline
\small{Part}
&\small{Score}\\
\hline
$C_1$ & $0.9$ \\
\hline
$C_2$ & $0.8$ \\
\hline
$C_3$ & $0.7$ \\
\hline
$C_1$ & $0.6$ \\
\hline
$C_2$ & $0.5$ \\
\hline
\end{tabular}
\begin{tabular}{c}
\small{Prob.} \\
\hline
$0.3$\\
$0.1$\\
$0.2$\\
$0.4$\\
$0.7$
\end{tabular}
\end{tabular}
\end{center}
Tuples are processed in the decreasing order of their scores, i.e., $t_1, t_2, \ldots, t_5$. Figure \ref{fig_rbevolution} illustrates each $DP_i$ table after the processing of tuple $t_i$. The annotation $(part\_id, prob)$ of each column is also illustrated. The entry in bold is the Global-Top$k$ probability of the corresponding tuple inducing the event relation.

\begin{figure}[htbp]
\subtable[$DP_1$]{
\begin{tabular}{|c|c|}
  \hline
  \backslashbox[8mm]{$k$}{col} & $\begin{array}{c} col_1 \\(1, 0.3) \end{array}$ \\
  \hline
  0 & $0$ \\
  \hline
  1 & $0.3$ \\
  \hline
  2 & $\mathbf{0.3}$ \\
  \hline
\end{tabular}
}
\subtable[$DP_2$]{
\begin{tabular}{|c|c|c|}
  \hline
  \backslashbox[8mm]{$k$}{col} & $\begin{array}{c} col_1 \\(1, 0.3) \end{array}$ & $\begin{array}{c} col_2 \\ (2, 0.1) \end{array}$ \\
  \hline
  0 & $0$ & $0$  \\
  \hline
  1 & $0.3$ & $0.07$  \\
  \hline
  2 & $0.3$ & $\mathbf{0.1}$ \\
  \hline
\end{tabular}
}
\subtable[$DP_3$]{
\begin{tabular}{|c|c|c|c|}
  \hline
  \backslashbox[8mm]{$k$}{col} & $\begin{array}{c} col_1 \\(1, 0.3) \end{array}$ & $\begin{array}{c} col_2 \\ (2, 0.1) \end{array}$ & $\begin{array}{c} col_3 \\ (3, 0.2) \end{array}$ \\
  \hline
  0 & $0$ & $0$ & $0$ \\
  \hline
  1 & $0.3$ & $0.07$ & $0.126$ \\
  \hline
  2 & $0.3$ & $0.1$ & $\mathbf{0.194}$\\
  \hline
\end{tabular}
}\\
\subtable[$DP_4$]{
\begin{tabular}{|c|c|c|c|}
  \hline
  \backslashbox[8mm]{$k$}{col} & $\begin{array}{c} col_1 \\(2, 0.1) \end{array}$ & $\begin{array}{c} col_2 \\ (3, 0.2) \end{array}$ & $\begin{array}{c} col_3 \\ (1, 0.4) \end{array}$ \\
  \hline
  0 & $0$ & $0$ & $0$\\
  \hline
  1 & $0.1$ & $0.18$  & $0.288$\\
  \hline
  2 & $0.1$ & $0.2$ & $\mathbf{0.392}$\\
  \hline
\end{tabular}
\label{fig_rbevolution_4}
}
\subtable[$DP_5$]{
\begin{tabular}{|c|c|c|c|}
  \hline
  \backslashbox[8mm]{$k$}{col} & $\begin{array}{c} col_1 \\(3, 0.2) \end{array}$ & $\begin{array}{c} col_2 \\ (1, 0.7) \end{array}$ & $\begin{array}{c} col_3 \\ (2, 0.7) \end{array}$ \\
  \hline
  0 & $0$ & $0$ & $0$\\
  \hline
  1 & $0.2$ & $0.56$  & $0.168$\\
  \hline
  2 & $0.2$ & $0.7$ & $\mathbf{0.602}$\\
  \hline
\end{tabular}
}
\caption{DP table evolution in \rollback}\label{fig_rbevolution}
\end{figure}

\noindent Take the processing of $t_3$ for example. Since $t_3$ is independent of $t_2$ and $t_1$, this is Case 2.
Therefore, the last column in $DP_2$ ($col_2$) needs to be reassociated with 
$t_{e_{C_{id(t_2)}}}=t_{e_{C_2}}$
in $E_3$. In $DP_3$, 
\begin{eqnarray*}
&&col_2.part\_id = id(t_2) = 2,\\
&&col_2.prob = \sum_{\begin{subarray}{l}t_{j''}\in C_2\\ 1\leq j'' \leq 2\end{subarray}} p(t_{j''})=p(t_2)=0.1.
\end{eqnarray*} 
The last column in $DP_3$ ($col_3$) is associated with the event tuple $t_{e_{t_3}}$ generated by Rule 1 in $E_3$:
\begin{eqnarray*}
&&col_3.part\_id = id(t_3) = 3,\\
&&col_3.prob = p(t_3) = 0.2.
\end{eqnarray*} Compared to $DP_2$, the first column with an annotation change in $DP_3$ is $col_3$. The DP table needs to be recomputed from $col_3$ (inclusive) upwards. In this case, it is only $col_3$.
Notice that, even though the annotation of $col_2$ does not change from $DP_2$ to $DP_3$, its meaning changes. In $DP_2$, $col_2$ is associated with $t_{e_{t_2}}$ in $E_2$ instead.
\end{example}

In Case 1 and Case 2, the event tuple which we want to ``erase'' from $E_i$, i.e., $t_{e_{t_i}}$, is associated with the last column in $DP_i$. In Case 3, by Equation (\ref{eqn_rbcase3}), we want to ``erase'' from $E_i$ the event tuple $t_{e_{C_{id(t_{i+1})}}}$ in addition to $t_{e_{t_i}}$. Assume $t_{e_{C_{id(t_{i+1})}}}$ is associated with $col_{j}$ in $DP_i$, and the columns in $DP_i$ are 
\[col_1,\ldots, col_{j-1}, col_j, col_{j+1}, \ldots, col_{cur-1}, col_{cur}\] 
which correspond to 
\[t_{e_{C_{i_1}}},\ldots, t_{e_{C_{i_{j-1}}}}, t_{e_{C_{i_j}}}, t_{e_{C_{i_{j+1}}}}, \ldots, t_{e_{C_{i_{cur-1}}}}, t_{e_{t_i}}\] 
in $E_i$ respectively. Obviously, $i_j=id(t_{i+1})$. By Equation (\ref{eqn_rbcase3}), 
\[E_{i+1}=\{t_{e_{C_{i_1}}},\ldots, t_{e_{C_{i_{j-1}}}}, t_{e_{C_{i_{j+1}}}}, \ldots, t_{e_{C_{i_{cur-1}}}}, t_{e_{C_{id(t_i)}}}, t_{e_{t_i}}\}.\] 
By {\em Fact 2}, as long as $t_{e_{t_i}}$ is associated with the last column in $DP_{i+1}$, the column association order of other tuples in $E_{i+1}$ does not matter in computing the Global-Top$k$ probability of $t_i$. By adopting a column association order such that 
\[t_{e_{C_{i_1}}},\ldots, t_{e_{C_{i_{j-1}}}}\]
is associated with 
\[col_1,\ldots, col_{j-1}\] 
respectively in $DP_{i+1}$, we can reuse the first $j-1$ columns already computed in $DP_i$. In our DP computation, the values in a column depend on the values in its previous column. Once we change the values in $col_j$, every $col_{j'}$, $j'>j$, needs to be recomputed regardless. Therefore, the recomputation cost is the same for any column association order of event tuples 
\[t_{e_{C_{i_{j+1}}}}, \ldots, t_{e_{C_{i_{cur-1}}}}, t_{e_{C_{id(t_i)}}}.\] 
In \rollback, we simply use this order above as the column association order.
In fact, the name of this optimization, \rollback, refers to the fact that we are ``rolling back'' the computation in the DP table until we hit $col_j$ and recompute all the columns with an index equal to or higher than $j$.

\begin{example}\label{expl_rollback2}
Continuing Example \ref{expl_rollback}, consider the processing of $t_5$. $t_5$ is independent of $t_4$, while $t_5$ and $t_2$ are exclusive. Therefore, this is Case 3. 
We first locate $col_j$ associated with $t_{e_{C_{id(t_5)}}}=t_{e_{C_2}}$ in $DP_4$. In this case, it is $col_1$. Then, we roll all the way back to $col_1$ in $DP_4$, erasing every column on the way including $col_1$. As $col_j=col_1$, there is no column from $DP_4$ that we can reuse in $DP_5$. We move on to recompute $col_{j'}$, $j\leq j'$, in $DP_5$ that are associated with $t_{e_{C_3}}$ and $t_{e_{C_{id(t_4)}}}=t_{e_{C_1}}$. In particular, $col_2$ in $DP_5$ is associated with $t_{e_{C_1}}$. Thus, 
\begin{eqnarray*}
col_2.part\_id &=& 1,\\
col_2.prob &=& \sum_{\begin{subarray}{l}t_{j''}\in C_1, 1\leq j'' \leq 4\end{subarray}} p(t_{j''})\\
&=&p(t_1)+p(t_4)\\
&=&0.3+0.4\\
&=&0.7.
\end{eqnarray*} 
The last column in $DP_5$ is again associated with the event tuple $t_{e_{t_5}}$ generated by Rule 1 in $E_5$.

Out of the five tuples, the processing of $t_1, t_2, t_3$ is of Case 2, and the processing of $t_4, t_5$ is of Case 3.
Whenever we compute/recompute the DP table, the event tuples associated with the columns are from the induced event relation, and therefore independent. Thus, every DP table computation progresses in the same fashion as that with the DP table in Example \ref{expl_dp}.

Finally, we keep the Global-Top$2$ probability of each tuple (from the original probabilistic relation) in a priority queue. 
When we finish processing all the tuples, we get the top-$2$ winners. 
In this example, the priority queue is updated every time we get an entry in bold.
The winners are $t_5$ and $t_4$ with the Global-Top$2$ probability $0.602$ and $0.392$ respectively.
\end{example}

\subsubsection{RollbackSort}

For the rollback operation in Case 3 of \rollback, define its \emph{depth} as the number of columns recomputed in rolling back excluding the last column. 
For example, when processing $t_5$ in Example \ref{expl_rollback2}, $col_1, col_2$ and $col_3$ are recomputed in $DP_5$. Therefore the \emph{depth} of this rollback operation is $3-1=2$.

Recall that in Case 3 of \rollback, we adopt an arbitrary order 
\[t_{e_{C_{i_{j+1}}}}, \ldots, t_{e_{C_{i_{cur-1}}}}, t_{e_{C_{id(t_i)}}}\] 
to process those event tuples in $DP_{i+1}$. The Global-Top$k$ computation in $E_{i+1}$ does not stipulate any particular order over those tuples. Any permutation of this order is equally valid. 
The intuition behind \rollbacksort\ is that we will be able to find a permutation that will reduce the {\em depth} of future rollback operations (if any), 
given additional statistics on the probabilistic relation $R^p$, namely the count of the tuples in each part of the partition.
Theoretically, it requires an extra pass over the relation to compute the statistics. In practice, however, this extra pass is often not needed because this statistics can be precomputed and stored.

In \rollbacksort, if the current situation is Case 3, we do a stable sort on 
\[t_{e_{C_{i_{j+1}}}}, \ldots, t_{e_{C_{i_{cur-1}}}}, t_{e_{C_{id(t_i)}}}\]
in the non-decreasing order of the number of unseen tuples in its corresponding part, and then use the resulting order to process those event tuples. The intuition is that each unseen tuple has the potential to trigger a rollback operation. By pushing the event tuple with the most unseen tuples close to the end of the current DP table, we could reduce the depth of future rollback operations.
In order to facilitate this sorting, we add one more component $unseen$ to the annotation of each column.

\begin{example}\label{expl_rollbacksort}
We redo the problem in Example \ref{expl_rollback} and Example \ref{expl_rollback2} using \rollbacksort. Now, the annotation of each column becomes $(part\_id, prob, unseen)$. The evolution of the DP table is shown in Figure \ref{fig_rbsortevolution}. In \rollbacksort, the statistics on all parts are available: $2$ tuples in $C_1$, $2$ tuples in $C_2$ and $1$ tuple in $C_3$.

\begin{figure}[htbp]
\subtable[$DP_1$]{
\begin{tabular}{|c|c|}
  \hline
  \backslashbox[6mm]{$k$}{col} & $\begin{array}{c} col_1 \\(1, 0.3, 1) \end{array}$ \\
  \hline
  0 & $0$ \\
  \hline
  1 & $0.3$ \\
  \hline
  2 & $\mathbf{0.3}$ \\
  \hline
\end{tabular}
}
\subtable[$DP_2$]{
\begin{tabular}{|c|c|c|}
  \hline
  \backslashbox[6mm]{$k$}{col} & $\begin{array}{c} col_1 \\(1, 0.3, 1) \end{array}$ & $\begin{array}{c} col_2 \\ (2, 0.1, 1) \end{array}$ \\
  \hline
  0 & $0$ & $0$  \\
  \hline
  1 & $0.3$ & $0.07$  \\
  \hline
  2 & $0.3$ & $\mathbf{0.1}$ \\
  \hline
\end{tabular}
}
\subtable[$DP_3$]{
\begin{tabular}{|c|c|c|c|}
  \hline
  \backslashbox[6mm]{$k$}{col} & $\begin{array}{c} col_1 \\(1, 0.3, 1) \end{array}$ & $\begin{array}{c} col_2 \\ (2, 0.1, 1) \end{array}$ & $\begin{array}{c} col_3 \\ (3, 0.2, 0) \end{array}$ \\
  \hline
  0 & $0$ & $0$ & $0$ \\
  \hline
  1 & $0.3$ & $0.07$ & $0.126$ \\
  \hline
  2 & $0.3$ & $0.1$ & $\mathbf{0.194}$\\
  \hline
\end{tabular}
}\\
\subtable[$DP_4$]{
\begin{tabular}{|c|c|c|c|}
  \hline
  \backslashbox[6mm]{$k$}{col} & $\begin{array}{c} col_1 \\(3, 0.2, 0) \end{array}$ & $\begin{array}{c} col_2 \\ (2, 0.1, 1) \end{array}$ & $\begin{array}{c} col_3 \\ (1, 0.4, 0) \end{array}$ \\
  \hline
  0 & $0$ & $0$ & $0$\\
  \hline
  1 & $0.2$ & $0.08$  & $0.288$\\
  \hline
  2 & $0.2$ & $0.1$ & $\mathbf{0.392}$\\
  \hline
\end{tabular}
}
\subtable[$DP_5$]{
\begin{tabular}{|c|c|c|c|}
  \hline
  \backslashbox[6mm]{$k$}{col} & $\begin{array}{c} col_1 \\(3, 0.2, 0) \end{array}$ & $\begin{array}{c} col_2 \\ (1, 0.7, 0) \end{array}$ & $\begin{array}{c} col_3 \\ (2, 0.7, 0) \end{array}$ \\
  \hline
  0 & $0$ & $0$ & $0$\\
  \hline
  1 & $0.2$ & $0.56$  & $0.168$\\
  \hline
  2 & $0.2$ & $0.7$ & $\mathbf{0.602}$\\
  \hline
\end{tabular}
}
\caption{DP table evolution in \rollbacksort}\label{fig_rbsortevolution}
\end{figure}

\noindent Consider the processing of $t_1$ in $DP_1$. As we just see one tuple $t_1$ from $C_1$, there is one more unseen tuple from $C_1$ coming in the future. Therefore, $col_1.unseen = 1$. All the other $unseen$ annotations are computed in the same way.

When processing $t_4$ (Case 3), the column associated with $t_{e_{C_{id(t_4)}}}$ in $DP_3$ is $col_1$.
We roll back to $col_1$ as before and recompute all the columns upwards in $DP_4$. Notice that, the recomputation is performed in the order $t_{e_{C_3}}$, $t_{e_{C_2}}$, in contrast to the order $t_{e_{C_2}}$, $t_{e_{C_3}}$ used in Example \ref{expl_rollback2} (Figure \ref{fig_rbevolution_4}). $C_2$ has one more unseen tuple which can trigger the rollback operation while there are no more unseen tuples from $C_3$. The benefit of this order becomes clear when we process $t_5$. We only need to rollback to $col_2$ in $DP_4$. The $depth$ of this rollback operation is $1$. Recall that the $depth$ of the same rollback operation is $2$ in Example \ref{expl_rollback2}. In other words, we save the computation of $1$ column by applying \rollbacksort.

\end{example}

\rollback\ and \rollbacksort\ significantly improve the performance in practice, as we will see in Section \ref{sec_exp}. 
The price we pay for this speedup is an increase in the space usage. The space complexity is $O(kn)$ for both optimization.
The quadratic theoretic bound on running time remains unchanged. 

\section{Global-Top$k$ under General Scoring Functions}\label{sec_WO}

\subsection{Semantics and Postulates}
\subsubsection{Global-Top$k$ Semantics with Allocation Policy}
Under a general scoring function, the Global-Top$k$ semantics remains the same. However, the definition of Global-Top$k$ probability in Definition \ref{def_topkprob_TO} needs to be generalized to handle {\em ties}.

Recall that under an injective scoring function $s$, there is a unique top-$k$ answer set $S$ in every possible world $W$. When the scoring function $s$ is non-injective, there may be multiple top-$k$ answer sets $S_1, \ldots, S_d$, each of which is returned nondeterministically. Therefore, for any tuple $t\in \cap S_i, i=1,\ldots,d$, the world $W$ contributes $Pr(W)$ to the Global-Top$k$ probability of $t$. On the other hand, for any tuple $t\in (\cup S_i-\cap S_i), i=1\ldots, d$, the world $W$ contributes only a {\em fraction} of $Pr(W)$ to the Global-Top$k$ probability of $t$. The {\em allocation policy} determines the value of this fraction, i.e., the {\em allocation coefficient}. Denote by $\alpha(t, W)$ the allocation coefficient of a tuple $t$ in a world $W$. 
Let $\ptopktuple{k}{s}(W)=\cup S_i, i=1,\ldots,d$.

\begin{definition}[Global-Top$k$ Probability under a General Scoring Function]\label{def_topkprob_WO} Assume a probabilistic relation $R^p=\langle R, p, \mathcal{C}\rangle$, a non-negative integer $k$ and a scoring function $s$ over $R^p$. 
For any tuple $t$ in $R$, the Global-Top$k$ probability of $t$, denoted by $P^{R^p}_{k,s}(t)$, is the sum of the (partial) probabilities of all possible worlds of $R^p$ whose top-$k$ answer set may contain $t$.

\begin{equation}\label{eqn_topkprob_WO}
P^{R^p}_{k,s}(t)=\sum_{
\begin{subarray}{l}
W\in pwd(R^p)\\
t\in \ptopktuple{k}{s}(W)
\end{subarray}}\alpha(t,W) Pr(W).
\end{equation}

\end{definition}

With no prior bias towards any tuple, it is natural to assume that each of $S_1, \ldots, S_d$ is returned nondeterministically with {\em equal} probability. Notice that this probability has nothing to do with tuple probabilities. Rather, it is determined by the number of equally qualified top-$k$ answer sets. Hence, we have the following {\em Equal} allocation policy. 

\begin{definition}[Equal Allocation Policy]
Assume a probabilistic relation $R^p=\langle R, p, \mathcal{C}\rangle$, a non-negative integer $k$ and a scoring function $s$ over $R^p$.
For a possible world $W\in pwd(R^p)$ and a tuple $t\in W$, let $a=|\{t'\in W| t' \succ_s t\}|$ and $b=|\{t'\in W| t' \sim_s t\}|$
\begin{displaymath}
\alpha(t,W) = \left\{ \begin{array}{ll}
1 & \textrm{~~~~if } a < k \textrm{ and } a+b \leq k\\
\dfrac{k-a}{b} & \textrm{~~~~if } a < k \textrm{ and } a+b > k
\end{array} \right.
\end{displaymath}
\end{definition}

This notion of {\em Equal} allocation policy is in the spirit of {\em uniform allocation policy} introduced in \cite{DBLP:journals/vldb/BurdickDJRV07} to handle imprecision in OLAP, although the specified goals are different. Note that \cite{DBLP:journals/vldb/BurdickDJRV07} also introduces other allocation policies based on additional information. In our application, it is also possible to design other allocation policies given additional information.

\subsubsection{Satisfaction of Postulates}
The semantic postulates in Section \ref{sec_postulates} are directly applicable to Global-Top$k$ with allocation policy.
In the Appendix A, we show that the {\em Equal} allocation policy preserves the semantic postulates of Global-Top$k$.

\subsection{Query Evaluation in Simple Probabilistic Relations}
\begin{definition} \label{def_T}
Let $R^p=\langle R, p, \mathcal{C}\rangle$ be a probabilistic relation, $k$ a non-negative integer and $s$ a general scoring function over $R^p$. Assume that $R=\{t_1, t_2, \ldots, t_n\}$, $t_1\succeq_s t_2\succeq_s \ldots \succeq_s t_n$. Let $T^{R^p}_{k, [i]}$, $k\leq i$, be the sum of the probabilities of all possible worlds of exactly $k$ tuples from $\{t_1, \ldots, t_i\}$:
\[
T^{R^p}_{k, [i]}=\sum_{\begin{subarray}{l}
W\in pwd(R^p)\\
|W\cap\{t_1,\ldots, t_i\}|= k
\end{subarray}}Pr(W)
\]
\end{definition}

As usual, we omit the superscript in $T^{R^p}_{k,[i]}$, i.e., $T_{k,[i]}$, when the context is unambiguous. 
Remark \ref{rmk_kworlds_TO} shows that in a simple probabilistic relation $T_{k,[i]}$ can be computed efficiently.

\begin{remark}\label{rmk_kworlds_TO}
Let $R^p=\langle R, p, \mathcal{C}\rangle$ be a simple probabilistic relation, $k$ a non-negative integer and $s$ a general scoring function over $R^p$. Assume that $R=\{t_1, t_2, \ldots, t_n\}$, $t_1\succeq_s t_2\succeq_s \ldots \succeq_s t_n$. 
For any $i$, $1\leq i\leq n-1$, $T^{R^p}_{k,[i]}$ can be computed using the DP table for computing the Global-Top$k$ probabilities in $R^p$ under
an order-preserving injective scoring function $s'$ such that $t_1\succ_{s'} t_2\succ_{s'} \ldots \succ_{s'} t_n$.
\end{remark}

\begin{proof} By case study,

\begin{itemize}
\item Case 1: If $k=0$, $1\leq i\leq n-1$, then 
\begin{equation}\label{eqn_bigT_case1}
T^{R^p}_{k, [i]}=\prod_{1\leq j\leq i}\overline{p}(t_j)=\frac{P^{R^p}_{1,s'}(t_{i+1})}{p(t_{i+1})}
\end{equation}
\item Case 2:
For every $1\leq k\leq i\leq n-1$, by the definition of $T^{R^p}_{k, [i]}$, we have
\begin{eqnarray*}
T^{R^p}_{k, [i]}&=&\sum_{\begin{subarray}{l}
W\in pwd(R^p)\\
|W\cap\{t_1,\ldots, t_i\}|\leq k
\end{subarray}}Pr(W)-
\sum_{\begin{subarray}{l}
W\in pwd(R^p)\\
|W\cap\{t_1,\ldots, t_i\}|\leq k-1
\end{subarray}}Pr(W)
\end{eqnarray*}

In the DP table computing the Global-Top$k$ probabilities in $R^p$ under function $s'$, we have
\begin{eqnarray*}
P^{R^p}_{k+1,s'}(t_{i+1})&=&\sum_{
\begin{subarray}{l}
W\in pwd(R^p)\\
t_{i+1}\in \topktuple{k+1}{s'}(W)
\end{subarray}}Pr(W)\hspace{0.9in}(s' \textrm{ is injective}) \\
&=&\sum_{
\begin{subarray}{l}
W\in pwd(R^p)\\
|W\cap\{t_1,\ldots, t_i\}|\leq k\\
t_{i+1}\in W
\end{subarray}}Pr(W)\\
&=&p(t_{i+1})\sum_{
\begin{subarray}{l}
W\in pwd(R^p)\\
|W\cap\{t_1,\ldots, t_i\}|\leq k
\end{subarray}}Pr(W)\hspace{0.4in}\textrm{(tuples are independent)}
\end{eqnarray*}

Therefore,
\begin{eqnarray}\label{eqn_bigT_case2}
T^{R^p}_{k, [i]}&=&\frac{P^{R^p}_{k+1,s'}(t_{i+1})}{p(t_{i+1})}-\frac{P^{R^p}_{k,s'}(t_{i+1})}{p(t_{i+1})}
\end{eqnarray}

Since $1\leq k\leq i\leq n-1$, both $P^{R^p}_{k+1,s'}(t_{i+1})$ and $P^{R^p}_{k,s'}(t_{i+1})$ can be computed by the DP table used to compute the Global-Top$k$ probabilities of tuples in $R^p$ under the injective scoring function $s'$.

\end{itemize}

\end{proof}

Remark \ref{rmk_tie_TO} shows that we can compute Global-Top$k$ probability under a general scoring function in polynomial time for an extreme case, where the probabilistic relation is simple and all tuples tie in scores. As we will see shortly, this special case plays an important role in our major result in Proposition \ref{prop_wo_simple}.

\begin{remark}\label{rmk_tie_TO}
Let $R^p=\langle R, p, \mathcal{C}\rangle$ be a simple probabilistic relation, $k$ a non-negative integer and $s$ a general scoring function over $R^p$. Assume that $R=\{t_1, \ldots, t_m\}$ and $t_1\sim_s t_2\sim_s \ldots \sim_s t_m$.  For any tuple $t_i, 1\leq i\leq m$, the Global-Top$k$ probability of $t_i$, i.e., $P^{R^p}_{k,s}(t_i)$, can be computed using Remark \ref{rmk_kworlds_TO}.

\end{remark}

\begin{proof} 
If $k>m$, it is trivial that $P^{R^p}_{k,s}(t_i)=p(t_i)$. Therefore, we only prove the case when $k\leq m$.
According to Equation (\ref{eqn_topkprob_WO}), for any $i$, $1\leq i\leq m$, 

\begin{eqnarray*}
P^{R^p}_{k,s}(t_i)&=&\sum_{j=1}^{m}\sum_{\begin{subarray}{l}
W\in pwd(R^p)\\
t_i\in \ptopktuple{k}{s}(W),|W|=j
\end{subarray}}\alpha(t_i, W)Pr(W)\nonumber\\
&=&\sum_{j=1}^{m}\sum_{\begin{subarray}{l}
W\in pwd(R^p)\\
t_i\in W, |W|= j
\end{subarray}}\alpha(t_i, W)Pr(W)\hspace{0.2in}(\textrm{Since all tuple tie }, \ptopktuple{k}{s}(W)=W)\nonumber\\
&=&\sum_{j=1}^{k}\sum_{\begin{subarray}{l}
W\in pwd(R^p)\\
t_i\in W, |W|=j
\end{subarray}}\alpha(t_i, W)Pr(W)+
\sum_{j=k+1}^{m}\sum_{\begin{subarray}{l}
W\in pwd(R^p)\\
t_i\in W, |W|=j
\end{subarray}}\alpha(t_i, W)Pr(W)\nonumber\\
&=&\sum_{j=1}^{k}\sum_{\begin{subarray}{l}
W\in pwd(R^p)\\
t_i\in W, |W|=j
\end{subarray}}Pr(W)+
\sum_{j=k+1}^{m}\frac{k}{j}\sum_{\begin{subarray}{l}
W\in pwd(R^p)\\
t_i\in W, |W|=j
\end{subarray}}Pr(W)
\end{eqnarray*} 

With out loss of generality, assume $i=m$, then the above equation becomes

\begin{eqnarray}\label{eqn_tie}
P^{R^p}_{k,s}(t_m)
&=&\sum_{j=1}^{k}\sum_{\begin{subarray}{l}
W\in pwd(R^p)\\
t_m\in W, |W|=j
\end{subarray}}Pr(W)+
\sum_{j=k+1}^{m}\frac{k}{j}\sum_{\begin{subarray}{l}
W\in pwd(R^p)\nonumber\\
t_m\in W, |W|=j
\end{subarray}}Pr(W)\\
&=&p(t_i)(\sum_{j=1}^{k}T^{R^p}_{j-1,[m-1]}+
\sum_{j=k+1}^{m}\frac{k}{j}T^{R^p}_{j-1,[m-1]}) 
\end{eqnarray} 



By Remark \ref{rmk_kworlds_TO}, every $T^{R^p}_{j-1,[m-1]}$, $0\leq j-1\leq m-1$, can be computed by the DP table computing Global-Top$k$ probabilities in $R^p$ under an order preserving injective scoring function $s'$, and Equation (\ref{eqn_bigT_case1}) or  (\ref{eqn_bigT_case2}). Therefore,
Equation (\ref{eqn_tie}) can be computed using Remark \ref{rmk_kworlds_TO}. 
\end{proof}

Based on Remark \ref{rmk_kworlds_TO} and Remark \ref{rmk_tie_TO}, we design Algorithm \ref{alg_ind_wo} and prove its correctness in Theorem \ref{thm_ind_wo} using Proposition \ref{prop_wo_simple}.

Assume $R^p=\langle R, p, \mathcal{C}\rangle$ where $R=\{t_1, t_2, \ldots, t_n\}$ and $t_1\succeq_s t_2\succeq_s \ldots \succeq_s t_n$. For any $t_l\in R$, $i_l$ is the largest index such that $t_{i_l}\succ_s t_l$, and $j_l$ is the largest index such that $t_{j_l}\succeq_s t_l$. 

Intuitively, Algorithm \ref{alg_ind_wo} and Proposition \ref{prop_wo_simple} convey the idea that, in a simple probabilistic relation, the computation of Global-Top$k$ under the {\em Equal} allocation policy can be simulated by the following procedure:

\begin{enumerate}
\item[(S1)] Independently flip a biased coin with probability $p(t_j)$ for each tuple $t_j\in R=\{t_1,t_2,\ldots, t_n\}$, which gives us a possible world $W\in pwd(R^p)$;
\item[(S2)] Return a top-$k$ answer set $S$ of $W$ nondeterministically (with equal probability in the presence of multiple top-$k$ sets). The Global-Top$k$ probability of $t_l$ is the probability that $t_l\in S$.
\end{enumerate}

The above Step (S1) can be further refined into:
\begin{enumerate}
\item[(S1.1)] Independently flip a biased coin with probability $p(t_j)$ for each tuple $t_j\in R_{A}=\{t_1,t_2\ldots, t_{i_l}\}$, which gives us a collection of tuples $W_{A}$;
\item[(S1.2)] Independently flip a biased coin with probability $p(t_j)$ for each tuple $t_j\in R_{B}=\{t_{i_l+1}, \ldots, t_n\}$, which gives us a collection of tuples $W_{B}$. $W=W_{A}\cup W_{B}$ is a possible world from $pwd(R^p)$;
\end{enumerate}

In order for $t_l$ to be in $S$, $W_{A}$ can have at most $k-1$ tuples. Let $|W_{A}|=k'$, then $k'< k$. Every top-$k$ answer set of $W$ contains all $k'$ tuples from $W_{A}$, plus the top-$(k-k')$ tuples from $W_{B}$. For $t_l$ to be in $S$, it has to be in the top-$(k-k')$ set of $W_{B}$.
Consequently, the probability of $t_l\in S$, i.e., the Global-Top$k$ probability of $t_l$, is the joint probability that $|W_{A}|=k'<k$ and $t_l$ belongs to the top-$(k-k')$ set of $W_{B}$. 
The former is $T_{k',[i_l]}$ and the latter is $P^{R^p_{B}}_{k-k',s}(t_l)$ , where $R^p_{B}$ is $R^p$ restricted to $R_{B}$.
Again, due to the independence among tuples, Step (S1.1) and Step (S1.2) are independent, and their joint probability is simply the product of the two.

Further notice that since $t_l$ has the highest score in $R_{B}$ and all tuples are independent in $R_{B}$, and any tuple with a score lower than that of $t_l$ does not have an influence on $P^{R^p_{B}}_{k-k',s}(t_l)$. In other words, $P^{R^p_{B}}_{k-k',s}(t_l)=P^{R^p_s(t_l)}_{k-k',s}(t_l)$, where $R^p_s(t_l)$ is $R^p$ restricted to all tuples tying with $t_l$ in $R$. Notice that the computation of $P^{R^p_s(t_l)}_{k-k',s}(t_l)$ is the extreme case addressed in Remark \ref{rmk_tie_TO}.

Algorithm \ref{alg_ind_wo} elaborates the algorithm based on the idea above, where $m=j_l-i_l$ is the number of tuples tying with $t_l$ (including $t_l$). 

Furthermore, Algorithm \ref{alg_ind_wo} exploits the overlapping among DP tables and makes the following two optimizations:

\begin{enumerate}
\item Use a single DP table to collect the information needed to compute all $T_{k',[i_l]}$, $k'=0,\ldots,k-1$, $l=1,\ldots,n$ and $k'\leq i_l$ (Line \ref{ln_wa_dptable}).

Notice that by definition, when $1\leq l \leq n$, $1\leq i_l\leq n-1$. It is easy to see that the DP table computing $T_{k-1,[n-1]}$ subsumes all other DP tables.

\item Use a single DP table to compute all $P^{R^p_s(t_l)}_{k-k',s}(t_l)$, $k'=0,\ldots,k-1$, for a tuple $t_l$ (Lines \ref{ln_wb_dptable}-\ref{ln_wb_compute_end}).

Notice that in Equation (\ref{eqn_tie}), for different $k'$, the computation of $P^{R^p_s(t_l)}_{k-k',s}(t_l)$  requires the same set of $T^{R^p_s(t_l)}_{j,[m-1]}$ values (Lines \ref{ln_wb_compute_bigT_start}-\ref{ln_wb_compute_bigT_end}). In Line \ref{ln_wb_compute}, $P^{R^p_s(t_l)}_{k-k',s}(t_l)$ is abbreviated as $P_l(k'')$, where $k''=k-k'$, to emphasize the changing parameter $k'$.
\end{enumerate}

Each DP table computation uses a call to Algorithm \ref{alg_indsub} (Line \ref{ln_wa_dptable} in Algorithm \ref{alg_ind_wo}, Line \ref{ln_sub_wb_dptable} in Algorithm \ref{alg_indsub_wo}).

\begin{algorithm}
\caption{\textbf{(Ind\_Topk\_Gen)} Evaluate Global-Top$k$ Queries in a Simple Probabilistic Relation under a General Scoring Function}
\label{alg_ind_wo}
\begin{algorithmic}[1]
\REQUIRE $R^p=\langle R, p, \mathcal{C}\rangle, k$
\ENSURE tuples in $R$ are sorted in the non-increasing order based on the scoring function $s$
\STATE Initialize a fixed cardinality $(k+1)$ priority queue $Ans$ of $\langle t, prob\rangle$ pairs, which compares pairs on $prob$, i.e., the Global-Top$k$ probability of $t$;
\STATE Get the DP table for computing $T_{k',[i]}, k'=0,\ldots k-1, i=1,\ldots,n-1$, $k'\leq i$  using Algorithm \ref{alg_indsub}, i.e., 
\[q(0\ldots k, 1\ldots |R|)=\textrm{Ind\_Topk\_Sub}(R^p, k);\]\label{ln_wa_dptable}
\FOR{$l=1$ to $|R|$}
\STATE $m=j_l-i_l$;
\IF{$m == 1$}
\STATE Add $\langle t_l, q(k, l)\rangle$ to $Ans$;\label{ln_notie}
\ELSE
\STATE Get the DP table for computing $P_{k-k',s}^{R_{s}^{p}(t_l)}(t_l)$, i.e., $P_l(k-k')$, $k'=0,\ldots, k-1$
\[q_{tie}(0\ldots m, 1\ldots m)=\textrm{Ind\_Topk\_Gen\_Sub}(R^p_s(t_l), t_l, m);\]\label{ln_wb_dptable}\label{ln_tie_start}
\FOR{$k''=0$ to $m-1$}\label{ln_wb_compute_bigT_start}
\STATE \[T^{R^p_s{(t_l)}}_{k'',[m-1]}=  \frac{q_{tie}(k''+1,m)-q_{tie}(k'',m)}{p(t_l)};\]\label{ln_wb_compute_bigT}
\ENDFOR\label{ln_wb_compute_bigT_end}
\FOR{$k''=1$ to $k$}\label{ln_wb_compute_start}
\STATE \[P_l(k'')=p(t_l)(\sum_{j=1}^{k''}T^{R^p_s(t_l)}_{j-1,[m-1]}+
\sum_{j=k''+1}^{m}\frac{k''}{j}T^{R^p_s(t_l)}_{j-1,[m-1]});\]\label{ln_wb_compute}
\ENDFOR \label{ln_wb_compute_end}
\STATE $P^{R^p}_{k,s}(t_l)=0$;\label{ln_wosimplemain_start}
\FOR{$k'=0$ to $k-1$}
\STATE \[T_{k',[i_l]}=  \frac{q(k'+1,i_l+1)-q(k',i_l+1)}{p(t_{i_l+1})};\]\label{ln_wa_compute}
\STATE
\[P^{R^p}_{k,s}(t_l) = P^{R^p}_{k,s}(t_l)+T_{k',[i_l]}\cdot P_l(k-k');\]
\ENDFOR\label{ln_wosimplemain_end}
\STATE Add $\langle t_l, P^{R^p}_{k,s}(t_l)\rangle$ to $Ans$;\label{ln_tie_end}
\ENDIF
\IF{$|Ans|>k$}
\STATE remove the pair with the smallest $prob$ value from $Ans$;
\ENDIF
\ENDFOR
\RETURN $\{t_i|\langle t_i,prob\rangle\in Ans\}$;
\end{algorithmic}
\end{algorithm}

\begin{algorithm}[H]
\caption{\textbf{(Ind\_Topk\_Gen\_Sub)} Compute the DP table for Global-Top$k$ probabilities in a Simple Probabilistic Relation under an All-Tie Scoring Function}
\label{alg_indsub_wo}
\begin{algorithmic}[1]
\REQUIRE $R^p_s(t_{target})=\langle R, p, \mathcal{C}\rangle, t_{target}, m$
\ENSURE $|R|=m$, $t_{target}\in R$
\STATE Rearrange tuples in $R$ such that $R=\{t_1,\ldots, t_{m-1}, t_{m}\}$ and $t_m=t_{target}$;
\STATE Assume the injective scoring function $s'$ is such that $t_1\succ_{s'}\ldots \succ_{s'}t_{m-1}\succ_{s'}t_{target}$;
\STATE Get the DP table
\[q_{tie}(0\ldots m, 1\ldots m)=\textrm{Ind\_Topk\_Sub}(R^p_s(t_{target}), m);\]\label{ln_sub_wb_dptable}
\RETURN $q_{tie}(0\ldots m, 1\ldots m)$;
\end{algorithmic}
\end{algorithm}

\begin{proposition}\label{prop_wo_simple}
Let $R^p=\langle R, p, \mathcal{C}\rangle$ be a simple probabilistic relation where $R=\{t_1, \ldots, t_n\}$, $t_1\succeq_s t_2\succeq_s \ldots \succeq_s t_n$, $k$ a non-negative integer and $s$ a scoring function. For every $t_l\in R$, the Global-Top$k$ probability of $t_l$ can be computed by the following equation:

\begin{equation}\label{eqn_wo_simple}
P^{R^p}_{k,s}(t_l)=\sum_{k'=0}^{k-1} T_{k',[i_l]}\cdot P_{k-k',s}^{R_{s}^{p}(t_l)}(t_l)
\end{equation}

\noindent where $R^p_s(t_l)$ is $R^p$ restricted to $\{t\in R|t\sim_s t_l\}$.
\end{proposition}

\noindent \emph{Proof}.~~~\emph{See} Appendix B.

\begin{theorem}[Correctness of Algorithm \ref{alg_ind_wo}]\label{thm_ind_wo}
Given a probabilistic relation $R^p=\langle R, p$, $\mathcal{C}\rangle$, a non-negative integer $k$ and a general scoring function $s$, Algorithm \ref{alg_ind_wo} correctly computes a Global-Top$k$ answer set of $R^p$ under the scoring function $s$. 
\end{theorem}

\begin{proof}
In Algorithm \ref{alg_ind_wo}, by Remark \ref{rmk_kworlds_TO}, Line \ref{ln_wa_dptable} and Line \ref{ln_wa_compute} correctly compute $T_{k',[i]}$ for $0\leq k'\leq k-1$, $1\leq i\leq n-1$, $k'\leq i$. 
The entries in Line \ref{ln_wb_dptable} serve to compute Line \ref{ln_wb_compute_bigT} by Equation (\ref{eqn_bigT_case2}). Recall that $R^p_s(t_l)$ is $R^p$ restricted to all tuples tying with $t_l$, which is the extreme case addressed in Remark \ref{rmk_tie_TO}.
By Remark \ref{rmk_tie_TO}, Line \ref{ln_wb_dptable} collects the information to compute $P^{R^p_s(t_l)}_{k-k',s}(t_l)$, i.e., $P_l(k'')$, $1\leq k''=k-k'\leq k$. Lines \ref{ln_wb_compute_start}-\ref{ln_wb_compute_end} correctly compute those values by Equation (\ref{eqn_tie}). 
Here, any non-existing $T^{R^p_s(t_l)}_{j-1,[m-1]}$, i.e., $j-1\not\in [0, m-1]$, is assumed to be zero.
By Proposition \ref{prop_wo_simple}, Lines \ref{ln_wosimplemain_start}-\ref{ln_wosimplemain_end} correctly compute the Global-Top$k$ probability of $t_l$. 
Also notice that in Line \ref{ln_notie}, the Global-Top$k$ probability of a tuple without tying tuples is retrieved directly. It is an optimization as the code handling the general case (i.e., $m>1$, Lines \ref{ln_tie_start}-\ref{ln_tie_end}) works for this special case as well. Again, the top-level structure with the priority queue in Algorithm \ref{alg_ind_wo} ensures that a Global-Top$k$ answer set is correctly computed.
\end{proof}

In Algorithm \ref{alg_ind_wo}, Line \ref{ln_wa_dptable} takes $O(kn)$, and for each tuple, there is one call to Algorithm \ref{alg_indsub_wo} in Line \ref{ln_wb_dptable}, which takes $O(m_{\max}^2)$, where $m_{\max}$ is the maximal number of tying tuples. Lines \ref{ln_wb_compute_bigT_start}-\ref{ln_wb_compute_bigT_end} take $O(m_{\max})$. Lines \ref{ln_wb_compute_start}-\ref{ln_wb_compute_end} take $O(km_{\max})$.
Therefore, Algorithm \ref{alg_ind_wo} takes $O(n\max(k,m_{\max}^2))$ altogether.

As before, the major space use is the computation of the two DP tables in Line \ref{ln_wa_dptable} and Line \ref{ln_wb_dptable}. A straightforward implementation leads to $O(kn)$ and $O(m_{\max}^2)$ space respectively. Therefore, the total space is $O(n\max(k,m_{\max}))$. Using a similar space optimization in Section \ref{topk_TO_Ind}, the space use for the two DP tables can be reduced to $O(k)$ and $O(m_{\max})$, respectively. Hence, the total space is $O(\max(k,m_{\max}))$.

\subsection{Query Evaluation in General Probabilistic Relations}

Recall that under an injective scoring function, every tuple $t$ in a general probabilistic relation $R^p=\langle R, p, \mathcal{C}\rangle $ induces a {\em simple} event relation $E^p$, and we reduce the computation of $t$'s Global-Top$k$ probability in $R^p$ to the computation of $t_{e_t}$'s Global-Top$k$ probability  in $E^p$. 

In the case of general scoring functions, we use the same reduction idea. However, now for each part $C_i\in \mathcal{C}, C_i\neq C_{id(t)}$, tuple $t$ induces in $E^p$ two {\em exclusive} tuples $t_{e_{C_i,\succ}}$ and $t_{e_{C_i,\sim}}$, corresponding to the {\em event} $e_{C_i,\succ}$ that ``there is a tuple from the part $C_i$ with a score {\em higher than} that of $t$'' and the {\em event} $e_{C_i,\sim}$ that ``there is a tuple from the part $C_i$ with a score {\em equal to} that of $t$'', respectively. In addition, in Definition \ref{def_ied_WO}, we allow the existence of tuples with probability $0$, in order to simplify the description of query evaluation algorithms. This is an artifact whose purpose will become clear in Theorem \ref{thm_recursion_WO}.

\begin{definition}[Induced Event Relation under General Scoring Functions]\label{def_ied_WO}
Given a probabilistic relation $R^p=\langle R, p, \mathcal{C}\rangle$, a scoring function $s$ over $R^p$ and a tuple $t\in C_{id(t)}\in \mathcal{C}$, the event relation induced by $t$, denoted by $E^p=\langle E, p^E, \mathcal{C}^E\rangle$, is a probabilistic relation whose support relation $E$ has only one attribute, $Event$. The relation $E$ and the probability function $p^E$ are defined by the following four generation rules and the postprocess step: 

\begin{itemize}
\item Rule 1.1:\hspace{0.2in} $t_{e_{t,\sim}}\in E \textrm{~and~} p^E(t_{e_{t,\sim}})=p(t)$;

\item Rule 1.2:\hspace{0.2in} $t_{e_{t,\succ}}\in E \textrm{~and~} p^E(t_{e_{t,\succ}})=0$;

\item Rule 2.1: \hfill \[\begin{array}{l}\forall C_i\in \mathcal{C}\wedge C_i\neq C_{id(t)}.
(t_{e_{C_i,\succ}}\in E) \textrm{ and } p^E(t_{e_{C_i},\succ})=\sum_{\begin{subarray}{l}
t'\in C_i\\
t'\succ_s t
\end{subarray}}p(t');\end{array}\]

\item Rule 2.2: \hfill \[\begin{array}{l}\forall C_i\in \mathcal{C}\wedge C_i\neq C_{id(t)}.
(t_{e_{C_i,\sim}}\in E) \textrm{ and } p^E(t_{e_{C_i},\sim})=\sum_{\begin{subarray}{l}
t'\in C_i\\
t'\sim_s t
\end{subarray}}p(t').\end{array}\]

\end{itemize}

Postprocess step: only when $p^E(t_{e_{C_i},\succ})$ and $p^E(t_{e_{C_i},\sim})$ are both $0$, delete both $t_{e_{C_i},\succ}$ and $t_{e_{C_i},\sim}$.
\end{definition}

\begin{proposition}\label{prop_iedscore_WO}
Given a probabilistic relation $R^p=\langle R, p, \mathcal{C}\rangle$ and a scoring function $s$, for any $t\in R^p$, the Global-Top$k$ probability of $t$ equals the Global-Top$k$ probability of $t_{e_t,\sim}$ when evaluating top-$k$ in the induced event relation $E^p=\langle E, p^E, \mathcal{C}^E\rangle$ under the scoring function $s^E:E\rightarrow \mathbb{R}$, $s^E(t_{e_t,\succ})=\frac{1}{2}$, $s^E(t_{e_t,\sim})=\frac{1}{2}$, $s^E(t_{e_{C_i},\sim})=\frac{1}{2}$ and $s^E(t_{e_{C_i,\succ}})=i$:

\[P^{R^p}_{k,s}(t)=P^{E^p}_{k,s^E}(t_{e_t,\sim}).\]
\end{proposition}

\noindent \emph{Proof}.~~~\emph{See} Appendix B.

Notice that the induced event relation $E^p$ in Definition \ref{def_ied_WO}, unlike its counterpart under an injective scoring function, is not simple. Therefore, we cannot utilize the algorithm in Proposition \ref{prop_wo_simple}. Rather, the induced relation $E^p$ is a special general probabilistic relation, where each part of the partition contains {\em exactly} two tuples. Recall that we allow tuples with probability $0$ now. 
For this special general probabilistic relation, the recursion in Theorem \ref{thm_recursion_WO} (Equation (\ref{eqn_recursion_WO_odd}), (\ref{eqn_recursion_WO_even})) collects enough information to compute the Global-Top$k$ probability of $t_{e_t,\sim}$ in $E^p$ (Equation (\ref{eqn_recursion_WO})). 

\begin{definition}[Secondary Induced Event Relations] Let $E^p=\langle E, p^E, \mathcal{C}^E\rangle$ be the event relation induced by tuple $t$ under a general scoring function $s$. Without loss of generality, assume 
\[E=\{t_{e_{C_1,\succ}},t_{e_{C_1,\sim}},\ldots,t_{e_{C_{m-1},\succ}},t_{e_{C_{m-1},\sim}},t_{e_{t,\succ}},t_{e_{t,\sim}}\},\]
and we can split $E$ into two non-overlapping subsets $E_{\succ}$ and $E_{\sim}$ such that
\[\begin{array}{l}E_{\succ}=\{t_{e_{C_1,\succ}},\ldots,t_{e_{C_{m-1},\succ}},t_{e_{t,\succ}}\},\\
E_{\sim}=\{t_{e_{C_1,\sim}},\ldots,t_{e_{C_{m-1},\sim}},t_{e_{t,\sim}}\}.\end{array}\]

The two {\em secondary induced event relation} $E^p_{\succ}$ and $E^p_{\sim}$ are $E^p$ restricted to $E^p_{\succ}$ and $E^p_{\sim}$ respectively. 
They are both simple probabilistic relations which are mutually related. 
For every $1\leq i\leq m-1$, the tuple $t_{i,\succ}$ ($t_{i,\sim}$ resp.) refers to $t_{e_{C_i,\succ}}$ ($t_{e_{C_i,\sim}}$ resp.). The tuple $t_{m,\succ}$ ($t_{m,\sim}$ resp.) refers to $t_{e_{t,\succ}}$ ($t_{e_{t,\sim}}$ resp.).
\end{definition}

In spirit, the recursion in Theorem \ref{thm_recursion_WO} is close to the recursion in Proposition \ref{prop_recursion}, even though they are not computing the same measure. The following table does a comparison between the measure $q$ in Proposition \ref{prop_recursion} and the measure $u$ in Theorem \ref{thm_recursion_WO}:

\begin{center}
\begin{tabular}{|l|l|c|}
\hline
Measure & $=\sum{Pr(W)}$ & $\begin{array}{l}|\{t_j|t_j\in W,\\ j\leq i, t_j\sim_{s} t\}|\end{array}$\\
\hline
$q(k, i)$ & \begin{tabular}{l}(1) $W$ contains $t_i$\\ (2) $W$ has {\em no more than $k$} tuples from $\{t_1, t_2, \ldots, t_i\}$\end{tabular} & -\\
\hline
$u_{\succ/\sim}(k, i, b)$ & \begin{tabular}{l}(1) $W$ contains $t_i$\\ (2) $W$ has {\em exactly $k$} tuples from $\{t_1, t_2, \ldots, t_i\}$\end{tabular} & $b$\\
\hline
\end{tabular}
\end{center}

Under the general scoring function $s^{E}$, a possible world of an induced relation $E^p$ may partially contribute to the tuple $t_{m,\sim}$'s Global-Top$k$ probability. The allocation coefficient depends on the combination of two factors: the number of tuples that are strictly better than $t_{m,\sim}$ and the number of tuples tying with $t_{m, \sim}$. Therefore, in the new measure $u$, first, we add one more dimension to keep track of $b$, i.e., the number of tying tuples of a subscript no more than $i$ in a world. Second, we keep track of distinct $(k, b)$ pairs.
Furthermore, the recursion on the measure $u$ differentiates between two cases: a non-tying tuple (handled by $u_{\succ}$) and a tying tuple (handled by $u_{\sim}$), since those two types of tuples have different influences on the values of $k$ and $b$.

Formally, let $u_{\succ}(k',i,b)$ ($u_{\sim}(k',i,b)$ resp.) be the sum of the probabilities of all the possible worlds $W$ of $E^p$ such that
\begin{enumerate}
\item $t_{i,\succ}\in W$ ($t_{i, \sim} \in W$ resp.)
\item $i$ is the $k'$th smallest tuple subscript in world $W$ 
\item the world $W$ contains $b$ tuples from $E^p_{\sim}$ with subscript less than or equal to $i$.
\end{enumerate}

The equations (\ref{eqn_recursion_WO_odd}) and (\ref{eqn_recursion_WO_even}) resemble Equation (\ref{eqn_recursion}), except that now, since we introduce tuples with probability $0$ to ensure that each part of $\mathcal{C}^{E}$ has exactly two tuples, we need to address the special cases when a divisor can be zero. Notice that, for any $i, 1\leq i\leq m$, at least one of $p^{E}(t_{i,\succ})$ and $p^{E}(t_{i,\sim})$ is non-zero, otherwise, they are not in $E^p$ by definition.

\begin{theorem}\label{thm_recursion_WO}
Given a probabilistic relation $R^p=\langle R, p, \mathcal{C}\rangle$, a scoring function $s$, $t\in R^p$,
and its induced event relation $E^p=\langle E, p^{E}, \mathcal{C}^{E}\rangle$, 
where $|E|=2m$,
the following recursion on $u_{\succ}(k',i,b)$ and $u_{\sim}(k',i,b)$ holds, where $b_{\max}$ is the number of tuples with a positive probability in $E^p_{\sim}$.

\noindent When $i=1, 0\leq k'\leq m$ and $0\leq b\leq b_{\max}$,
\begin{equation*}
u_{\succ}(k',1,b)=\left\{\begin{array}{lr}
p^{E}(t_{1,\succ})\hspace{0.5in}& k'=1, b=0\\
0 & \textrm{ otherwise}
\end{array}
\right.
\end{equation*}

\begin{equation*}
u_{\sim}(k',1,b)=\left\{\begin{array}{lr}
p^{E}(t_{1,\sim})\hspace{0.5in}& k'=1, b=1\\
0 & \textrm{ otherwise}
\end{array}
\right.
\end{equation*}

\noindent For every $i$, $2\leq i\leq m$, $0\leq k'\leq m$ and $0\leq b\leq b_{\max}$,

\begin{eqnarray}\label{eqn_recursion_WO_odd}
&&u_{\succ}(k',i,b)=\\
&&\begin{tabular}{|c|c|}
\hline
Condition & Formula\\
\hline
$k'=0$  & $0$ \\
\hline
\begin{tabular}{l}
$1\leq k' \leq m$, $p^{E}(t_{i-1,\succ})>0$
\end{tabular}
& 
\begin{tabular}{l}
\\[1pt]
$(u_{\succ}(k',i-1,b)\dfrac{1-p^{E}(t_{i-1,\succ})-p^{E}(t_{i-1,\sim})}{p^{E}(t_{i-1,\succ})}$ \\
$+u_{\succ}(k'-1,i-1,b)$\\
$+u_{\sim}(k'-1,i-1,b))p^{E}(t_{i,\succ})$
\end{tabular}
\\
\hline
\begin{tabular}{l}
$1\leq k' \leq m$, $p^{E}(t_{i-1,\succ})=0$\\
and $0\leq b<b_{\max}$
\end{tabular}
&
\begin{tabular}{l}
\\[1pt]
$(u_{\sim}(k',i-1,b+1)\dfrac{1-p^{E}(t_{i-1,\succ})-p^{E}(t_{i-1,\sim})}{p^{E}(t_{i-1,\sim})}$\\
$+u_{\succ}(k'-1,i-1,b)$\\
$+u_{\sim}(k'-1,i-1,b))p^{E}(t_{i,\succ})$
\end{tabular}
\\ 
\hline
\begin{tabular}{l}
$1\leq k' \leq m$, $p^{E}(t_{i-1,\succ})=0$\\
and $b=b_{\max}$
\end{tabular}
&
\begin{tabular}{l}
\\[1pt]
$(u_{\succ}(k'-1,i-1,b)+u_{\sim}(k'-1,i-1,b))p^{E}(t_{i,\succ})$
\end{tabular}
\\
\hline
\end{tabular}\nonumber
\end{eqnarray}

\begin{eqnarray}\label{eqn_recursion_WO_even}
&&u_{\sim}(k',i,b)=\\
&&\begin{tabular}{|c|c|}
\hline
Condition & Formula \\
\hline
 $k'=0$ or $b=0$ & $0$\\
\hline
\begin{tabular}{l}
$1\leq k' \leq m$, $1\leq b\leq b_{\max}$\\
and $p^{E}(t_{i-1,\sim})>0$
\end{tabular}
&
\begin{tabular}{l}
\\[1pt]
$(u_{\sim}(k',i-1,b)\dfrac{1-p^{E}(t_{i-1,\succ})-p^{E}(t_{i-1,\sim})}{p^{E}(t_{i-1,\sim})}$ \\
$+u_{\succ}(k'-1,i-1,b-1)$\\
$+u_{\sim}(k'-1,i-1,b-1))p^{E}(t_{i,\sim})$
\end{tabular}
\\
\hline
\begin{tabular}{l}
$1\leq k' \leq m$, $1\leq b\leq b_{\max}$\\
and $p^{E}(t_{i-1,\sim})=0$
\end{tabular}
&
\begin{tabular}{l}
\\[1pt]
$(u_{\succ}(k',i-1,b-1)\dfrac{1-p^{E}(t_{i-1,\succ})-p^{E}(t_{i-1,\sim})}{p^{E}(t_{i-1,\succ})}$\\
$+u_{\succ}(k'-1,i-1,b-1)$\\
$+u_{\sim}(k'-1,i-1,b-1))p^{E}(t_{i,\sim})$
\end{tabular}
\\
\hline
\end{tabular}\nonumber
\end{eqnarray}


The Global-Top$k$ probability of $t_{e_t,\sim}$ in $E^p$ under the scoring function $s^E$ can be computed by the following equation:

\begin{eqnarray}
P^{E^p}_{k,s^E}(t_{e_t,\sim})&=&P^{E^p}_{k,s^E}(t_{m,\sim})\nonumber\\
&=&\sum^{b_{\max}}_{b=1}(\sum_{k'=1}^{k}u_{\sim}(k',m,b)+\sum_{k'=k+1}^{k+b-1}\frac{k-(k'-b)}{b}u_{\sim}(k',m,b))\label{eqn_recursion_WO}
\end{eqnarray}

\end{theorem}

\noindent \emph{Proof}.~~~\emph{See} Appendix B.

Recall that we design Algorithm \ref{alg_ind} based on the recursion in Proposition \ref{prop_recursion}. Similarly, a DP algorithm based on the mutual recursion in Theorem \ref{thm_recursion_WO} is available. We are going to skip the details. Instead, we show how the algorithm works using Example \ref{expl_mutualrecursion} below.

The time complexity of the recursion in Theorem \ref{thm_recursion_WO} determines the complexity of the algorithm. It takes $O(b_{\max}n^2)$ for one tuple, and $O(m_{\max}n^3)$ for computing all $n$ tuples. Recall that $m_{\max}$ is the maximal number of tying tuples in $R$, and thus $b_{\max}\leq m_{\max}$. Again, the priority queue takes $O(n\log k)$. Altogether, the algorithm takes $O(m_{\max}n^3)$ time.

The space complexity of this algorithm is $O(b_{\max}n^2)$ in a straightforward implementation and $O(b_{\max}n)$ if space optimized as in Section \ref{topk_TO_Ind}.
\begin{example}\label{expl_mutualrecursion}
When evaluating a top-$2$ query in $R^p=\langle R, p, \mathcal{C}\rangle$, consider a tuple $t\in R$ and its induced event relation $E^p=\langle E, p^{E}, \mathcal{C}^{E}\rangle$
\begin{center}
\begin{tabular}{c c}
\begin{tabular}{|c||c|c|c|c|}
  \hline
  $E_{\succ}$ & $t_{e_{C_1,\succ}}$ & $t_{e_{C_2,\succ}}$ & $t_{e_{C_3,\succ}}$ & $t_{e_{t,\succ}}$\\
   & $(t_1)$ & $(t_3)$ & $(t_5)$ & $(t_7)$ \\
  \hline
  $p^E$ & $0.6$ & $0.5$ & $0.2$ & $0$ \\
  \hline
\end{tabular}
&
\begin{tabular}{|c||c|c|c|c|}
  \hline
  $E_{\sim}$ & $t_{e_{C_1,\sim}}$ & $t_{e_{C_2,\sim}}$ & $t_{e_{C_3,\sim}}$ & $t_{e_{t,\sim}}$\\
   & $(t_2)$ & $(t_4)$ & $(t_6)$ & $(t_8)$ \\
  \hline
  $p^E$ & $0$ & $0.25$ & $0.6$ & $0.4$ \\
  \hline
\end{tabular}
\end{tabular}
\end{center}
In order to compute the Global-Top$k$ probability of $t_8$ (i.e., $t_{e_{t},\sim}$) in $E^p$, Theorem \ref{thm_recursion_WO} leads to the following DP tables, each for a distinct combination of a value of $b$ and a secondary induced relation, where $b_{\max}=3$.

\begin{figure}[htbp]\centering
\subtable[($b=0$, $E^p_{\succ}$)]{
\begin{tabular}{|c||c|c|c|c|}
  \hline
  $k\backslash t$ & $t_1$ & $t_3$ & $t_5$ & $t_7$ \\
  \hline
  0 & $0$ & $0$ & $0$ & $0$  \\
  \hline
  1 & $0.6$ & $0.2$ & $0.02$ & $0$ \\
  \hline
  2 & $0$ & $0.3$ & $0.07$ & $0$ \\
  \hline
  3 & $0$ & $0$ & $0.06$ & $0$\\
  \hline
  4 & $0$ & $0$ & $0$ &  $0$\\
  \hline
\end{tabular}
}
\subtable[($b=1$, $E^p_{\succ}$)]{
\begin{tabular}{|c||c|c|c|c|}
  \hline
  $k\backslash t$ & $t_1$ & $t_3$ & $t_5$ & $t_7$\\
  \hline
  0 & $0$ & $0$ & $0$ & $0$ \\
  \hline
  1 & $0$ & $0$ & $0$ & $0$ \\
  \hline
  2 & $0$ & $0$ & $0.02$ & $0$ \\
  \hline
  3 & $0$ & $0$ & $0.03$ & $0$ \\
  \hline
  4 & $0$ & $0$ & $0$ & $0$ \\
  \hline
\end{tabular}
}
\subtable[($b=2$, $E^p_{\succ}$)]{
\begin{tabular}{|c||c|c|c|c|}
  \hline
  $k\backslash t$ & $t_1$ & $t_3$ & $t_5$ & $t_7$\\
  \hline
  0 & $0$ & $0$ & $0$ & $0$ \\
  \hline
  1 & $0$ & $0$ & $0$ & $0$ \\
  \hline
  2 & $0$ & $0$ & $0$ & $0$ \\
  \hline
  3 & $0$ & $0$ & $0$ & $0$ \\
  \hline
  4 & $0$ & $0$ & $0$ & $0$ \\
  \hline
\end{tabular}
}
\subtable[($b=3$, $E^p_{\succ}$)]{
\begin{tabular}{|c||c|c|c|c|}
  \hline
  $k\backslash t$ & $t_1$ & $t_3$ & $t_5$ & $t_7$\\
  \hline
  0 & $0$ & $0$ & $0$ & $0$ \\
  \hline
  1 & $0$ & $0$ & $0$ & $0$ \\
  \hline
  2 & $0$ & $0$ & $0$ & $0$ \\
  \hline
  3 & $0$ & $0$ & $0$ & $0$ \\
  \hline
  4 & $0$ & $0$ & $0$ & $0$ \\
  \hline
\end{tabular}
}\\
\subtable[($b=0$, $E^p_{\sim}$)]{
\begin{tabular}{|c||c|c|c|c|}
  \hline
  $k\backslash t$ & $t_2$ & $t_4$ & $t_6$ & $t_8$\\
  \hline
  0 & $0$ & $0$ & $0$ & $0$ \\
  \hline
  1 & $0$ & $0$ & $0$ & $0$ \\
  \hline
  2 & $0$ & $0$ & $0$ & $0$ \\
  \hline
  3 & $0$ & $0$ & $0$ & $0$ \\
  \hline
  4 & $0$ & $0$ & $0$ & $0$ \\
  \hline
\end{tabular}
}
\subtable[($b=1$, $E^p_{\sim}$)]{
\begin{tabular}{|c||c|c|c|c|}
  \hline
  $k\backslash t$ & $t_2$ & $t_4$ & $t_6$ & $t_8$\\
  \hline
  0 & $0$ & $0$ & $0$ & $0$ \\
  \hline
  1 & $0$ & $0.1$ & $0.06$ & $\mathbf{0.008}$ \\
  \hline
  2 & $0$ & $0.15$ & $0.21$ & $\mathbf{0.036}$ \\
  \hline
  3 & $0$ & $0$ & $0.18$ & $0.052$ \\
  \hline
  4 & $0$ & $0$ & $0$ & $0.024$ \\
  \hline
\end{tabular}
}
\subtable[($b=2$, $E^p_{\sim}$)]{
\begin{tabular}{|c||c|c|c|c|}
  \hline
  $k\backslash t$ & $t_2$ & $t_4$ & $t_6$ & $t_8$\\
  \hline
  0 & $0$ & $0$ & $0$ & $0$ \\
  \hline
  1 & $0$ & $0$ & $0$ & $\mathbf{0}$ \\
  \hline
  2 & $0$ & $0$ & $0.06$ & $\mathbf{0.032}$ \\
  \hline
  3 & $0$ & $0$ & $0.09$ & $\mathbf{0.104}$ \\
  \hline
  4 & $0$ & $0$ & $0$ & $0.084$ \\
  \hline
\end{tabular}
}
\subtable[($b=3$, $E^p_{\sim}$)]{
\begin{tabular}{|c||c|c|c|c|}
  \hline
  $k\backslash t$ & $t_2$ & $t_4$ & $t_6$ & $t_8$\\
  \hline
  0 & $0$ & $0$ & $0$ & $0$ \\
  \hline
  1 & $0$ & $0$ & $0$ & $\mathbf{0}$ \\
  \hline
  2 & $0$ & $0$ & $0$ & $\mathbf{0}$ \\
  \hline
  3 & $0$ & $0$ & $0$ & $\mathbf{0.024}$ \\
  \hline
  4 & $0$ & $0$ & $0$ & $\mathbf{0.036}$ \\
  \hline
\end{tabular}
}
\caption{Mutual Recursion in Example \ref{expl_mutualrecursion}}
\label{fig_mutualrecursion}
\end{figure}
The computation of each entry follows the mutual recursion in Theorem \ref{thm_recursion_WO}, for example, 
\begin{eqnarray*}
u_{\succ}(2,5,0)&=&(u_{\succ}(1,3,0)+u_{\sim}(1,4,0)+u_{\succ}(2,3,0)\frac{1-p^{E}(t_3)-p^{E}(t_4)}{p^{E}(t_3)})p^{E}(t_5)\\
&=&(0.2+0+0.3\frac{1-0.5-0.25}{0.5})0.2 = 0.07
\end{eqnarray*}
\begin{eqnarray*}
u_{\sim}(2,6,1)&=&(u_{\succ}(1,3,0)+u_{\sim}(1,4,0)+u_{\sim}(2,4,1)\frac{1-p^{E}(t_3)-p^{E}(t_4)}{p^{E}(t_4)})p^{E}(t_6)\\
&=&(0.2+0+0.15\frac{1-0.5-0.25}{0.25})0.6 = 0.21
\end{eqnarray*}

Finally, under the scoring function $s^E$ defined in Proposition \ref{prop_iedscore_WO}

\begin{eqnarray*}
P^{E^p}_{k,s^E}(t_{e_{t},\sim})&=&P^{E^p}_{2,s^E}(t_8)\\
&=&\sum^{3}_{b=1}(\sum_{k'=1}^{2}u_{\sim}(k',8,b)+\sum_{k'=2+1}^{2+b-1}\frac{2-(k'-b)}{b}u_{\sim}(k',8,b))\\
&=&u_{\sim}(1,8,1)+u_{\sim}(2,8,1)\\
& &+u_{\sim}(1,8,2)+u_{\sim}(2,8,2)+\frac{1}{2}u_{\sim}(3,8,2)\\
& &+u_{\sim}(1,8,3)+u_{\sim}(2,8,3)+\frac{2}{3}u_{\sim}(3,8,3)+\frac{1}{3}u_{\sim}(3,8,4)\\
&=&0.008+0.036+0+0.032+\frac{1}{2}0.104+0+0+\frac{2}{3}0.024+\frac{1}{3}0.036\\
&=&0.156
\end{eqnarray*}
Bold entries in Figure \ref{fig_mutualrecursion} are involved in the above equation.
\end{example}

\section{Experiments}\label{sec_exp}

We report here an empirical study on various optimization techniques proposed in Section \ref{sec_TA} and Section \ref{sec_r_rs}, as the behavior of the straightforward implementation of our algorithms is pretty much predicted by the aforementioned theoretical analysis. 
We implement all the algorithms in C++ and run experiments on a machine with Intel Core2 1.66G CPU running Cygwin on Windows XP with 1GB memory.

Each synthetic dataset has a uniform random score distribution and a uniform random probability distribution. There is no correlation between the score and the probability.
The size ($n$) of the dataset varies from $5$K up to $1$M. In a dataset of a general probabilistic relation, $x$ is the percentage of exclusive tuples and $s$ is the max number of exclusive tuples in a part from the partition. In other words, in a general probabilistic relation of size $n$, there are $\lceil nx\rceil$ tuples involved in a non-trivial part from the partition. The size of each part is a random number from $[2, s]$. Unless otherwise stated, $x$ defaults to $0.1$ and $d$ defaults to $20$. The default value of $k$ in a top-$k$ query is $100$.

For simple relations, the baseline algorithm \simplebasic\ is the space optimized version of Algorithm \ref{alg_ind} and \ref{alg_indsub} mentioned in Section \ref{topk_TO_Ind}. \simpleta\ integrates the TA optimization technique in Section \ref{sec_TA}.
For general relations, the baseline algorithm \generalbasic\ is a straightforward implementation of Algorithm \ref{alg_indEx} and \ref{alg_indExsub}. \rollback\ and \rollbacksort\ implements the two optimization techniques in Section \ref{sec_r_rs} respectively.

\subsubsection{Summary of experiments} We draw the following conclusions from the forthcoming experimental results:

\begin{itemize}
\item[$\bullet$] Optimizations such as \simpleta, \rollback\ and \rollbacksort\ are effective and significantly reduce the running time. On average, \simpleta\ saves about half of the computation cost in simple relations.
Compared to \generalbasic, \rollback\ and \rollbacksort\ improve the running time up to 2 and 3 orders of magnitude respectively.

\item[$\bullet$] Decreasing the percentage of exclusive tuples ($x$) improves the running time of \rollback\ and \rollbacksort. When $x$ is fixed, increasing the max number of tuples in each part ($s$) improves the running time of \rollback\ and \rollbacksort.

\item[$\bullet$] For general probabilistic relations, \rollbacksort\ scales well to large datasets.
\end{itemize}

\begin{figure}[htbp]
\subfigure[Simple Prob.~Relation]{
 \includegraphics[width=\figwidth]{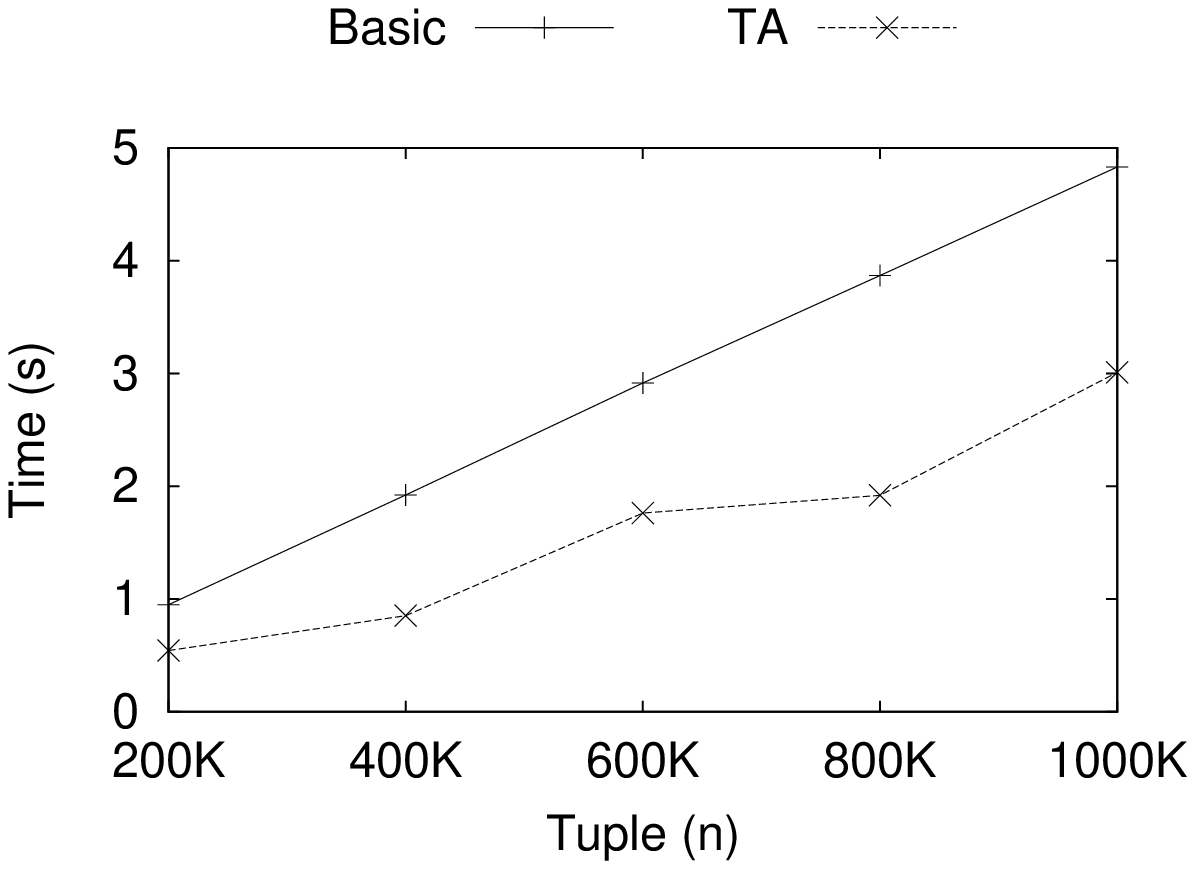}
 \label{exp_simple_time_naive_vs_ta}
}%
\subfigure[General Prob.~Relation]{
 \includegraphics[width=\figwidth]{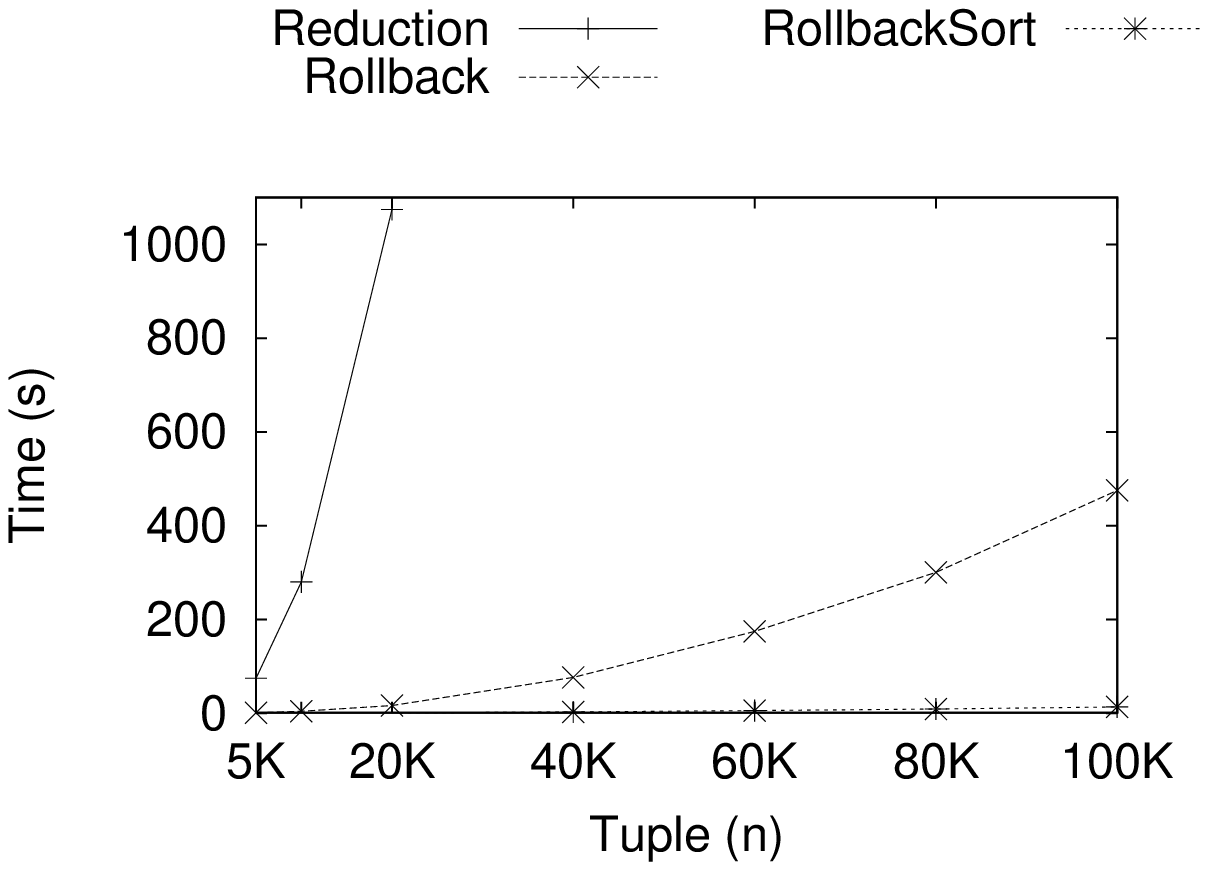}
 \label{exp_algcmp}
}%
\caption{Performance of Optimizations}\label{fig_optimizations}
\end{figure}

\begin{figure}[htbp]
\subfigure[Running time vs $x$ (\rollback)]{
 \includegraphics[width=\figwidth]{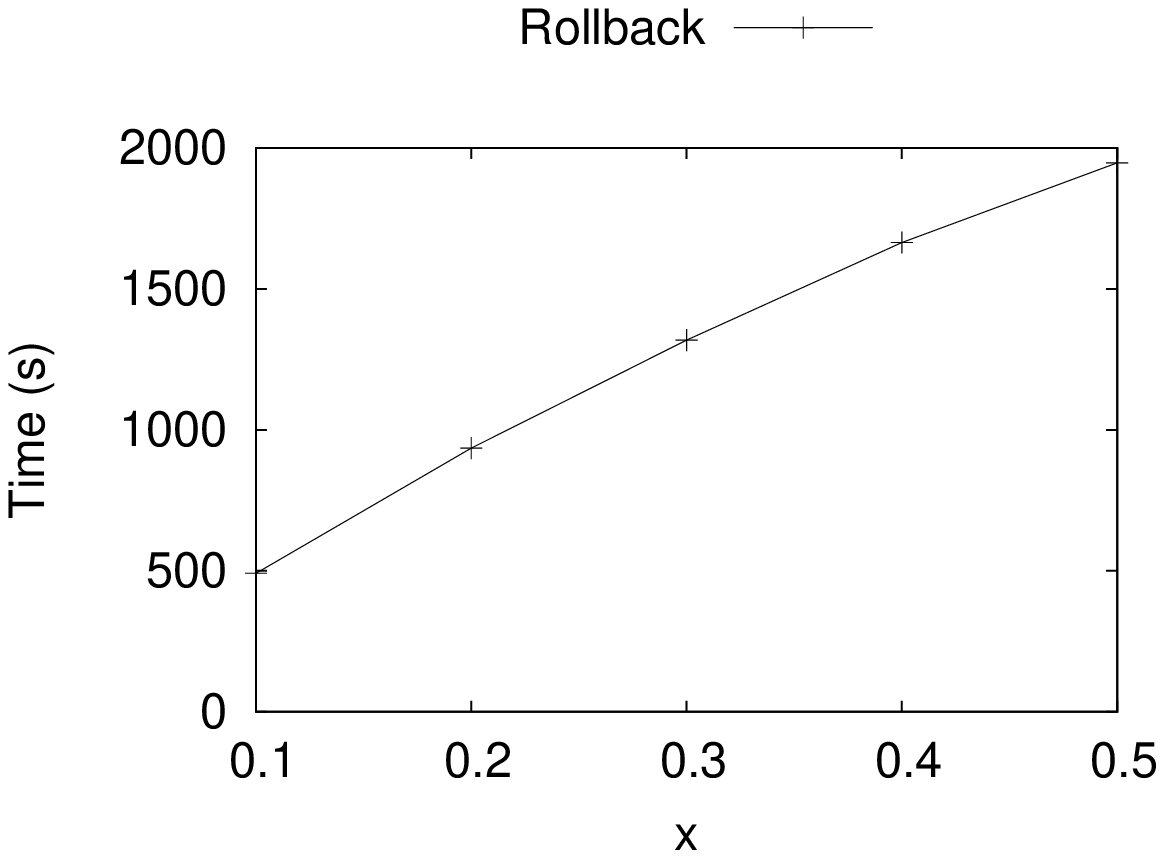}
 \label{exp_sen_r_percentage}
}%
\subfigure[Running time vs $x$ (\rollbacksort)]{
 \includegraphics[width=\figwidth]{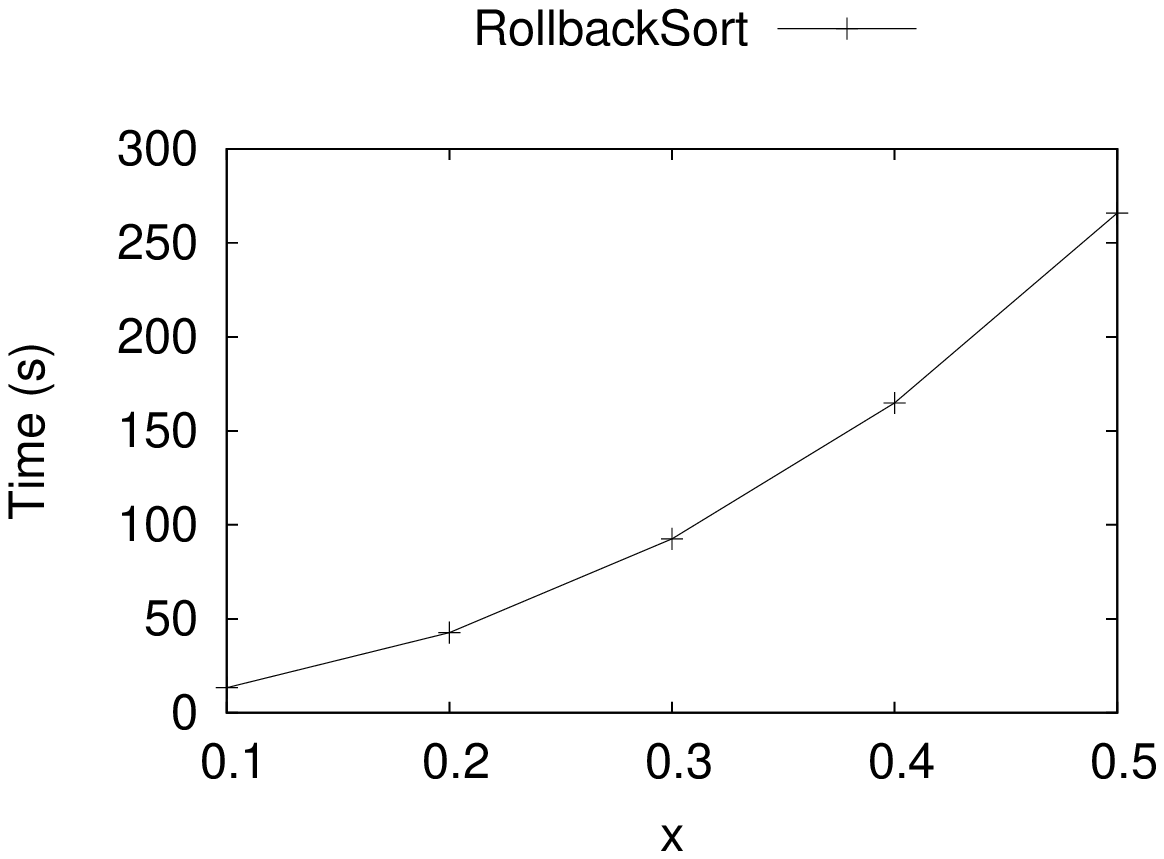}
 \label{exp_sen_rs_percentage}
}\\
\subfigure[Running time vs $s$ (\rollback)]{
 \includegraphics[width=\figwidth]{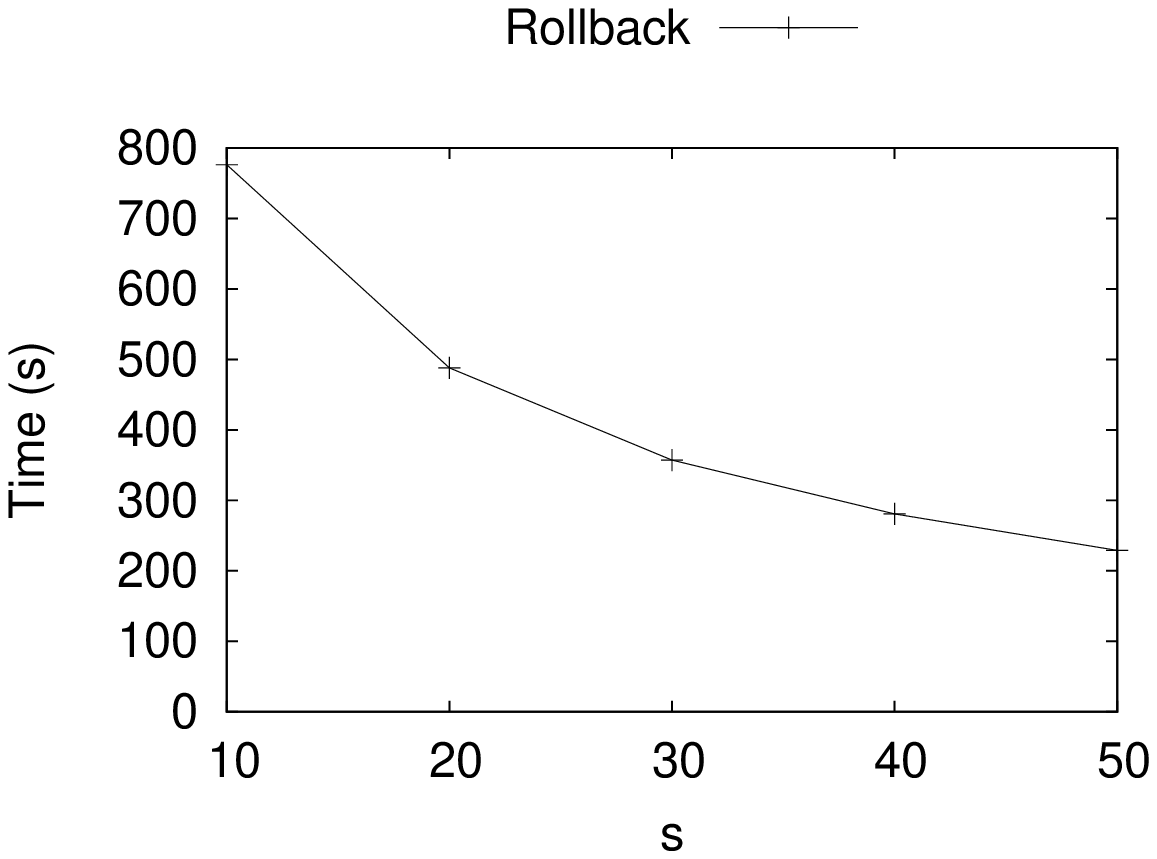}
 \label{exp_sen_r_maxclique}
}
\subfigure[Running time vs $s$ (\rollbacksort)]{
 \includegraphics[width=\figwidth]{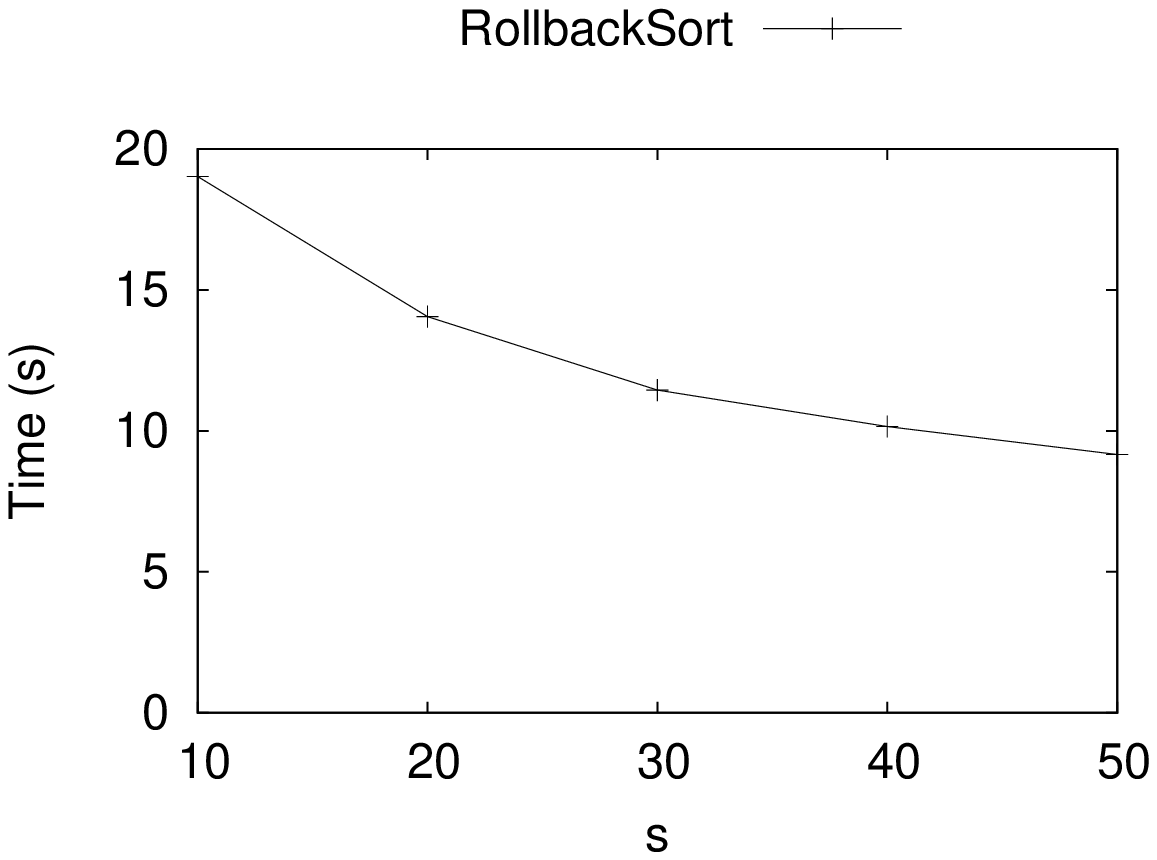}
 \label{exp_sen_rs_maxclique}
}%
\caption{Sensitivity to Parameters}\label{fig_sen_r_rs}
\end{figure}

\subsection{Performance of Optimizations}
Figure \ref{exp_simple_time_naive_vs_ta} illustrates the improvement of \simpleta\ over \simplebasic\ for simple probabilistic relations. While \simplebasic\ is already linear in terms of $n$, \simpleta\ still saves a significant amount of computation, i.e., a little less than half. It worths emphasizing that there is no correlation between the score and the probability in our datasets. It is well-known that TA optimization has a better performance when there is a positive correlation between attributes, and a worse performance when there is a negative correlation between attributes. Therefore, the dataset we show, i.e., with no correlation, should represent an average case. 

For general probabilistic relations, Figure \ref{exp_algcmp} illustrates the performance of \generalbasic, \rollback\ and \rollbacksort\ when $n$ varies from 5K to 100K. For the baseline algorithm \generalbasic, we show only the first three data points, as the rest are off the chart. The curve of \generalbasic\ reflects the quadratic theoretical bound. From Figure \ref{exp_algcmp}, it is clear that the heuristic \rollback\ and \rollbacksort\ greatly reduce the running time over the quadratic bound. The improvement is up to 2 and 3 orders of magnitude for \rollback\ and \rollbacksort\ respectively.

\subsection{Sensitivity to Parameters}
Our second set of experiments studies the influence of various parameters on \rollback\ and \rollbacksort. The results are shown in Figure \ref{fig_sen_r_rs}.
Notice the difference between the scale of y-axis of Figure \ref{exp_sen_r_percentage} (resp.~Figure \ref{exp_sen_r_maxclique}) and that of Figure \ref{exp_sen_rs_percentage} (resp.~Figure \ref{exp_sen_rs_maxclique}). \rollbacksort\ outperforms \rollback\ by one order of magnitude.

 Figure \ref{exp_sen_r_percentage} and \ref{exp_sen_rs_percentage} show the impact of varying the percentage of exclusive tuples ($x$) in the dataset.
It is to be expected that with the increase of the percentage of exclusive tuples, more rollback operations are needed in both \rollback\ and \rollbacksort. However, \rollback\ shows a linear increase, while \rollbacksort\ shows a trend more than linear but less than quadratic. 

Figure \ref{exp_sen_r_maxclique} and \ref{exp_sen_rs_maxclique} illustrate the impact of the size of the parts in the partition. In these two sets of experiments, we fix the total number of exclusive tuples, and vary the max size of a part ($s$). A large $s$ suggests fewer but relatively larger parts in the partition, as compared to a small $s$. For both \rollback\ and \rollbacksort, we see a similar trend that as $s$ increases, the running time decreases. The relative decrease in \rollback\ is larger than that of \rollbacksort, which can be explained by the fact that \rollbacksort\ is already optimized for repetitive occurrences of tuples from the same part, and therefore it should be less subjective to the size of parts.

\subsection{Scalability}
\begin{figure}[htbp]
\subfigure[Running time vs $n$]{
 \includegraphics[width=\figwidth]{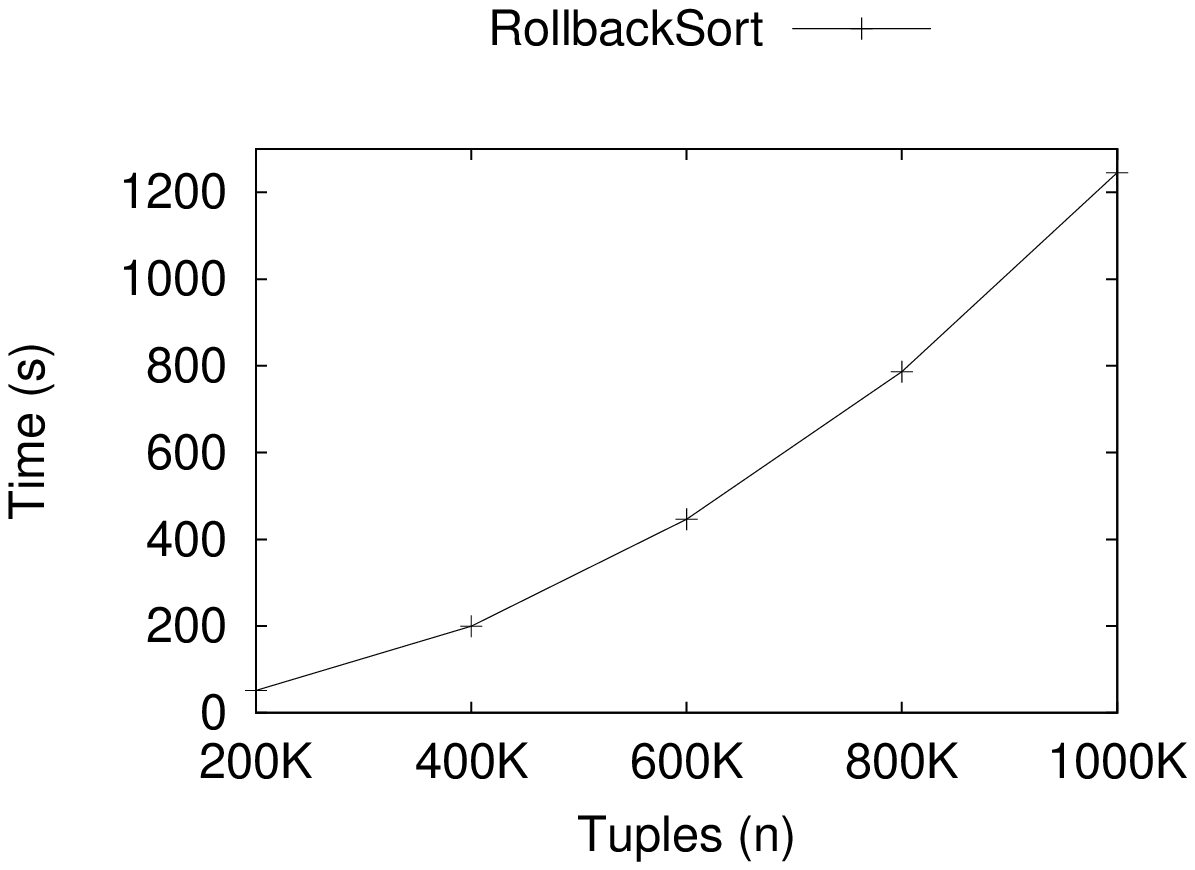}
 \label{exp_scale_rs_n}
}%
\subfigure[Running time vs $k$]{
 \includegraphics[width=\figwidth]{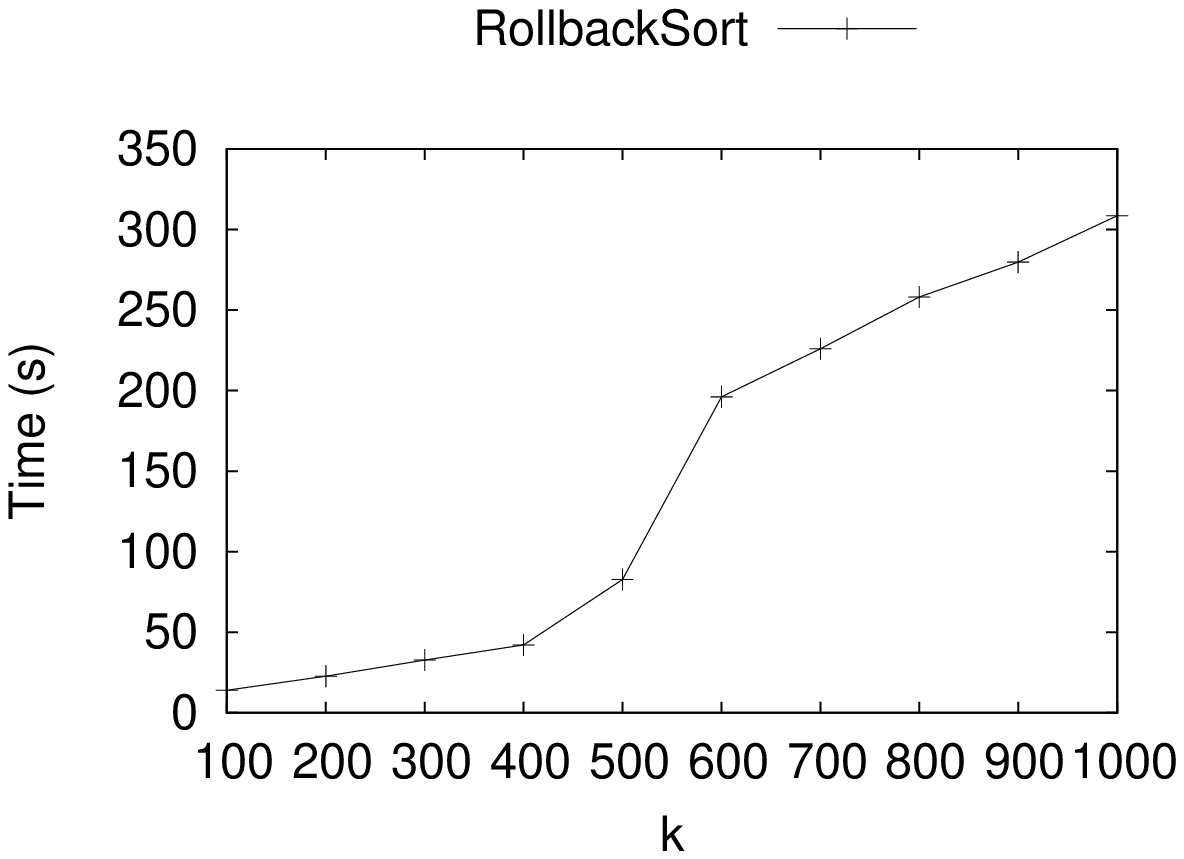}
 \label{exp_scale_rs_k}
}%
\caption{Scalability of \rollbacksort}\label{fig_scale}
\end{figure}

As we have already seen analytically in Section \ref{topk_TO_Ind} and empirically in Figure \ref{exp_simple_time_naive_vs_ta}, the algorithm for simple probabilistic relations scales linearly to large datasets. \simpleta\ can further improve the performance.

For general probabilistic databases, Figure \ref{fig_scale} shows that \rollbacksort\ scales well to large datasets. Figure \ref{exp_scale_rs_n} illustrates the running time of \rollbacksort\ when $n$ increases to $1$M tuples. The trend is more than linear, but much slower than quadratic.
Figure \ref{exp_scale_rs_k} shows the impact of $k$ on the running time. Notice that, the general trend in Figure \ref{exp_scale_rs_k} is linear except there is a ``step-up'' when $k$ is about $500$. We conjecture that this is due to the non-linear maintenance cost of the priority queue used in the algorithm.

\section{Conclusion}

We study the semantic and computational problems for top-$k$ queries in probabilistic databases. We propose three postulates to categorize top-$k$ semantics in probabilistic databases and discuss their satisfaction by the semantics in the literature. Those postulates are the first step to analyze different semantics. We do not think that a single semantics is superior/inferior to other semantics just because of postulate satisfaction. Rather, we deem that the choice of the semantics should be guided by the application.
The postulates help to create a profile of each semantics.
We propose a new top-$k$ semantics, namely Global-Top$k$, which satisfies the postulates to a large degree. We study the computational problem of query evaluation under Global-Top$k$ semantics for simple and general probabilistic relations when the scoring function is injective. For the former, we propose a dynamic programming algorithm and effectively optimize it with Threshold Algorithm. For the latter, we show a polynomial reduction to the simple case, and design \rollback\ and \rollbacksort\ optimizations to speed up the computation. We conduct an empirical study to verify the effectiveness of those optimizations.
Furthermore, we extend the Global-Top$k$ semantics to general scoring functions and introduce the concept of allocation policy to handle ties in score. To the best of our knowledge, this is the first attempt to address the tie problem rigorously. Previous work either does not consider ties or uses an arbitrary tie-breaking mechanism.
Advanced dynamic programming algorithms are proposed for query evaluation under general scoring functions for both simple and general probabilistic relations. We provide theoretical analysis following every algorithm proposed.

For completeness, we list in Table \ref{tab_complexity} the complexity of the best known algorithm for the semantics in the literature. Since no other work addresses general scoring functions in a systematical way, those results are restricted to injective scoring functions.
\begin{center}
\begin{table}[htbp]\center
  \begin{tabular}{|l|c|c|}
\hline
     Semantics & Simple Probabilistic DB & General Probabilistic DB\\
\hline
    Global-Top$k$ & $O(kn)$ & $O(kn^2)$\\
    PT-$k$ & $O(kn)$ & $O(kn^2)$\\
    U-Top$k$ & $O(n\log k)$ & $O(n\log k)$\\
    U-$k$Ranks & $O(kn)$ & $O(kn^2)$\\
\hline
  \end{tabular}
\caption{Time Complexity of Different Semantics}\label{tab_complexity}
\end{table}
\end{center}

\section{Future Work}

Several variants of the existing semantics have been proposed in the literature \cite{DBLP:journals/tods/SolimanIC08}, their postulate satisfaction deserves further study.
So far, the research reported in the literature has primarily focused on indepedent and exclusive relationships among tuples \cite{DBLP:conf/icde/SolimanIC07,DBLP:journals/tods/SolimanIC08,DBLP:conf/sigmod/HuaPZL08,DBLP:conf/icde/YiLKS08}. It will be interesting to investigate other complex relationships between tuples. 
Other possible directions include top-$k$ evaluation in other uncertain database models proposed in the literature \cite{DBLP:journals/tcs/OlteanuKA08} and more general preference queries in probabilistic databases.

\section{Acknowledgment}
We acknowledge the input of Graham Cormode who showed that Faithfulness in general probabilistic relations is problematic. Jan Chomicki acknowledges the discussions with Sergio Flesca. Xi Zhang acknowledges the discussions with Hung Q.~Ngo.

\section{Appendix A: Semantic Postulates}\label{sec_app_postulates}
\begin{center}

\begin{threeparttable}
\begin{tabular}{|l|c|c|c|c|}
  \hline
  Semantics & Exact $k$ & Faithfulness & Stability \\
  \hline
  \tnote{$\dagger$}~~Global-Top$k$ & \checkmark (1)& \checkmark/$\times$  (5)& \checkmark (9)\\
  PT-$k$ & $\times$ (2)& \checkmark/$\times$ (6)& \checkmark  (10)\\
  U-Top$k$ & $\times$ (3)& \checkmark/$\times$  (7)& \checkmark (11)\\
  U-$k$Ranks & $\times$ (4)& $\times$ (8) & $\times$ (12)\\
  \hline
\end{tabular}
\begin{tablenotes}
\item[$\dagger$] \footnotesize{Postulates of Global-Top$k$ semantics are proved under general scoring functions with {\em Equal} allocation policy.}
\end{tablenotes}
\caption{Postulate Satisfaction for Different Semantics in Table \ref{tab_postulates}}
\end{threeparttable}

\end{center}

The following proofs correspond to the numbers next to each entry in the above table. 
Assume that we are given a probabilistic relation $R^p=\langle R, p, \mathcal{C}\rangle$, a non-negative integer $k$ and an injective scoring function $s$.

\subsection{Exact $k$}
\begin{enumerate}
\item[(1)]Global-Top$k$ satisfies \emph{Exact $k$}.

We compute the Global-Top$k$ probability for each tuple in $R$. If there are at least $k$ tuples in $R$, we are always able to pick the $k$ tuples with the highest Global-Top$k$ probability. In case when there are more than $k-r+1$ tuple(s) with the $r$th highest Global-Top$k$ probability, where $r=1,2\ldots,k$, only $k-r+1$ of them will be picked nondeterministically.

\item[(2)]PT-$k$ violates \emph{Exact $k$}.

Example \ref{expl_twosemantics} illustrates a counterexample in a simple probabilistic relation.

\item[(3)]U-Top$k$ violates \emph{Exact $k$}.

Example \ref{expl_twosemantics} illustrates a counterexample in a simple probabilistic relation.

\item[(4)]U-$k$Ranks violates \emph{Exact $k$}.

Example \ref{expl_twosemantics} illustrates a counterexample in a simple probabilistic relation.
\end{enumerate}

\subsection{Faithfulness}
\begin{enumerate}
\item[(5)]Global-Top$k$ satisfies \emph{Faithfulness} in simple probabilistic relations while it violates \emph{Faithfulness} in general probabilistic relations.
\begin{itemize}
\item[(5a)]{Simple Probabilistic Relations}

By the assumption, $t_1\succ_s t_2$ and $p(t_1)>p(t_2)$, so we need to show that $P_{k,s}(t_1)>P_{k,s}(t_2)$.

For every $W\in pwd(R^p)$ such that $t_2\in \ptopktuple{k}{s}(W)$ and
$t_1\not\in \ptopktuple{k}{s}(W)$, obviously $t_1\not\in W$.
Otherwise, since $t_1\succ_s t_2$, $t_1$ would be in $\ptopktuple{k}{s}(W)$. 
Since all tuples are independent, there is always a world $W'\in pwd(R^p)$, $W'=(W\backslash\{t_2\})\cup\{t_1\}$ and $Pr(W')=Pr(W)\frac{p(t_1)\bar{p}(t_2)}{\bar{p}(t_1) p(t_2)}$.
Since $p(t_1)> p(t_2)$, $Pr(W')> Pr(W)$.
Moreover, $t_1$ will substitute for $t_2$ in the top-$k$ answer set to $W'$. 
It is easy to see that $\alpha(t_1, W')=1$ in $W'$ and also in any world $W$ such that both $t_1$ and $t_2$ are in $\ptopktuple{k}{s}(W)$, $\alpha(t_1, W)=1$.

Therefore, for the Global-Top$k$ probability of $t_1$ and $t_2$, we have

\begin{eqnarray*}
P_{k,s}(t_2)&=&\sum_{\begin{subarray}{l}
W\in pwd(R^p)\\
t_1\in \ptopktuple{k}{s}(W)\\
t_2\in \ptopktuple{k}{s}(W)
\end{subarray}
}\alpha(t_2, W)Pr(W)+
\sum_{\begin{subarray}{l}
W\in pwd(R^p)\\
t_1\not\in \ptopktuple{k}{s}(W)\\
t_2\in \ptopktuple{k}{s}(W)
\end{subarray}
}\alpha(t_2, W)Pr(W)\\
&<&\sum_{\begin{subarray}{l}
W\in pwd(R^p)\\
t_1\in \ptopktuple{k}{s}(W)\\
t_2\in \ptopktuple{k}{s}(W)
\end{subarray}
}Pr(W)+
\sum_{\begin{subarray}{l}
W'\in pwd(R^p)\\
t_1\in \ptopktuple{k}{s}(W')\\
t_2\not\in W'
\end{subarray}
}Pr(W')\\
&=&\sum_{\begin{subarray}{l}
W\in pwd(R^p)\\
t_1\in \ptopktuple{k}{s}(W)\\
t_2\in \ptopktuple{k}{s}(W)
\end{subarray}
}\alpha(t_1, W)Pr(W)+
\sum_{\begin{subarray}{l}
W'\in pwd(R^p)\\
t_1\in \ptopktuple{k}{s}(W')\\
t_2\not\in W'
\end{subarray}
}\alpha(t_1, W')Pr(W')\\
&\leq&\sum_{\begin{subarray}{l}
W\in pwd(R^p)\\
t_1\in \ptopktuple{k}{s}(W)\\
t_2\in \ptopktuple{k}{s}(W)
\end{subarray}
}\alpha(t_1, W)Pr(W)+
\sum_{\begin{subarray}{l}
W'\in pwd(R^p)\\
t_1\in \ptopktuple{k}{s}(W')\\
t_2\not\in W'
\end{subarray}
}\alpha(t_1, W')Pr(W')\\
&&+\sum_{\begin{subarray}{l}
W''\in pwd(R^p)\\
t_1\in \ptopktuple{k}{s}(W'')\\
t_2 \in W''\\
t_2 \not\in \ptopktuple{k}{s}(W'')
\end{subarray}
}\alpha(t_1, W'')Pr(W'')\\
&=&P_{k,s}(t_1).
\end{eqnarray*}

The equality in $\leq$ holds when $s(t_2)$ is among the $k$ highest scores and there are at most $k$ tuples (including $t_2$) with higher or equal scores. Since there is at least one inequality in the above equation, we have
\[P_{k,s}(t_1)>P_{k,s}(t_2).\]

\item[(5b)]{General Probabilistic Relations}

The following is a counterexample.

Say $k=1$, $R=\{t_1, \ldots, t_9\}$, $t_1\succ_s \ldots \succ_s t_9$, $\{t_1,\ldots,t_7, t_9\}$ are exclusive. $p(t_i)=0.1, i=1\ldots 7$, $p(t_8)=0.4$, $p(t_9)=0.3$.

By Global-Top$k$, the top-$1$ answer is $\{t_9\}$, while $t_8\succ_s t_9$ and $p(t_8)>p(t_9)$, which violates \emph{Faithfulness}.
\end{itemize}
\item[(6)]PT-$k$ satisfies \emph{Faithfulness} in simple probabilistic relations while it violates \emph{Faithfulness} in general probabilistic relations.

For simple probabilistic relations, we can use the same proof in (5) to show that PT-$k$ satisfies \emph{Faithfulness}. The only change would be that we need to show $P_{k,s}(t_1)>p_{\tau}$ as well. Since $P_{k,s}(t_2)>p_{\tau}$ and $P_{k,s}(t_1)>P_{k,s}(t_2)$, this is obviously true. For general probabilistic relations, we can use the same counterexample in (5) and set threshold $p_{\tau}=0.15$.

\item[(7)] U-Top$k$ satisfies \emph{Faithfulness} in simple probabilistic relations while it violates \emph{Faithfulness} in general probabilistic relations.
\begin{itemize}
\item[(7a)]{Simple Probabilistic Relations}

By contradiction. If U-Top$k$ violates \emph{Faithfulness} in a simple probabilistic relation, there exists $R^p=\langle R, p, \mathcal{C}\rangle$ and exists $t_i, t_j\in R, t_i\succ_s t_j, p(t_i)> p(t_j)$, and by U-Top$k$,
$t_j$ is in the top-$k$ answer set to $R^p$ under the scoring function $s$ while $t_i$ is not.

$S$ is a top-$k$ answer set to $R^p$ under the function $s$ by the U-Top$k$ semantics, $t_j\in S$ and $t_i\not\in S$. Denote by $Q_{k,s}(S)$ the probability of $S$ under the U-Top$k$ semantics. That is,
\[Q_{k,s}(S)=\sum_{\begin{subarray}{l} W\in pwd(R^p)\\
S=\topktuple{k}{s}(W)\end{subarray}}Pr(W).\]
For any world $W$ contributing to $Q_{k,s}(S)$, $t_i\not \in W$. Otherwise, since $t_i\succ_s t_j$, $t_i$ would be in $\topktuple{k}{s}(W)$, which is $S$. Define a world $W'=(W\backslash\{t_j\})\cup\{t_i\}$. Since $t_i$ is independent of any other tuple in $R$, $W'\in pwd(R^p)$ and $Pr(W')=Pr(W)\frac{p(t_i)\bar{p}(t_j)}{\bar{p}(t_i)p(t_j)}$. Moreover, $\topktuple{k}{s}(W')=(S\backslash\{t_j\})\cup\{t_i\}$. Let $S'=(S\backslash\{t_j\})\cup\{t_i\}$, then $W'$ contributes to $Q_{k,s}(S')$.

\begin{eqnarray*}
Q_{k,s}(S')&=&\sum_{
\begin{subarray}{l}
W\in pwd(R^p)\\
S'=\topktuple{k}{s}(W)
\end{subarray}
}Pr(W)\\
&\geq&\sum_{
\begin{subarray}{l}
W\in pwd(R^p)\\
S=\topktuple{k}{s}(W)
\end{subarray}
}Pr((W\backslash\{t_j\})\cup\{t_i\})\\
&=&\sum_{
\begin{subarray}{l}
W\in pwd(R^p)\\
S=\topktuple{k}{s}(W)
\end{subarray}
}Pr(W)\frac{p(t_i)\bar{p}(t_j)}{\bar{p}(t_i)p(t_j)}\\
&=&\frac{p(t_i)\bar{p}(t_j)}{\bar{p}(t_i)p(t_j)}\sum_{
\begin{subarray}{l}
W\in pwd(R^p)\\
S=\topktuple{k}{s}(W)
\end{subarray}
}Pr(W)\\
&=&\frac{p(t_i)\bar{p}(t_j)}{\bar{p}(t_i)p(t_j)}Q_{k,s}(S)\\
&>&Q_{k,s}(S),
\end{eqnarray*}

which is a contradiction.

\item[(7b)]{General Probabilistic Relations}

The following is a counterexample. 

Say $k=2$, $R=\{t_1, t_2, t_3, t_4\}$, $t_1\succ_s t_2\succ_s t_3\succ_s t_4$, $t_1$ and $t_2$ are exclusive, $t_3$ and $t_4$ are exclusive. $p(t_1)=0.5$, $p(t_2)=0.45$, $p(t_3)=0.4$, $p(t_4)=0.3$.

By U-Top$k$, the top-$2$ answer is $\{t_1, t_3\}$, while $t_2\succ_s t_3$ and $p(t_2)>p(t_3)$, which violates \emph{Faithfulness}.
\end{itemize}
\item[(8)]U-$k$Ranks violates \emph{Faithfulness}.

The following is a counterexample. 

Say $k=2$, $R^p$ is simple. $R=\{t_1, t_2, t_3\}$, $t_1\succ_s t_2\succ_s t_3$,
$p(t_1)=0.48, p(t_2)=0.8, p(t_3)=0.78$.

The probabilities of each tuple at each rank are as follows:

\begin{center}
\begin{tabular}{c c c c}
  \hline
    & $t_1$ & $t_2$ & $t_3$ \\
  \hline
  rank 1 & 0.48 & 0.416 & 0.08112 \\
  rank 2 & 0 & 0.384 & 0.39936\\
  rank 3 & 0 & 0 & 0.29952\\
  \hline
\end{tabular}
\end{center}

By U-$k$Ranks, the top-$2$ answer set is $\{t_1, t_3\}$ while $t_2\succ t_3$ and $p(t_2)>p(t_3)$, which contradicts \emph{Faithfulness}. 

\end{enumerate}

\subsection{Stability}
\begin{enumerate}
\item[(9)]Global-Top$k$ satisfies \emph{Stability}.

In the rest of this proof, let $A$ be the set of all winners under the Global-Top$k$ semantics.

\textbf{Part I}: Probability.
\begin{itemize}
\item[Case 1:] Winners.

For any winner $t\in A$, if we only raise the probability of $t$, we have a new probabilistic relation $(R^p)'=\langle R,p',\mathcal{C}\rangle$, where the new probability function $p'$ is such that $p'(t)>p(t)$ and for any $t'\in R, t'\neq t, p'(t')=p(t')$. Note that $pwd(R^p)=pwd((R^p)')$.
In addition, assume $t\in C_t$, where $C_t\in \mathcal{C}$. By Global-Top$k$,

\begin{eqnarray*}
P^{R^p}_{k,s}(t)&=&\sum_{
\begin{subarray}{l}
W\in pwd(R^p)\\
t\in \ptopktuple{k}{s}(W)
\end{subarray}
}\alpha(t, W)Pr(W)
\end{eqnarray*}
and
\begin{eqnarray*}
P^{(R^p)'}_{k,s}(t)&=&\sum_{
\begin{subarray}{l}
W\in pwd(R^p)\\
t\in \ptopktuple{k}{s}(W)
\end{subarray}
}\alpha(t, W)Pr(W)\frac{p'(t)}{p(t)}\\
&=&\frac{p'(t)}{p(t)}P^{R^p}_{k,s}(t).
\end{eqnarray*}

For any other tuple $t'\in R, t'\neq t$, we have the following equation:

\begin{eqnarray*}
P^{(R^p)'}_{k,s}(t')&=&\sum_{
\begin{subarray}{l}
W\in pwd(R^p)\\
t'\in \ptopktuple{k}{s}(W),t\in W
\end{subarray}}\alpha(t', W)Pr(W)\frac{p'(t)}{p(t)}\\
& &+\sum_{
\begin{subarray}{l}
W\in pwd(R^p)\\
t'\in \ptopktuple{k}{s}(W),~t\not\in W\\
(C_t\backslash\{t\})\cap W = \emptyset
\end{subarray}}\alpha(t', W)Pr(W)\frac{c-p'(t)}{c-p(t)}\\
&&+\sum_{
\begin{subarray}{l}
W\in pwd(R^p)\\
t'\in \ptopktuple{k}{s}(W),~t\not\in W\\
(C_t\backslash\{t\})\cap W \neq \emptyset
\end{subarray}}\alpha(t', W)Pr(W)\nonumber\\
&\leq &\frac{p'(t)}{p(t)}(\sum_{
\begin{subarray}{l}
W\in pwd(R^p)\\
t'\in \ptopktuple{k}{s}(W)\\
t\in W
\end{subarray}}\alpha(t', W)Pr(W)\\
& &+\sum_{
\begin{subarray}{l}
W\in pwd(R^p)\\
t'\in \ptopktuple{k}{s}(W),~t\not\in W\\
(C_t\backslash\{t\})\cap W = \emptyset
\end{subarray}}\alpha(t', W)Pr(W)\\
&&+\sum_{
\begin{subarray}{l}
W\in pwd(R^p)\\
t'\in \ptopktuple{k}{s}(W),~t\not\in W\\
(C_t\backslash\{t\})\cap W \neq \emptyset
\end{subarray}}\alpha(t', W)Pr(W)
)\nonumber\\
&=&\frac{p'(t)}{p(t)}P^{R^p}_{k,s}(t'),
\end{eqnarray*}

where $c=1-\sum_{t''\in C_t\backslash\{t\}}p(t'')$.

Now we can see that, $t$'s Global-Top$k$ probability in $(R^p)'$ will be raised to \emph{exactly} $\frac{p'(t)}{p(t)}$ times of that in $R^p$ under the same weak order scoring function $s$, and for any tuple other than $t$, its Global-Top$k$ probability in $(R^p)'$ can be raised to \textit{as much as} $\frac{p'(t)}{p(t)}$ times of that in $R^p$ under the same scoring function $s$. As a result, $P^{(R^p)'}_{k,s}(t)$ is still among the highest $k$ Global-Top$k$ probabilities in $(R^p)'$ under the function $s$, and therefore still a winner.

\item[Case 2:] Losers.

This case is similar to \emph{Case 1}.
\end{itemize}
\textbf{Part II}: Score.
\begin{itemize}
\item[Case 1:] Winners.

For any winner $t\in A$, we evaluate $R^p$ under a new general scoring function $s'$. Comparing to $s$, $s'$ only raises the score of $t$. That is, $s'(t)>s(t)$ and for any $t'\in R, t'\neq t, s'(t')=s(t')$. Then, in addition to all the worlds already \emph{totally} (i.e., $\alpha(t,W)=1$) or \emph{partially} (i.e., $\alpha(t, W)<1$) contributing to $t$'s Global-Top$k$ probability when evaluating $R^p$ under $s$, some other worlds may now totally or partially contribute to $t$'s Global-Top$k$ probability. Because, under the function $s'$, $t$ might climb high enough to be in the top-$k$ answer set of those worlds. Moreover, if a possible world $W$ contributes partially under scoring function $s$, it is easy to see that it contributes totally under scoring function $s'$.

For any tuple $t''$ other than $t$ in $R$, 
\begin{enumerate}
\item[(i)] If $s(t'')\neq s(t)$, then its Global-Top$k$ probability under the function $s'$ either stays the same (if the ``climbing'' of $t$ does not knock that tuple out of the top-$k$ answer set in some possible world) or decreases (otherwise); 
\item[(ii)] If $s(t'')=s(t)$, then for any possible world $W$ contributing to $t''$'s Global-Top$k$ under scoring function $s$, $\alpha(t'', W)=\frac{k-a}{b}$, and now under scoring function $s'$, $\alpha'(t'', W)=\frac{k-a-1}{b-1}<\frac{k-a}{b}=\alpha(t'', W)$. Therefore the Global-Top$k$ of $t''$ under scoring function $s'$ is less than that under scoring function $s$.
\end{enumerate}

Consequently, $t$ is still a winner when evaluating $R^p$ under the function $s'$.

\item[Case 2:] Losers.

This case is similar to \emph{Case 1}.
\end{itemize}

\item[(10)]PT-$k$ satisfies \emph{Stability}.

In the rest of this proof, let $A$ be the set of all winners under the PT-$k$ semantics.

\textbf{Part I}: Probability.
\begin{itemize}
\item[Case 1:] Winners.

For any winner $t\in A$, if we only raise the probability of $t$, we have a new probabilistic relation $(R^p)'=\langle R,p',\mathcal{C}\rangle$, where the new probability function $p'$ is such that $p'(t)>p(t)$ and for any $t'\in R, t'\neq t, p'(t')=p(t')$. Note that $pwd(R^p)=pwd((R^p)')$.
In addition, assume $t\in C_t$, where $C_t\in \mathcal{C}$. The Global-Top$k$ probability of $t$ is such that

\begin{eqnarray*}
P^{R^p}_{k,s}(t)&=&\sum_{
\begin{subarray}{l}
W\in pwd(R^p)\\
t\in \topktuple{k}{s}(W)
\end{subarray}
}Pr(W)\geq p_{\tau}
\end{eqnarray*}
and
\begin{eqnarray*}
P^{(R^p)'}_{k,s}(t)&=&\sum_{
\begin{subarray}{l}
W\in pwd(R^p)\\
t\in \topktuple{k}{s}(W)
\end{subarray}
}Pr(W)\frac{p'(t)}{p(t)}\\
&=&\frac{p'(t)}{p(t)}P^{R^p}_{k,s}(t)>P^{R^p}_{k,s}(t)\geq p_{\tau}.
\end{eqnarray*}

Therefore, $P^{(R^p)'}_{k,s}(t)$ is still above the threshold $p_{\tau}$, and $t$ still belongs to the top-$k$ answer set of $(R^p)'$ under the function $s$.

\item[Case 2:] Losers.

This case is similar to \emph{Case 1}.
\end{itemize}
\textbf{Part II}: Score.
\begin{itemize}
\item[Case 1:] Winners.

For any winner $t\in A$, we evaluate $R^p$ under a new scoring function $s'$. Comparing to $s$, $s'$ only raises the score of $t$. 
Use a similar argument as that in (9) Part II Case 1 but under injective scoring functions, we can show that the Global-Top$k$ probability of $t$ is non-decreasing and is still above the threshold $p_{\tau}$. Therefore, tuple $t$ still belongs to the top-$k$ answer set under the function $s'$.

\item[Case 2:] Losers.

This case is similar to \emph{Case 1}.
\end{itemize}
\item[(11)]U-Top$k$ satisfies \emph{Stability}.

In the rest of this proof, let $A$ be the set of all winners under U-Top$k$ semantics.

\textbf{Part I}: Probability.
\begin{itemize}
\item[Case 1:] Winners. 

For any winner $t\in A$, if we only raise the probability of $t$, we have a new probabilistic relation $(R^p)'=\langle R,p',\mathcal{C}\rangle$, where the new probabilistic function $p'$ is such that $p'(t)>p(t)$ and for any $t'\in R, t'\neq t, p'(t')=p(t')$. In the following discussion, we use superscript to indicate the probability in the context of $(R^p)'$. Note that $pwd(R^p)=pwd((R^p)')$.

Recall that $Q_{k,s}(A_t)$ is the probability of a top-$k$ answer set $A_t\subseteq A$ under U-Top$k$ semantics, where $t\in A_t$. Since $t\in A_t$, $Q'_{k,s}(A_t)=Q_{k,s}(A_t)\frac{p'(t)}{p(t)}$.

For any candidate top-$k$ answer set $B$ other than $A_t$, i.e., $\exists W\in pwd(R^p), \topktuple{k}{s}(W)=B$ and $B\neq A_t$. By definition, 
\[Q_{k,s}(B)\leq Q_{k,s}(A_t).\] 
For any world $W$ contributing to $Q_{k,s}(B)$, its probability either increase $\frac{p'(t)}{p(t)}$ times (if $t\in W$), or stays the same (if $t\not\in W$ and $\exists t'\in W, t'$ and $t$ are exclusive), or decreases (otherwise). Therefore,
\[Q'_{k,s}(B) \leq Q_{k,s}(B)\frac{p'(t)}{p(t)}.\]
Altogether, 
\[Q'_{k,s}(B)\leq Q_{k,s}(B)\frac{p'(t)}{p(t)}\leq Q_{k,s}(A_t)\frac{p'(t)}{p(t)}= Q'_{k,s}(A_t).\] 

Therefore, $A_t$ is still a top-$k$ answer set to $(R^p)'$ under the function $s$ and $t\in A_t$ is still a winner.

\item[Case 2:] Losers.

It is more complicated in the case of losers. 
We need to show that for any loser $t$, if we decrease its probability, no top-$k$ candidate answer set $B_t$ containing $t$ will be a new top-$k$ answer set under the U-Top$k$ semantics. The procedure is similar to that in \emph{Case 1}, except that when we analyze the new probability of any original top-$k$ answer set $A_i$, we need to differentiate between two cases: 
\begin{enumerate}
\item[(a)] $t$ is exclusive with some tuple in $A_i$;
\item[(b)] $t$ is independent of all the tuples in $A_i$.
\end{enumerate}
It is easier with (a), where all the worlds contributing to the probability of $A_i$ do not contain $t$. In (b), some worlds contributing to the probability of $A_i$ contain $t$, while others do not. And we calculate the new probability for those two kinds of worlds differently. As we will see shortly, the probability of $A_i$ stays unchanged in either (a) or (b).

For any loser $t\in R, t\not\in A$, by applying the technique used in \emph{Case 1}, we have a new probabilistic relation $(R^p)'=\langle R, p', \mathcal{C}\rangle$, where the new probabilistic function $p'$ is such that $p'(t)<p(t)$ and for any $t'\in R, t'\neq t, p'(t')=p(t')$. Again, $pwd(R^p)=pwd((R^p)')$.

For any top-$k$ answer set $A_i$ to $R^p$ under the function $s$, $A_i\subseteq A$.
Denote by $S_{A_i}$ all the possible worlds contributing to $Q_{k,s}(A_i)$. Based on the membership of $t$, $S_{A_i}$ can be partitioned into two subsets $S_{A_i}^t$ and $S_{A_i}^{\bar{t}}$.
\[\begin{array}{l}
S_{A_i} = \{W|W\in pwd(R^p), \topktuple{k}{s}(W)=A_i\};\\
S_{A_i}=S_{A_i}^{t}\cup S_{A_i}^{\bar{t}}, S_{A_i}^{t}\cap S_{A_i}^{\bar{t}}=\emptyset,\\
\forall W\in S_{A_i}^{t}, t\in W\textrm{ and }\forall W\in S_{A_i}^{\bar{t}}, t\not\in W.
\end{array}\]

If $t$ is exclusive with some tuple in $A_i$, $S_{A_i}^{t}=\emptyset$. In this case, any world $W\in S_{A_i}^{\bar{t}}$ contains one of $t$'s exclusive tuples, therefore $W$'s probability will not be affected by the change in $t$'s probability. In this case,

\begin{eqnarray*}
Q'_{k,s}(A_i)&=&\sum_{
\begin{subarray}{l}
W\in pwd(R^p)\\
W\in S_{A_i}^{\bar{t}}
\end{subarray}
}Pr'(W)
=\sum_{
\begin{subarray}{l}
W\in pwd(R^p)\\
W\in S_{A_i}^{\bar{t}}
\end{subarray}
}Pr(W)\\
&=&Q_{k,s}(A_i).
\end{eqnarray*}

Otherwise, $t$ is independent of all the tuples in $A_i$. In this case, 

\[\frac{
\sum_{
\begin{subarray}{l}
W\in pwd(R^p)\\
W\in S_{A_i}^{t}
\end{subarray}
}Pr(W)}{\sum_{
\begin{subarray}{l}
W\in pwd(R^p)\\
W\in S^{\bar{t}}_{A_i}
\end{subarray}
}
Pr(W)}=\frac{p(t)}{1-p(t)}\]
and
\begin{eqnarray*}
Q'_{k,s}(A_i)&=&\sum_{
\begin{subarray}{l}
W\in pwd(R^p)\\
W\in S_{A_i}^{t}
\end{subarray}}Pr(W)\frac{p'(t)}{p(t)}\\
&&+\sum_{
\begin{subarray}{l}
W\in pwd(R^p)\\
W\in S_{A_i}^{\bar{t}}
\end{subarray}}Pr(W)\frac{1-p'(t)}{1-p(t)}\\
&=&\sum_{
\begin{subarray}{l}
W\in pwd(R^p)\\
W\in S_{A_i}
\end{subarray}}Pr(W)\\
&=&Q_{k,s}(A_i).
\end{eqnarray*}

We can see that in both cases, $Q'_{k,s}(A_i)=Q_{k,s}(A_i)$. 

Now for any top-$k$ candidate answer set containing $t$, say $B_t$ such that $B_t\not\subseteq A$, by definition, $Q_{k,s}(B_t)< Q_{k,s}(A_i)$. Moreover, 
\[Q'_{k,s}(B_t)=Q_{k,s}(B_t)\frac{p'(t)}{p(t)}<Q_{k,s}(B_t).\] 
Therefore, 
\[Q'_{k,s}(B_t)<Q_{k,s}(B_t)< Q_{k,s}(A_i)=Q'_{k,s}(A_i).\] 
Consequently, $B_t$ is still not a top-$k$ answer set to $(R^p)'$ under the function $s$. Since no top-$k$ candidate answer set containing $t$ can be a top-$k$ answer set to $(R^p)'$ under the function $s$, $t$ is still a loser.
\end{itemize}
\textbf{Part II}: Score.

Again, $A_i\subseteq A$ is a top-$k$ answer set to $R^p$ under the function $s$ by U-Top$k$ semantics.
\begin{itemize}
\item[Case 1:] Winners.

For any winner $t\in A_i$, we evaluate $R^p$ under a new scoring function $s'$. Comparing to $s$, $s'$ only raises the score of $t$. That is, $s'(t)>s(t)$ and for any $t'\in R, t'\neq t, s'(t')=s(t')$. In some possible world such that $W\in pwd(R^p)$ and $\topktuple{k}{s}(W)\neq A_i$, $t$ might climb high enough to be in $\topktuple{k}{s'}(W)$. Define $T$ to the set of such top-$k$ candidate answer sets.
\[T=\{\topktuple{k}{s'}(W) | W\in pwd(R^p), t\not \in \topktuple{k}{s}(W)\wedge t\in \topktuple{k}{s'}(W)\}.\]
Only a top-$k$ candidate set $B_j\in T$ can possibly end up with a probability higher than that of $A_i$ across all possible worlds, and thus substitute for $A_i$ as a new top-$k$ answer set to $R^p$ under the function $s'$. In that case, $t\in B_j$, so $t$ is still a winner.

\item[Case 2:] Losers.

For any loser $t\in R, t\not\in A$. Using a similar technique to \emph{Case 1}, the new scoring function $s'$ is such that $s'(t)<s(t)$ and for any $t'\in R, t'\neq t, s'(t')=s(t')$. When evaluating $R^p$ under the function $s'$, for any world $W\in pwd(R^p)$ such that $t\not \in \topktuple{k}{s}(W)$, the score decrease of $t$ will not effect its top-$k$ answer set, i.e., $\topktuple{k}{s'}(W)=\topktuple{k}{s}(W)$. For any world $W\in pwd(R^p)$ such that $t\in \topktuple{k}{s}(W)$, $t$ might go down enough to drop out of $\topktuple{k}{s'}(W)$. In this case, $W$ will contribute its probability to a top-$k$ candidate answer set without $t$, instead of the original one with $t$. In other words, under the function $s'$, comparing to the evaluation under the function $s$, the probability of a top-$k$ candidate answer set with $t$ is non-increasing, while the probability of a top-$k$ candidate answer set without $t$ is non-decreasing\footnote{Here, any subset of $R$ with cardinality at most $k$ that is not a top-$k$ candidate answer set under the function $s$ is conceptually regarded as a top-$k$ candidate answer set with probability zero under the function $s$.}. 

Since any top-$k$ answer set to $R^p$ under the function $s$ does not contain $t$, it follows from the above analysis that any top-$k$ candidate answer set containing $t$ will not be a top-$k$ answer set to $R^p$ under the new function $s'$, and thus $t$ is still a loser.
\end{itemize}
\item[(12)]U-$k$Ranks violates \emph{Stability}.

The following is a counterexample. 

Say $k=2$, $R^p$ is simple. $R=\{t_1, t_2, t_3\}$, $t_1\succ_s t_2\succ_s t_3$. $p(t_1)=0.3, p(t_2)=0.4, p(t_3)=0.3$.
\begin{center}
\begin{tabular}{c c c c}
  \hline
   & $t_1$ & $t_2$ & $t_3$ \\
  \hline
  rank 1 & 0.3 & 0.28 & 0.126 \\
  rank 2 & 0 & 0.12 & 0.138\\
  rank 3 & 0 & 0 & 0.036\\
  \hline
\end{tabular}
\end{center}
By U-$k$Ranks, the top-$2$ answer set is $\{t_1, t_3\}$. 

Now raise the score of $t_3$ such that $t_1\succ_{s'} t_3 \succ_{s'} t_2$. 
\begin{center}
\begin{tabular}{c c c c}
  \hline
   & $t_1$ & $t_3$ & $t_2$ \\
  \hline
  rank 1 & 0.3 & 0.21 & 0.196 \\
  rank 2 & 0 & 0.09 & 0.168\\
  rank 3 & 0 & 0 & 0.036\\
  \hline
\end{tabular}
\end{center}
By U-$k$Ranks, the top-$2$ answer set is $\{t_1, t_2\}$. By raising the score of $t_3$, we actually turn the winner $t_3$ to a loser, which contradicts \emph{Stability}.

\end{enumerate}

\section{Appendix B: Proofs}
\subsection{Proof for Proposition \ref{prop_recursion}}
\noindent \textbf{Proposition \ref{prop_recursion}.}\textit{
Given a simple probabilistic relation $R^p=\langle R, p, \mathcal{C}\rangle$ and an injective scoring function $s$ over $R^p$, if $R=\{t_1, t_2, \ldots, t_n\}$ and $t_1\succ_s t_2\succ_s \ldots \succ_s t_n$, the following recursion on Global-Top$k$ queries holds.
\begin{equation*}
q(k,i) = \left\{ \begin{array}{lr}
0  & k=0 \\
p(t_i) & 1\leq i\leq k \\
(q(k,i-1)\dfrac{\bar{p}(t_{i-1})}{p(t_{i-1})}+q(k-1,i-1))p(t_i)&\textrm{ otherwise}
\end{array} \right.
\end{equation*}
where $q(k,i)=P_{k,s}(t_i)$ and $\bar{p}(t_{i-1})=1-p(t_{i-1})$.}

\begin{proof}
By induction on $k$ and $i$.
\begin{itemize}
\item Base case.
\begin{itemize}
\item $k=0$ 

For any $W\in pwd(R^p)$, $\topktuple{0}{s}(W)=\emptyset$. Therefore, for any $t_i\in R$, the Global-Top$k$ probability of $t_i$ is $0$.
\item $k>0$ and $i=1$

$t_1$ has the highest score among all tuples in $R$. As long as tuple $t_1$ appears in a possible world $W$, it will be in the $\topktuple{k}{s}(W)$. So the Global-Top$k$ probability of $t_i$ is the probability that $t_1$ appears in possible worlds, i.e., $q(k,1)=p(t_1)$.
\end{itemize}

\item Inductive step.

Assume the theorem holds for $0\leq k\leq k_0$ and $1\leq i\leq i_0$.
For any $W\in pwd(R^p)$, $t_{i_0}\in \topktuple{k_0}{s}(W)$ iff $t_{i_0}\in W$ and 
there are at most $k_0-1$ tuples with a higher score in $W$.
Note that any tuple with score lower than the score of $t_{i_0}$ does not have any influence on $q(k_0,i_0)$, because its presence/absence in a possible world will not affect the presence of $t_{i_0}$ in the top-$k$ answer set of that world.

Since all the tuples are independent, 
\[q(k_0, i_0)=p(t_{i_0}) \sum_{
\begin{subarray}{c}
W\in pwd(R^p)\\
|\{t|t\in W \wedge t\succ_s t_{i_0}\}|<k_0
\end{subarray}}Pr(W).\]

\begin{enumerate}
\item[(1)] $q(k_0,i_0+1)$ is the Global-Top$k_0$ probability of tuple $t_{i_0+1}$.
\begin{eqnarray*}
q(k_0, i_0+1)&=&\sum_{\begin{subarray}{l}
W\in pwd(R^p)\\
t_{i_0+1}\in \topktuple{k_0}{s}(W)\\ 
t_{i_0}\in \topktuple{k_0}{s}(W)
\end{subarray}}Pr(W)\\
&+&\sum_{\begin{subarray}{l}
W\in pwd(R^p)\\
t_{i_0+1}\in \topktuple{k_0}{s}(W)\\
t_{i_0}\in W,~t_{i_0}\not\in \topktuple{k_0}{s}(W)
\end{subarray}}Pr(W)\\
&+&\sum_{\begin{subarray}{l}
W\in pwd(R^p)\\
t_{i_0+1}\in \topktuple{k_0}{s}(W)\\
t_{i_0}\not \in W
\end{subarray}}Pr(W).
\end{eqnarray*}

For the first part of the left hand side,

\[\sum_{\begin{subarray}{l}
W\in pwd(R^p)\\
t_{i_0+1}\in \topktuple{k_0}{s}(W)\\
t_{i_0}\in \topktuple{k_0-1}{s}(W)
\end{subarray}
}Pr(W)
=p(t_{i_0+1})q(k_0-1, i_0).\]

The second part is zero. Since $t_{i_0}\succ_s t_{i_0+1}$,
if $t_{i_0+1}\in \topktuple{k_0}{s}(W)$ and  $t_{i_0}\in W$, then $t_{i_0} \in \topktuple{k_0}{s}(W)$.

The third part is the sum of the probabilities of all possible worlds such that $t_{i_0+1}\in W,t_{i_0}\not\in W$ and there are at most $k_0-1$ tuples with score higher than the score of $t_{i_0}$ in $W$.
So it is equivalent to 

\begin{eqnarray*}
&&p(t_{i_0+1})\overline{p}(t_{i_0}) \sum_{\begin{subarray}{l}
|\{t|t\in W \wedge t\succ_s t_{i_0}\}|<k_0
\end{subarray}}Pr(W)\\
&=&p(t_{i_0+1})\overline{p}(t_{i_0})\frac{q(k_0,i_0)}{p(t_{i_0})}.\end{eqnarray*}

Altogehter, we have

\begin{eqnarray*}
&&q(k_0, i_0+1)\\
&=&p(t_{i_0+1})q(k_0-1, i_0)+p(t_{i_0+1})\overline{p}(t_{i_0})\frac{q(k_0,i_0)}{p(t_{i_0})}\\
&=&(q(k_0-1, i_0)+q(k_0,i_0)\frac{\overline{p}(t_{i_0})}{p(t_{i_0})})p(t_{i_0+1}).
\end{eqnarray*}

\item[(2)] $q(k_0+1,i_0)$ is the Global-Top$(k_0+1)$ probability of tuple $t_{i_0}$.
Use a similar argument as above, it can be shown that this case is correctly computed by Equation (\ref{eqn_recursion}) as well. 
\end{enumerate}
\end{itemize}
\end{proof}

\subsection{Proof for Theorem \ref{thm_ta}}
\noindent \textbf{Theorem \ref{thm_ta} (Correctness of Algorithm $1^{\textit{TA}}$).}\textit{
Given a simple probabilistic relation $R^p=\langle R,p,\mathcal{C}\rangle$, a non-negative integer $k$ and an injective scoring function $s$ over $R^p$, the above TA-based algorithm correctly finds a Global-Top$k$ top-$k$ answer set.}

\begin{proof} 
In every iteration of Step (2), say $\underline{t}=t_i$, for any unseen tuple $t$, $s'$ is an injective scoring function over $R^p$, which only differs from $s$ in the score of $t$. Under the function $s'$, $t_i\succ_{s'}t\succ_{s'}t_{i+1}$. If we evaluate the top-$k$ query in $R^p$ under $s'$ instead of $s$, $P_{k,s'}(t)=\frac{p(t)}{\underline{p}}UP$. On the other hand, for any $W\in pwd(R^p)$, $W$ contributing to $P_{k,s}(t)$ implies that $W$ contributes to $P_{k,s'}(t)$, while the reverse is not necessarily true. So, we have $P_{k,s'}(t)\geq P_{k,s}(t)$. Recall that $\underline{p}\geq p(t)$, therefore $UP\geq \frac{p(t)}{\underline{p}}UP = P_{k,s'}(t)\geq P_{k,s}(t)$.
The conclusion follows from the correctness of the original TA algorithm and Algorithm \ref{alg_ind}.
\end{proof}

\subsection{Proof for Lemma \ref{thm_ied}}
\noindent \textbf{Lemma \ref{thm_ied}.} \textit{
Let $R^p=\langle R, p, \mathcal{C}\rangle$ be a probabilistic relation, $s$ an injective scoring function, $t\in R$, and $E^p=\langle E,p^E,\mathcal{C}^E\rangle$ the event relation induced by $t$.
Define $Q^p=\langle E-\{t_{e_t}\},p^E,\mathcal{C}^E-\{\{t_{e_t}\}\}\rangle$.
Then, the Global-Top$k$ probability of $t$ satisfies the following:}
\[
P^{R^p}_{k,s}(t)=p(t)\sum_{\begin{subarray}{l}
W_e\in pwd(Q^p)\\
|W_e|< k
\end{subarray}}Pr(W_e).\]

\begin{proof}
Given $t\in R$, $k$ and $s$, let $A$ be a subset of $pwd(R^p)$ such that $W \in A\Leftrightarrow t\in \topktuple{k}{s}(W)$. If we group all the possible worlds in $A$ by the set of parts whose tuple in $W$ has higher score than the score of $t$, then we will have the following partition:
\[A=A_1\cup A_2 \cup \ldots \cup A_q, A_i\cap A_j=\emptyset, i\neq j\]
and
\[\begin{array}{l}
\forall A_i, \forall W_1, W_2 \in A_i,i=1,2,\ldots,q, \\
\{C_j|\exists t'\in W_1\cap C_j, t'\succ_s t\} = \{C_j|\exists t'\in W_2\cap C_j, t'\succ_s t\}.
\end{array}\]
Moreover, denote $CharParts(A_i)$ to $A_i$'s characteristic set of parts.

\smallskip

Now, let $B$ be a subset of $pwd(Q^p)$, such that $W_e\in B\Leftrightarrow |W_e|< k$.
There is a bijection $g: \{A_i|A_i\in A\}\rightarrow B$, mapping each part $A_i$ in $A$ to a possible world in $B$ which contains only tuples corresponding to the parts in $A_i$ 's characteristic set.
\[g(A_i)=\{t_{e_{C_j}}|C_j \in CharParts(A_i)\}.\]

The following equation holds from the definition of an induced event relation and Proposition \ref{ppt_iedind}.
\begin{eqnarray*}
\sum_{W\in A_i}Pr(W)&=&p(t)\prod_{C_i\in CharParts(A_i)}p(t_{e_{C_i}})\prod_{
\begin{subarray}{l}
C_i\in\mathcal{C}-\{C_{id(t)}\}\\
C_i\not\in CharParts(A_i)
\end{subarray}}(1-p(t_{e_{C_i}}))\\
&=&p(t)Pr(g(A_i)).
\end{eqnarray*}
Therefore,
\begin{eqnarray*}
P^{R^p}_{k,s}(t)&=&\sum_{W\in A}Pr(W)=\sum^q_{i=1}(\sum_{W\in A_i}Pr(W))\\
&=&\sum^q_{i=1}p(t) Pr(g(A_i))=p(t)\sum^q_{i=1}Pr(g(A_i))\\
&=&p(t)\sum_{W_e\in B}Pr(W_e)\\
&=&p(t)(\sum_{\begin{subarray}{l}
W_e\in pwd(Q^p)\\
|W_e|< k
\end{subarray}}Pr(W_e)).
\end{eqnarray*} 
\end{proof}

\subsection{Proof for Proposition \ref{prop_iedscore}}

\noindent \textbf{Proposition \ref{prop_iedscore} (Correctness of Algorithm \ref{alg_indExsub}).}\textit{
Given a probabilistic relation $R^p=\langle R, p, \mathcal{C}\rangle$ and an injective scoring function $s$, for any $t\in R^p$, the Global-Top$k$ probability of $t$ equals the Global-Top$k$ probability of $t_{e_t}$ when evaluating top-$k$ in the induced event relation $E^p=\langle E, p^E, \mathcal{C}^E\rangle$ under the injective scoring function $s^E:E\rightarrow \mathbb{R}, s^E(t_{e_t})=\frac{1}{2}$ and $s^E(t_{e_{C_i}})=i$:}
\[P^{R^p}_{k,s}(t)=P^{E^p}_{k,s^E}(t_{e_t}).\]

\begin{proof}
Since $t_{e_t}$ has the lowest score under $s^E$, for any $W_e\in pwd(E^p)$, the only chance $t_{e_t}\in \topktuple{k}{s^E}(W_e)$ is when there are at most $k$ tuples in $W_e$, including $t_{e_t}$.
\[\begin{array}{l}
\forall W_e\in pwd(E^p), \\
t_{e_t}\in \topktuple{k}{s}(W_e) \Leftrightarrow (t_{e_t}\in W_e \wedge |W_e|\leq k).
\end{array}\]
Therefore, 
\[P^{E^p}_{k,s^E}(t_{e_t})=\sum_{t_{e_t}\in W_e\wedge |W_e|\leq k}Pr(W_e).\]

In the proof of Lemma \ref{thm_ied}, $B$ contains all the possible worlds having at most $k-1$ tuples from $E-\{t_{e_t}\}$. By Proposition \ref{ppt_iedind}, 
\[\sum_{t_{e_t}\in W_e\wedge |W_e|\leq k}Pr(W_e)=p(t)\sum_{W'_e\in B}Pr(W'_e).\]
By Lemma \ref{thm_ied},
\[p(t)\sum_{W'_e\in B}Pr(W'_e)=P^{R^p}_{k,s}(t).\]
Consequently,
\[P^{R^p}_{k,s}(t)=P^{E^p}_{k,s^E}(t_{e_t}). 
\]
\end{proof}

\subsection{Proof for Proposition \ref{prop_wo_simple}}
\noindent \textbf{Proposition \ref{prop_wo_simple} (Correctness of Algorithm \ref{alg_ind_wo}).}\textit{
Let $R^p=\langle R, p, \mathcal{C}\rangle$ be a simple probabilistic relation where $R=\{t_1, \ldots, t_n\}$, $t_1\succeq_s t_2\succeq_s \ldots \succeq_s t_n$, $k$ a non-negative integer and $s$ a scoring function. For every $t_l\in R$, the Global-Top$k$ probability of $t_l$ can be computed by the following equation:
\begin{equation*}
P^{R^p}_{k,s}(t_l)=\sum_{k'=0}^{k-1} T_{k',[i_l]}\cdot P_{k-k',s}^{R_{s}^{p}(t_l)}(t_l)
\end{equation*}
\noindent where $R^p_s(t_l)$ is $R^p$ restricted to $\{t\in R|t\sim_s t_l\}$.
}

\begin{proof}
Given a tuple $t_l\in R$, let $R_{\theta}$ be the support relation $R$ restricted to $\{t\in R|t~\theta~t_l\}$, and $R^p_{\theta}$ be $R^p$ restricted to $R_{\theta}$, where $\theta\in \{\succ,\sim,\prec,\preceq\}$ (subscript $s$ omitted). Similarly, for each possible world $W\in pwd(R^p)$, $W_{\theta}=W\cap R_{\theta}$.
 
Each possible world $W\in pwd(R^p)$ such that $t_l\in \ptopktuple{k}{s}(W)$ contributes \\$\min(1,\frac{k-a}{b})Pr(W)$ to $P^{R^p}_{k,s}(t_l)$, where $a=|W_{\succ}|$ and $b=|W_{\sim}|$.

\begin{eqnarray*}
P^{R^p}_{k,s}(t_l)&=&\sum_{\begin{subarray}{l}
W\in pwd(R^p), t_l\in W\\
|W_{\succ}|=a, 0\leq a \leq k-1\\
|W_{\sim}|=b, 1\leq b\leq m
\end{subarray}}\min(1, \frac{k-a}{b})Pr(W)\\
&=&\sum_{a=0}^{k-1}\sum_{b=1}^{m}\min(1,\frac{k-a}{b})(\sum_{
\begin{subarray}{l}
W\in pwd(R^p), t_l\in W\\
|W_{\succ}|=a, |W_{\sim}|=b
\end{subarray}}Pr(W))\\
&=&\sum_{a=0}^{k-1}\sum_{b=1}^{m}\min(1, \frac{k-a}{b})(\sum_{\begin{subarray}{l}
W_{\succ}\in pwd(R^p_{\succ})\\
|W_{\succ}|=a
\end{subarray}}Pr(W_{\succ}) \sum_{
\begin{subarray}{l}
W_{\preceq}\in pwd(R^p_{\preceq}), t_l\in W_{\preceq}\\
|W_{\sim}|=b
\end{subarray}}Pr(W_{\preceq}))\\
&=&\sum_{a=0}^{k-1}(\sum_{\begin{subarray}{l}
W_{\succ}\in pwd(R^p_{\succ})\\
|W_{\succ}|=a
\end{subarray}}Pr(W_{\succ})
\sum_{b=1}^{m}\min(1,\frac{k-a}{b})
(\sum_{
\begin{subarray}{l}
W_{\preceq}\in pwd(R^p_{\preceq}), t_l\in W_{\preceq}\\
|W_{\sim}|=b
\end{subarray}}Pr(W_{\preceq})))\\
&=&\sum_{a=0}^{k-1}(T_{a,[i_l]}
\sum_{b=1}^{m}\min(1,\frac{k-a}{b})
(\sum_{
\begin{subarray}{l}
W_{\sim}\in pwd(R^p_{\sim}), t_l\in W_{\sim}\\
|W_{\sim}|=b
\end{subarray}}Pr(W_{\sim})
\sum_{
\begin{subarray}{l}
W_{\prec}\in pwd(R^p_{\prec})
\end{subarray}}Pr(W_{\prec})
))\\
&=&\sum_{a=0}^{k-1}(T_{a,[i_l]}
\sum_{b=1}^{m}\min(1,\frac{k-a}{b})
(\sum_{
\begin{subarray}{l}
W_{\sim}\in pwd(R^p_{\sim}), t_l\in W_{\sim}\\
|W_{\sim}|=b
\end{subarray}}Pr(W_{\sim})))\\
&=&\sum_{a=0}^{k-1} T_{a,[i_l]}\cdot P_{k-a,s}^{R_{s}^{p}(t_l)}(t_l)
\end{eqnarray*}
where $m$ is the number of tying tuples with $t_l$ (inclusive), i.e., $m=|R_{s}^{p}(t_l)|$.

\end{proof}

\subsection{Proof for Proposition \ref{prop_iedscore_WO}}
\noindent \textbf{Proposition \ref{prop_iedscore_WO}.}\textit{
Given a probabilistic relation $R^p=\langle R, p, \mathcal{C}\rangle$ and a scoring function $s$, for any $t\in R^p$, the Global-Top$k$ probability of $t$ equals the Global-Top$k$ probability of $t_{e_t,\sim}$ when evaluating top-$k$ in the induced event relation $E^p=\langle E, p^E, \mathcal{C}^E\rangle$ under the scoring function $s^E:E\rightarrow \mathbb{R}$, $s^E(t_{e_t,\succ})=\frac{1}{2}$, $s^E(t_{e_t,\sim})=\frac{1}{2}$, $s^E(t_{e_{C_i},\sim})=\frac{1}{2}$ and $s^E(t_{e_{C_i,\succ}})=i$:
\[P^{R^p}_{k,s}(t)=P^{E^p}_{k,s^E}(t_{e_t,\sim}).\]
}

\begin{proof}
Similar to what we did in the Proof for Lemma \ref{thm_ied}. We are trying to create a bijection.

Given $t\in R$, $k$ and $s$, let $A$ be a subset of $pwd(R^p)$ such that $W \in A\Leftrightarrow t\in \ptopktuple{k}{s}(W)$. If we group all the possible worlds in $A$ by the set of parts whose tuple in $W$ has a score higher than or equal to that of $t$, then we will have the following partition:
\[A=A_1\cup A_2 \cup \ldots \cup A_q, A_i\cap A_j=\emptyset, i\neq j\]
and
\[\begin{array}{l}
\forall A_i, \forall W_1, W_2 \in A_i,i=1,2,\ldots,q, \\
\{C_{j,\succ}|\exists t'\in W_1\cap C_j, t'\succ_s t\} = \{C_{j, \succ}|\exists t'\in W_2\cap C_j, t'\succ_s t\}\\
\textrm{and}\\
\{C_{j, \sim}|\exists t'\in W_1\cap C_j, t'\sim_s t\} = \{C_{j, \sim}|\exists t'\in W_2\cap C_j, t'\sim_s t\}.
\end{array}\]
Moreover, denote $CharParts(A_i)$ to $A_i$'s characteristic set of parts. Note that all $W\in A_i$ have the same allocation coefficient $\alpha(t, W)$, denoted by $\alpha_i$.

\smallskip

Now, let $B$ be a subset of $pwd(E^p)$, such that $W_e\in B\Leftrightarrow t_{e_t,\sim} \in \ptopktuple{k}{s}(W_e)$.
There is a bijection $g: \{A_i|A_i\in A\}\rightarrow B$, mapping each part $A_i$ in $A$ to the a possible world in $B$ which contains only tuples corresponding to parts in $A_i$ 's characteristic set.
\begin{eqnarray*}
g(A_i)&=&\{t_{e_{C_j}, \succ}|C_{j, \succ} \in CharParts(A_i)\}\cup\{t_{e_{C_j}, \sim}|C_{j, \sim} \in CharParts(A_i)\}
\end{eqnarray*}

Furthermore, the allocation coefficient $\alpha_i$ of $A_i$ equals to the allocation coefficient $\alpha(t_{e_t, \sim}, g(A_i))$ under the function $s^E$.

The following equation holds from the definition of an induced event relation under general scoring functions.
\begin{eqnarray*}
\sum_{W\in A_i}Pr(W)&=&\prod_{C_{i, \succ}\in CharParts(A_i)}p(t_{e_{C_i}, \succ}) \prod_{C_{i, \sim}\in CharParts(A_i)}p(t_{e_{C_i}, \sim})\\
& & \prod_{\begin{subarray}{l}C_i\in \mathcal{C}\\
C_{i, \sim}\not\in CharParts(A_i)\\
C_{i, \succ}\not\in CharParts(A_i)
\end{subarray}}(1-p(t_{e_{C_i}, \succ})-p(t_{e_{C_i}, \sim}))\\
&=&Pr(g(A_i)).
\end{eqnarray*}
Therefore,
\begin{eqnarray*}
P^{R^p}_{k,s}(t)&=&\sum_{W\in A}\alpha(t, W)Pr(W)=\sum^q_{i=1}(\alpha_i\sum_{W\in A_i}Pr(W))\\
&=&\sum^q_{i=1}\alpha_i Pr(g(A_i))=\sum^q_{i=1}\alpha(t_{e_t, \sim}, g(A_i))Pr(g(A_i))\\
&=&\sum_{W_e\in B}\alpha(t_{e_t, \sim}, W_e)Pr(W_e)\hspace{0.2in} (g \textrm{ is a bijection})\\
&=&P^{E^p}_{k,s^E}(t_{e_t,\sim}).
\end{eqnarray*} 
\end{proof}

\subsection{Proof for Theorem \ref{thm_recursion_WO}}
\noindent \textbf{Theorem \ref{thm_recursion_WO}.}\textit{
Given a probabilistic relation $R^p=\langle R, p, \mathcal{C}\rangle$, a scoring function $s$, $t\in R^p$,
and its induced event relation $E^p=\langle E, p^{E}, \mathcal{C}^{E}\rangle$, 
where $|E|=2m$,
the following recursion on $u_{\succ}(k',i,b)$ and $u_{\sim}(k',i,b)$ holds, where $b_{\max}$ is the number of tuples with a positive probability in $E^p_{\sim}$.\\
When $i=1, 0\leq k'\leq m$ and $0\leq b\leq b_{\max}$,
\begin{equation*}
u_{\succ}(k',1,b)=\left\{\begin{array}{lr}
p^{E}(t_{1,\succ})\hspace{0.5in}& k'=1, b=0\\
0 & \textrm{ otherwise}
\end{array}
\right.
\end{equation*}
\begin{equation*}
u_{\sim}(k',1,b)=\left\{\begin{array}{lr}
p^{E}(t_{1,\sim})\hspace{0.5in}& k'=1, b=1\\
0 & \textrm{ otherwise}
\end{array}
\right.
\end{equation*}
For every $i$, $2\leq i\leq m$, $0\leq k'\leq m$ and $0\leq b\leq b_{\max}$,\\
$u_{\succ}(k',i,b)=$\hfill (\ref{eqn_recursion_WO_odd})
\begin{eqnarray*}
&&\begin{tabular}{|c|c|}
\hline
Condition & Formula\\
\hline
$k'=0$  & $0$ \\
\hline
\begin{tabular}{l}
$1\leq k' \leq m$, $p^{E}(t_{i-1,\succ})>0$
\end{tabular}
& 
\begin{tabular}{l}
\\[1pt]
$(u_{\succ}(k',i-1,b)\dfrac{1-p^{E}(t_{i-1,\succ})-p^{E}(t_{i-1,\sim})}{p^{E}(t_{i-1,\succ})}$ \\
$+u_{\succ}(k'-1,i-1,b)$\\
$+u_{\sim}(k'-1,i-1,b))p^{E}(t_{i,\succ})$
\end{tabular}
\\
\hline
\begin{tabular}{l}
$1\leq k' \leq m$, $p^{E}(t_{i-1,\succ})=0$\\
and $0\leq b<b_{\max}$
\end{tabular}
&
\begin{tabular}{l}
\\[1pt]
$(u_{\sim}(k',i-1,b+1)\dfrac{1-p^{E}(t_{i-1,\succ})-p^{E}(t_{i-1,\sim})}{p^{E}(t_{i-1,\sim})}$\\
$+u_{\succ}(k'-1,i-1,b)$\\
$+u_{\sim}(k'-1,i-1,b))p^{E}(t_{i,\succ})$
\end{tabular}
\\ 
\hline
\begin{tabular}{l}
$1\leq k' \leq m$, $p^{E}(t_{i-1,\succ})=0$\\
and $b=b_{\max}$
\end{tabular}
&
\begin{tabular}{l}
\\[1pt]
$(u_{\succ}(k'-1,i-1,b)+u_{\sim}(k'-1,i-1,b))p^{E}(t_{i,\succ})$
\end{tabular}
\\
\hline
\end{tabular}
\end{eqnarray*}
$u_{\sim}(k',i,b)=$ \hfill(\ref{eqn_recursion_WO_even})
\begin{eqnarray*}
&&\begin{tabular}{|c|c|}
\hline
Condition & Formula \\
\hline
 $k'=0$ or $b=0$ & $0$\\
\hline
\begin{tabular}{l}
$1\leq k' \leq m$, $1\leq b\leq b_{\max}$\\
and $p^{E}(t_{i-1,\sim})>0$
\end{tabular}
&
\begin{tabular}{l}
\\[1pt]
$(u_{\sim}(k',i-1,b)\dfrac{1-p^{E}(t_{i-1,\succ})-p^{E}(t_{i-1,\sim})}{p^{E}(t_{i-1,\sim})}$ \\
$+u_{\succ}(k'-1,i-1,b-1)$\\
$+u_{\sim}(k'-1,i-1,b-1))p^{E}(t_{i,\sim})$
\end{tabular}
\\
\hline
\begin{tabular}{l}
$1\leq k' \leq m$, $1\leq b\leq b_{\max}$\\
and $p^{E}(t_{i-1,\sim})=0$
\end{tabular}
&
\begin{tabular}{l}
\\[1pt]
$(u_{\succ}(k',i-1,b-1)\dfrac{1-p^{E}(t_{i-1,\succ})-p^{E}(t_{i-1,\sim})}{p^{E}(t_{i-1,\succ})}$\\
$+u_{\succ}(k'-1,i-1,b-1)$\\
$+u_{\sim}(k'-1,i-1,b-1))p^{E}(t_{i,\sim})$
\end{tabular}
\\
\hline
\end{tabular}
\end{eqnarray*}
The Global-Top$k$ probability of $t_{e_t,\sim}$ in $E^p$ under the scoring function $s^E$ can be computed by the following equation:
\begin{eqnarray*}
P^{E^p}_{k,s^E}(t_{e_t,\sim})&=&P^{E^p}_{k,s^E}(t_{m,\sim})\\
&=&\sum^{b_{\max}}_{b=1}(\sum_{k'=1}^{k}u_{\sim}(k',m,b)+\sum_{k'=k+1}^{k+b-1}\frac{k-(k'-b)}{b}u_{\sim}(k',m,b)) (\ref{eqn_recursion_WO})
\end{eqnarray*}
}

\begin{proof}
Equation (\ref{eqn_recursion_WO}) follows Equation (\ref{eqn_recursion_WO_odd}) and Equation (\ref{eqn_recursion_WO_even}) as it is a simple enumeration based on Definition \ref{def_topkprob_WO}.
We are going to prove Equation (\ref{eqn_recursion_WO_odd}) and Equation (\ref{eqn_recursion_WO_even}) by an induction on $i$.
\begin{itemize}
\item Base case: $i=1, 0\leq k'\leq m$ and $0\leq b\leq b_{\max}$ 

When $i=1$, based on the definition of $u$, the only non-zero entries are $u_{\succ}(1, 1, 0)$ and $u_{\sim}(1, 1, 1)$. The former is the probability sum of all possible worlds which contain $t_{1, \succ}$ and do not contain $t_{1, \sim}$. The second requirement is redundant since those two tuples are exclusive. Therefore, it is simply the probability of $t_{1, \succ}$. Similarly, the latter is the probability sum of all possible worlds which contain $t_{1, \sim}$ and do not contain $t_{1, \succ}$. Again, it is simply the probability of $t_{1, \sim}$. It is easy to check that no possible worlds satisfy other combinations of $k'$ and $b$ when $i=1$, therefore their probabilities are $0$.

\item Inductive step.

Assume the theorem holds for $i\leq i_0$, $0\leq k'\leq m$ and $0\leq b\leq b_{\max}$, where $1\leq i_0\leq m-1$.

Denote $E_{\succ,[i]}$ and $E_{\sim,[i]}$ to the set of the first $i$ tuples in $E_{\succ}$ and $E_{\sim}$ respectively.

For any $W\in pwd(E^p)$, by definition, $W$ contributes to $u_{\succ/\sim}(k', i_0, b)$ iff $t_{i_0, \succ/\sim}\in W$ and $|W\cap (E_{\succ, [i_0]}\cup E_{\sim, [i_0]})|=k'$ and $|W\cap E_{\sim, [i_0]}|=b$. Since $E_{\succ, [i_0]}\cap E_{\sim, [i_0]}=\emptyset$, we have:

$W\textrm{ contributes to }u_{\succ/\sim}(k', i_0, b)$ $\Leftrightarrow$ $t_{i_0,\succ/\sim}\in W \textrm{ and }|W\cap E_{\succ, [i_0]}|=k'-b \textrm{ and } |W\cap E_{\sim, [i_0]}|=b.$

\begin{enumerate}
\item[(1)] $u_{\succ}(k',i_0+1,b)$ is the probability sum of all possible worlds $W$ such that $t_{i_0+1, \succ}\in W$, $|W\cap E_{\succ, [i_0+1]}|=k'-b$ and $|W\cap E_{\sim, [i_0+1]}|=b$.
\begin{eqnarray*}
u_{\succ}(k', i_0+1, b)&=&\sum_{\begin{subarray}{l}
W\in pwd(E^p), t_{i_0+1, \succ}\in W\\
|W\cap E_{\succ, [i_0+1]}|=k'-b\\
|W\cap E_{\sim, [i_0+1]}|=b
\end{subarray}}Pr(W)\\
&=&\sum_{\begin{subarray}{l}
W\in pwd(E^p), t_{i_0+1, \succ}\in W\\
|W\cap E_{\succ, [i_0]}|=k'-1-b\\
|W\cap E_{\sim, [i_0]}|=b
\end{subarray}}Pr(W)\hspace{0.1in} \begin{array}{l}\textrm{(Since }t_{i_0+1,\succ}\in W,\\t_{i_0+1,\sim}\not\in W)\end{array}\\
&=&\sum_{\begin{subarray}{l}
W\in pwd(E^p)\\
t_{i_0+1, \succ}\in W, t_{i_0, \succ}\in W\\
|W\cap E_{\succ, [i_0]}|=k'-1-b\\
|W\cap E_{\sim, [i_0]}|=b
\end{subarray}}Pr(W)\\
&+&\sum_{\begin{subarray}{l}
W\in pwd(E^p)\\
t_{i_0+1, \succ}\in W, t_{i_0, \sim}\in W\\
|W\cap E_{\succ, [i_0]}|=k'-1-b \\
|W\cap E_{\sim, [i_0]}|=b
\end{subarray}}Pr(W)\\
&+&\sum_{\begin{subarray}{l}
W\in pwd(E^p)\\
t_{i_0+1, \succ}\in W, t_{i_0, \succ}\not\in W, t_{i_0, \sim}\not\in W\\
|W\cap E_{\succ, [i_0]}|=k'-1-b \\
|W\cap E_{\sim, [i_0]}|=b
\end{subarray}}Pr(W)
\end{eqnarray*}

For the first part of the left hand side,
\[
\sum_{\begin{subarray}{l}
W\in pwd(E^p)\\
t_{i_0+1, \succ}\in W, t_{i_0, \succ}\in W\\
|W\cap E_{\succ, [i_0]}|=k'-1-b\\
|W\cap E_{\sim, [i_0]}|=b
\end{subarray}}Pr(W)
=p(t_{i_0+1})\sum_{\begin{subarray}{l}
W\in pwd(E^p), t_{i_0, \succ}\in W\\
|W\cap E_{\succ, [i_0]}|=k'-1-b\\
|W\cap E_{\sim, [i_0]}|=b
\end{subarray}}Pr(W)
=p(t_{i_0+1})u_{\succ}(k'-1, i_0, b).\]

For the second part of the left hand side,
\[
\sum_{\begin{subarray}{l}
W\in pwd(E^p)\\
t_{i_0+1, \succ}\in W, t_{i_0, \sim}\in W\\
|W\cap E_{\succ, [i_0]}|=k'-1-b\\
|W\cap E_{\sim, [i_0]}|=b
\end{subarray}}Pr(W)
=p(t_{i_0+1})\sum_{\begin{subarray}{l}
W\in pwd(E^p), t_{i_0, \sim}\in W\\
|W\cap E_{\succ, [i_0]}|=k'-1-b\\
|W\cap E_{\sim, [i_0]}|=b
\end{subarray}}Pr(W)
=p(t_{i_0+1})u_{\sim}(k'-1, i_0, b).\]

For the third part of the left hand side, if $p(t_{i_0,\succ})+p(t_{i_0,\sim})=1$, then there is no possible world satisfying this condition, therefore it is zero. Otherwise,
\begin{eqnarray}
\sum_{\begin{subarray}{l}
W\in pwd(E^p)\\
t_{i_0+1, \succ}\in W \\
t_{i_0, \succ}\not\in W, t_{i_0, \sim}\not\in W\\
|W\cap E_{\succ, [i_0]}|=k'-1-b\\
|W\cap E_{\sim, [i_0]}|=b
\end{subarray}}Pr(W)
&=&p(t_{i_0+1})
\sum_{\begin{subarray}{l}
W\in pwd(E^p)\\
t_{i_0, \succ}\not\in W, t_{i_0, \sim}\not\in W\\
|W\cap E_{\succ, [i_0]}|=k'-1-b\\
|W\cap E_{\sim, [i_0]}|=b
\end{subarray}}Pr(W)\label{eqn_target}
\end{eqnarray}

Equation (\ref{eqn_target}) can be computed either by Equation (\ref{eqn_way1}) when $p(t_{i_0}, \succ)>0$ or by Equation (\ref{eqn_way2}) when $p(t_{i_0}, \sim)>0$ and $b<b_{\max}$. Notice that at least one of $p(t_{i_0}, \succ)$ and $p(t_{i_0}, \sim)$ is positive, otherwise neither tuple is in the induced event relation $E^p$ according to Definition \ref{def_ied_WO}.

\begin{eqnarray}
\sum_{\begin{subarray}{l}
W\in pwd(E^p)\\
t_{i_0, \succ}\not\in W, t_{i_0, \sim}\not\in W\\
|W\cap E_{\succ, [i_0]}|=k'-1-b\\
|W\cap E_{\sim, [i_0]}|=b
\end{subarray}}Pr(W)
&=&
\frac{1-p(t_{i_0, \succ})-p(t_{i_0, \sim})}{p(t_{i_0, \succ})}
\sum_{\begin{subarray}{l}
W\in pwd(E^p), t_{i_0, \succ}\in W\\
|W\cap E_{\succ, [i_0]}|=k'-b\\
|W\cap E_{\sim, [i_0]}|=b
\end{subarray}}Pr(W)\nonumber\\
&=&\frac{1-p(t_{i_0, \succ})-p(t_{i_0, \sim})}{p(t_{i_0, \succ})}u_{\succ}(k', i_0, b).\label{eqn_way1}
\end{eqnarray}

\begin{eqnarray}
\sum_{\begin{subarray}{l}
W\in pwd(E^p)\\
t_{i_0, \succ}\not\in W, t_{i_0, \sim}\not\in W\\
|W\cap E_{\succ, [i_0]}|=k'-1-b\\
|W\cap E_{\sim, [i_0]}|=b
\end{subarray}}Pr(W)
&=&
\frac{1-p(t_{i_0, \succ})-p(t_{i_0, \sim})}{p(t_{i_0, \sim})}
\sum_{\begin{subarray}{l}
W\in pwd(E^p), t_{i_0, \sim}\in W\\
|W\cap E_{\succ, [i_0]}|=k'-1-b\\
|W\cap E_{\sim, [i_0]}|=b+1
\end{subarray}}Pr(W)\nonumber\\
&=&\frac{1-p(t_{i_0, \succ})-p(t_{i_0, \sim})}{p(t_{i_0, \sim})}u_{\sim}(k', i_0, b+1).\label{eqn_way2}
\end{eqnarray}

A subtlety is that when $p(t_{i_0},\succ)=0$ and $b=b_{\max}$, neither Equation (\ref{eqn_way1}) nor Equation (\ref{eqn_way2}) applies. However, in this case, one of the conditions in Equation (\ref{eqn_target}) is that $|W\cap E_{\sim, [i_0]}|=b=b_{\max}$, which implies $i_0=m$. Otherwise, the world $W$ does not have enough tuples from $E_{\sim}$. On the other hand, we know that $i_0\leq m-1$. Therefore, there are simply no possible worlds satisfying the condition in Equation (\ref{eqn_target}), and Equation (\ref{eqn_target}) equals $0$.

Altogether, we show that this case can be correctly computed by Equation (\ref{eqn_recursion_WO_odd}).
\item[(2)] $u_{\sim}(k',i_0+1,b)$ is the probability sum of all possible worlds $W$ such that $t_{i_0+1, \sim}\in W$, $|W\cap E_{\succ, [i_0+1]}|=k'-b$ and $|W\cap E_{\sim, [i_0+1]}|=b$.
Using a similar argument as above, it can be shown that this case is correctly computed by Equation (\ref{eqn_recursion_WO_even}) as well. 
\end{enumerate}
\end{itemize}
\end{proof}

\bibliographystyle{splncs}
\bibliography{mytopk}

\begin{thebibliography}{10}

\bibitem{DBLP:conf/icde/ZhangC08}
Zhang, X., Chomicki, J.:
\newblock On the semantics and evaluation of top-k queries in probabilistic
  databases.
\newblock In: ICDE Workshops. (2008)  556--563

\bibitem{DBLP:journals/jacm/ImielinskiL84}
Imielinski, T., Lipski, W.:
\newblock Incomplete information in relational databases.
\newblock J. ACM \textbf{31}(4) (1984)  761--791

\bibitem{DBLP:conf/vldb/CavalloP87}
Cavallo, R., Pittarelli, M.:
\newblock The theory of probabilistic databases.
\newblock In: VLDB. (1987)

\bibitem{DBLP:journals/ai/Halpern90}
Halpern, J.Y.:
\newblock An analysis of first-order logics of probability.
\newblock Artif. Intell. \textbf{46}(3) (1990)  311--350

\bibitem{abiteboul94databases}
Abiteboul, S., Hull, R., Vianu, V.:
\newblock Foundations of Databases : The Logical Level.
\newblock Addison Wesley (1994)

\bibitem{DBLP:journals/tois/FuhrR97}
Fuhr, N., R{\"o}lleke, T.:
\newblock A probabilistic relational algebra for the integration of information
  retrieval and database systems.
\newblock ACM Trans. Inf. Syst. \textbf{15}(1) (1997)  32--66

\bibitem{DBLP:journals/tcs/Zimanyi97}
Zim{\'a}nyi, E.:
\newblock Query evaluation in probabilistic relational databases.
\newblock Theor. Comput. Sci. \textbf{171}(1-2) (1997)  179--219

\bibitem{DBLP:journals/tods/LakshmananLRS97}
Lakshmanan, L.V.S., Leone, N., Ross, R.B., Subrahmanian, V.S.:
\newblock Probview: A flexible probabilistic database system.
\newblock ACM Trans. Database Syst. \textbf{22}(3) (1997)  419--469

\bibitem{DBLP:journals/vldb/DalviS07}
Dalvi, N.N., Suciu, D.:
\newblock Efficient query evaluation on probabilistic databases.
\newblock VLDB J. \textbf{16}(4) (2007)  523--544

\bibitem{DBLP:conf/vldb/BenjellounSHW06}
Benjelloun, O., Sarma, A.D., Halevy, A.Y., Widom, J.:
\newblock Uldbs: Databases with uncertainty and lineage.
\newblock In: VLDB. (2006)

\bibitem{DBLP:conf/cidr/Widom05}
Widom, J.:
\newblock Trio: A system for integrated management of data, accuracy, and
  lineage.
\newblock In: CIDR. (2005)

\bibitem{MayBMS}

\newblock http://www.infosys.uni-sb.de/projects/maybms/

\bibitem{DBLP:journals/tcs/OlteanuKA08}
Olteanu, D., Koch, C., Antova, L.:
\newblock World-set decompositions: Expressiveness and efficient algorithms.
\newblock Theor. Comput. Sci. \textbf{403}(2-3) (2008)  265--284

\bibitem{DBLP:journals/jcss/Fagin99}
Fagin, R.:
\newblock Combining fuzzy information from multiple systems.
\newblock J. Comput. Syst. Sci. \textbf{58}(1) (1999)  83--99

\bibitem{DBLP:conf/pods/FaginLN01}
Fagin, R., Lotem, A., Naor, M.:
\newblock Optimal aggregation algorithms for middleware.
\newblock In: PODS. (2001)

\bibitem{DBLP:conf/vldb/NatsevCSLV01}
Natsev, A., Chang, Y.C., Smith, J.R., Li, C.S., Vitter, J.S.:
\newblock Supporting incremental join queries on ranked inputs.
\newblock In: VLDB. (2001)

\bibitem{DBLP:journals/tods/MarianBG04}
Marian, A., Bruno, N., Gravano, L.:
\newblock Evaluating top-{\it } queries over web-accessible databases.
\newblock ACM Trans. Database Syst. \textbf{29}(2) (2004)  319--362

\bibitem{DBLP:conf/vldb/GuhaKMS04}
Guha, S., Koudas, N., Marathe, A., Srivastava, D.:
\newblock Merging the results of approximate match operations.
\newblock In: VLDB. (2004)  636--647

\bibitem{DBLP:conf/vldb/IlyasAE02}
Ilyas, I.F., Aref, W.G., Elmagarmid, A.K.:
\newblock Joining ranked inputs in practice.
\newblock In: VLDB. (2002)

\bibitem{DBLP:conf/vldb/IlyasAE03}
Ilyas, I.F., Aref, W.G., Elmagarmid, A.K.:
\newblock Supporting top-k join queries in relational databases.
\newblock In: VLDB. (2003)

\bibitem{DBLP:conf/icde/SolimanIC07}
Soliman, M.A., Ilyas, I.F., Chang, K.C.C.:
\newblock Top-k query processing in uncertain databases.
\newblock In: ICDE. (2007)

\bibitem{DBLP:journals/tods/SolimanIC08}
Soliman, M.A., Ilyas, I.F., Chang, K.C.C.:
\newblock Probabilistic top-{\it } and ranking-aggregate queries.
\newblock ACM Trans. Database Syst. \textbf{33}(3) (2008)

\bibitem{DBLP:conf/icde/ReDS07}
R\'e, C., Dalvi, N.N., Suciu, D.:
\newblock Efficient top-k query evaluation on probabilistic data.
\newblock In: ICDE. (2007)

\bibitem{DBLP:conf/sigmod/HuaPZL08}
Hua, M., Pei, J., Zhang, W., Lin, X.:
\newblock Ranking queries on uncertain data: a probabilistic threshold
  approach.
\newblock In: SIGMOD Conference. (2008)  673--686

\bibitem{DBLP:journals/tkde/BrunoW07}
Bruno, N., Wang, H.:
\newblock The threshold algorithm: From middleware systems to the relational
  engine.
\newblock IEEE Trans. Knowl. Data Eng. \textbf{19}(4) (2007)  523--537

\bibitem{DBLP:journals/vldb/BurdickDJRV07}
Burdick, D., Deshpande, P.M., Jayram, T.S., Ramakrishnan, R., Vaithyanathan,
  S.:
\newblock {OLAP} over uncertain and imprecise data.
\newblock VLDB J. \textbf{16}(1) (2007)  123--144

\bibitem{DBLP:conf/icde/YiLKS08}
Yi, K., Li, F., Kollios, G., Srivastava, D.:
\newblock Efficient processing of top-k queries in uncertain databases.
\newblock In: ICDE. (2008)  1406--1408

\end{thebibliography}

\end{document}